\documentclass[sigconf,nonacm]{acmart}

\renewcommand\footnotetextcopyrightpermission[1]{} 
\AtBeginDocument{%
  }
    
\usepackage[utf8]{inputenc}
\usepackage{color}
\usepackage{xcolor}

\usepackage{subfigure}

\definecolor{Plum}{rgb}{0.56, 0.27, 0.52}
\definecolor{MidnightBlue}{rgb}{0.1, 0.1, 0.44}
\definecolor{Gray}{rgb}{0.5, 0.5, 0.5}
\definecolor{ForestGreen}{rgb}{0.0, 0.27, 0.13}
\definecolor{Orange}{rgb}{1.0, 0.5, 0.0}

\usepackage{centernot}
\usepackage{hhline}
\usepackage{mdframed}

\usepackage{amsmath}
\usepackage{bm} 

\usepackage[scr=boondoxo]{mathalfa} 

\usepackage[group-separator={,},table-format=4.2]{siunitx}
\usepackage{textcomp}
\usepackage{caption}
\usepackage{pst-func}

\usepackage{graphicx}
\usepackage{paralist}
\usepackage{environ}

\usepackage{pgfplots}
\usepackage{pgfplotstable}
\usepackage{filecontents}
\usepackage{booktabs}
\usepackage[l3]{csvsimple}

\pgfplotsset{select coords between index/.style 2 args={
    x filter/.code={
        \ifnum\coordindex<#1\fi
        \ifnum\coordindex>#2\fi
    }
}}

\usepackage{tikz}
\usetikzlibrary{shadows}
\usetikzlibrary{shapes.misc} 
\usepackage{mathtools}
\usepackage{array,multirow}
\usepackage{algorithm}
\usepackage{url}
\usepackage{pgfplotstable}
\usepackage{threeparttable}
\usepackage{extarrows}

\usepackage{etoolbox}
\usepackage{arydshln}

\usepackage{xspace} 
 
\usepackage{dirtytalk} 
\usepackage{stmaryrd} 

\usepackage[inline,shortlabels]{enumitem} 
\newlist{inlinelist}{enumerate*}{1}
\setlist*[inlinelist,1]{%
  label=(\roman*),
}

\usepackage{xifthen}  
\usepackage{ifthen}

\usepackage{nicefrac}

\usepackage{cleveref}

\crefname{lstlisting}{listing}{listings}
\Crefname{lstlisting}{Listing}{Listings}

\usepackage{bigints}

\usepackage{lscape}

\usetikzlibrary{pgfplots.dateplot}
\usetikzlibrary{matrix,chains,positioning,decorations.pathreplacing}
\usetikzlibrary{patterns,calc,fit,arrows}
\usetikzlibrary{shapes.geometric}

\usepackage{relsize}

\usepackage[final,nomargin,inline,index]{fixme} 
\fxusetheme{color}

\FXRegisterAuthor{bart}{anbart}{\color{magenta} {\underline{bart}}}
\FXRegisterAuthor{ric}{anric}{\color{red} {\underline{ric}}}
\FXRegisterAuthor{zun}{anzun}{\color{blue} {\underline{zun}}}

\usepackage[disable,textsize=scriptsize, backgroundcolor=red!20!white,bordercolor=red,linecolor=red]{todonotes}

\hypersetup{
  breaklinks   = true,
  colorlinks   = true, 
  urlcolor     = blue, 
  linkcolor    = blue, 
  citecolor    = red   
}
\hypersetup{final} 

\usepackage{listings}
\usepackage{lstautogobble} 
\definecolor{listingBG}{HTML}{FFFFCB}%
\definecolor{listingFrame}{HTML}{BBBB98}%
\definecolor{listingLineno}{rgb}{0.5,0.5,1.0}%

\usepackage{tcolorbox}

\definecolor{LightGrey}{rgb}{0.975,0.975,0.975}

\lstdefinelanguage{illum}{
	commentstyle=\color{Gray},
	morecomment=[l]{//},
	morecomment=[s]{/*}{*/},
	classoffset=0,
        escapechar=\$,
        tabsize=2,
	morekeywords={precond_wallet,after,precond_if,auth,process},
	keywordstyle=\color{Plum}\bfseries,
	classoffset=1,
	morekeywords={clause,call,send},
	keywordstyle=\color{NavyBlue}\bfseries,
	classoffset=2,
	morekeywords={int,string,bool,address,uint,mapping},
	keywordstyle=\color{MidnightBlue}\bfseries,
	basicstyle=\fontseries{m}\normalsize\ttfamily
	\lst@ifdisplaystyle\scriptsize\fi,
}

\lstdefinelanguage{hellum}{
	commentstyle=\color{Gray},
	morecomment=[l]{//},
	morecomment=[s]{/*}{*/},
	classoffset=0,
        escapechar=\$,
        tabsize=2,
        literate={\ \ }{{\ }}1,
	morekeywords={contract,constructor,function,if,else,require},
	keywordstyle=\color{red},
	classoffset=1,
	morekeywords={balance,after,input,next,auth,transfer,view},
	keywordstyle=\color{red},
	classoffset=2,
	morekeywords={int,string,bool,address,uint,mapping},
	keywordstyle=\color{blue},
 	classoffset=3,
  	keywordstyle=\color{Plum}\bfseries,
	basicstyle=\fontseries{m}\normalsize\ttfamily
        \lst@ifdisplaystyle\scriptsize\fi,
}

\lstdefinelanguage{rulecode}{
	autogobble=true,
	commentstyle=\color{Gray},
	morecomment=[l]{//},
	morecomment=[s]{/*}{*/},
	classoffset=0,
        escapechar=\$,
        tabsize=2,
	morekeywords={after,before,send,require,receive},
	keywordstyle=\color{red},
	classoffset=1,
	morekeywords={contract,bal,if,else,then,and,or,not,hash,len,substr,toStr,validFrom,validTo,signedBy},
	keywordstyle=\color{blue},
	classoffset=2,
	morekeywords={Rule,Precond,Effects},
	keywordstyle=\color{Plum}\bfseries,
 	classoffset=3,
	morekeywords = {InitialState,InitialBalance},
  	keywordstyle=\color{ForestGreen}\bfseries,
	classoffset=4,
	morekeywords={int,bitstring,map,var,bool,address,const},
	keywordstyle=\color{blue},
    classoffset=5, 
    keywordstyle = \color{MidnightBlue}\bfseries, 
	basicstyle=\fontseries{m}\normalsize\ttfamily
        \lst@ifdisplaystyle\footnotesize\fi,
}

\newcommand{\ifempty}[3]{%
  \ifthenelse{\isempty{#1}}{#2}{#3}%
}

\newcommand{\ifdots}[3]{%
  \ifthenelse{\equal{#1}{...}}{#2}{#3}%
}

\newcommand{\hidden}[1]{}

\newcommand{\rulelang}{hURF\xspace}
\newcommand{\toolname}{\textsc{DiSCo\_sim}\xspace}

\newcommand{\hutxo}{hUTXO\xspace}

\newcommand{\ruleinline}[1]{\mbox{\lstinline[language=rulecode,basicstyle=\fontseries{m}\small\ttfamily]{#1}}}
\newcommand{\ruleinlinesmall}[1]{\mbox{\lstinline[language=rulecode,basicstyle=\fontseries{m}\scriptsize\ttfamily]{#1}}}

\renewcommand{\vec}[1]{\boldsymbol{#1}}

\newcommand{\Real}[1]{\mathrm{Real}}

\newcommand{\codefont}{\fontsize{9}{9}\selectfont}
\newcommand{\code}[1]{{\tt\codefont {#1}}}

\newcommand{\Eg}{E.g.\@\xspace}
\newcommand{\eg}{e.g.\@\xspace}
\newcommand{\ie}{i.e.\@\xspace}
\newcommand{\wrt}{w.r.t.\@\xspace}

\newcommand{\emptyseq}{\varepsilon}

  {}

\newcommand{\BTC}{\textup{%
  \leavevmode
  \vtop{\offinterlineskip 
    \setbox0=\hbox{B}%
    \setbox2=\hbox to\wd0{\hfil\hskip-.03em
    \vrule height .3ex width .15ex\hskip .08em
    \vrule height .3ex width .15ex\hfil}
    \vbox{\copy2\box0}\box2}}\xspace}

\def\pmvColor{\color{ForestGreen}}
\newcommand{\pmvFmt}[1]{{\pmvColor{\sf #1}}}

\newcommand{\pmv}[2][]{\pmvFmt{#2}_{\pmvColor{#1}}\xspace}
\newcommand{\pmvA}[1][]{\pmv[{#1}]{A}}

\def\fieldColor{\color{Plum}}
\newcommand{\txTag}[3][]{{\fieldColor\sf #3}\ifempty{#1}{\ifempty{#2}{}{: {#2}}}{[{#1}]\ifempty{#2}{}{: {#2}}}}

\newcommand{\txOut}[2][]{\txTag[{#1}]{#2}{out}}

\newcommand{\txf}[1]{\mathit{#1}}

\newcommand{\txscript}{\txTag{}{script}}

\def\txColor{\color{MidnightBlue}}
\def\fieldColor{\color{Plum}}
\newcommand{\txFmt}[1]{{\txColor{\sf #1}}}

\newcommand{\tx}[2][]{\txFmt{#2}_{\txColor{#1}}}
\newcommand{\txT}[1][]{\tx[#1]{T}}

\newcommand{\txTi}[1][]{\txFmt{T'_{\txColor{{\it #1}}}}}

\DeclareMathAlphabet{\mathbfsf}{\encodingdefault}{\sfdefault}{bx}{n}

\newcommand{\bcB}{\mathcal{B}}
\newcommand{\accA}[1][]{\mathcal{A}\ifempty{#1}{}{({#1})}}
\newcommand{\accAi}[1][]{\mathcal{A'}\ifempty{#1}{}{({#1})}}

\newcommand{\eqdef}{\triangleq}

\definecolor{LightGrey}{rgb}{0.95,0.95,0.95}
\definecolor{keyword}{HTML}{7F0055}

\def\tokColor{\color{Orange}}
\newcommand{\tokFmt}[1]{{ \mathrm{\tokColor{#1}} }}
\newcommand{\tok}[2][]{\tokFmt{#2}_{\tokColor{#1}}\xspace}
\newcommand{\tokT}[1][]{\tok[{#1}]{T}}

\DeclareMathSymbol{:}{\mathpunct}{operators}{"3A}

\newcommand{\waltok}[2]{#1:#2}
\newcommand{\walenum}[1]{[#1]}

\newlength\replength
\newcommand\repfrac{.1}

\newcommand\rulewidth{.6pt}
\newcommand\tdashfill[1][\repfrac]{\cleaders\hbox to \replength{%
  \smash{\rule[\arraystretch\ht\strutbox]{\repfrac\replength}{\rulewidth}}}\hfill}

\newcommand\tdotfill[1][\repfrac]{\cleaders\hbox to \replength{%
  \smash{\raisebox{\arraystretch\dimexpr\ht\strutbox-.1ex\relax}{.}}}\hfill}

\newcommand{\rtx}{{\sf rtx}}

\newcommand{\txo}{{\fieldColor{\sf o}}} 
\newcommand{\txoi}{{\fieldColor{\sf o}'}} 

\newcommand{\true}{\mathit{true}}
\newcommand{\false}{\mathit{false}}

\newcommand{\contrAdvC}[2]{\mathcal{C}} 

\newcommand{\confG}[1][]{\Gamma_{#1}}

\newcommand{\confContr}[3][]{\langle {#2}, {#3} \rangle_{#1}}
\newcommand{\confDep}[3][]{\langle {#2}, {#3} \rangle_{#1}}

\newcommand{\cstate}[2][]{\sigma_{{#1}}\ifempty{#2}{}{({#2})}}

\newcommand{\cstatei}[2][]{\sigma'_{{#1}}\ifempty{#2}{}{({#2})}}

\newcommand{\ctrId}[1][]{\gamma_{#1}}
\newcommand{\cstateOut}[2][]{\Sigma_{{#1}}\ifempty{#2}{}{({#2})}}
\newcommand{\cstateOuti}[2][]{\Sigma'_{{#1}}\ifempty{#2}{}{({#2})}}

\newcommand{\bal}[1][]{w_{#1}}
\newcommand{\bali}[1][]{w'_{#1}}
\newcommand{\ruleR}{\mathrm{R}}

\newcommand{\minHash}{\mathrm{min}_H}
\newcommand{\maxHash}{\mathrm{max}_H}

\newcommand{\outMap}{\mathfrak{Out}}
\newcommand{\outMapi}{\mathfrak{Out}^*}

\newcommand{\type}[1]{ {\sf \fieldColor {#1} } } 
\renewcommand{\eqdef}{\stackrel{\text{def}}{=}}

\usepackage{amsthm}
\theoremstyle{plain}

\theoremstyle{definition}

\newtoggle{anonymous}
\togglefalse{anonymous}

\newtoggle{arxiv}
\toggletrue{arxiv}

\iftoggle{arxiv}{\fancyfoot{}}{}

\newtoggle{itasec25}
\togglefalse{itasec25}

\iftoggle{arxiv}{
  \usepackage{hyperref}
  \usepackage[hyphenbreaks]{breakurl}
}{}

\makeatletter
\renewcommand\paragraph{\@startsection{paragraph}{4}{\z@}%
  {2.25ex \@plus 1ex \@minus .2ex}%
  {-0.75em}%
  {\normalfont\normalsize\bfseries}}
\makeatother

\begin{document}

\iftoggle{arxiv}{\thispagestyle{empty}}{}


\title{\!\!Scalable UTXO Smart Contracts via Fine-Grained Distributed State}

\iftoggle{anonymous}{
\author{Anonymous authors}
\affiliation{\institution{Anonymous institution} \city{Anonymous city} \country{Anonymous country}}
}{
\author{Massimo Bartoletti}
\affiliation{%
  \institution{University of Cagliari}
  \city{Cagliari}
  \country{Italy}
}
\email{bart@unica.it}
\author{Riccardo Marchesin}
\affiliation{%
  \institution{University of Trento}
  \city{Trento}
  \country{Italy}
}
\email{riccardo.marchesin@unitn.it}
\author{Roberto Zunino}
\affiliation{%
  \institution{University of Trento}
  \city{Trento}
  \country{Italy}
}
\email{roberto.zunino@unitn.it}

}



\begin{abstract}

UTXO-based smart contract platforms face an efficiency bottleneck, in that any transaction sent to a contract must specify the entire updated contract state.
This requirement becomes particularly burdensome when the contract state contains dynamic data structures, as needed in many use cases to track interactions between users and the contract.
The problem is twofold: on the one hand, a large state in transactions implies a large transaction fee; on the other hand, a large centralized state is detrimental to the parallelization of transactions --- a feature that is often cited as a key advantage of UTXO-based blockchains over account-based ones. 
We propose a technique to efficiently execute smart contracts on an extended UTXO blockchain, which allows the contract state to be distributed across multiple UTXOs. 
In this way, transactions only need to specify the part of the state they need to access, reducing their size (and fees). 
We show how to exploit our model to parallelize the validation of transactions on multi-core CPUs.
We implement our technique and provide an empirical validation of its effectiveness. 
\end{abstract}



\maketitle

\iftoggle{arxiv}{
\section{Introduction}
\label{sec:intro}

Current smart contract platforms can be roughly partitioned in two main classes: account-based (\eg, Ethereum) and UTXO-based (\eg,  Cardano).
Besides the well-known differences in the programming style of their smart contracts~\cite{Brunjes20isola}, these two paradigms have fundamental differences that affect the performance (\eg, the transaction throughput), costs and scalability of smart contracts.
A well-known weakness of the account-based model \emph{\`a la} Ethereum is that it hinders the parallelization of validator nodes~\cite{Garamvolgyi22icse}. 
Indeed, detecting when two transactions are parallelizable is a hard problem, and approaches that try to improve the node performance require complex techniques based on optimistic execution~\cite{Dickerson17podc,Anjana19pdp,Androulaki18eurosys,Gelashvili23ppopp,Zhang23icpp} or static analysis~\cite{BGM21lmcs,Pirlea21pldi}. 
Not being able to efficiently use the computational resources of validators is a problem, since validators perform the vast majority of work in contract executions.
Parallel validation is instead a significant selling point of UTXO-based blockchains.
There, detecting when two transactions are parallelizable is trivial: just check that their inputs satisfy the Bernstein conditions~\cite{BGM21lmcs}.
In other words, when one transaction spends an output, we need to ensure that the same output must not be accessed (either spent or read) by the other transaction.
However, in smart contract platforms based on the UTXO model, like Cardano, smart contracts are typically implemented as a \emph{single} transaction output,
encoding both the contract logic and the current contract state.
In this way, actions on the same contract are \emph{never} parallelizable, since they attempt to spend the same output.
Another problem is that any transactions acting on a contract must include the \emph{whole} updated contract state. When this gets large (\eg, for contracts storing user data in maps), it becomes a performance bottleneck. Besides that, transaction size is usually capped (\eg, Cardano has a 16KB hard cap~\cite{CardanoFeeEstimator}), so one can easily reach a point when further contract updates are forbidden (this could also be exploited by DoS attacks). 
Although the blockchain size could be reduced by storing on-chain only the hash of the contract state~\cite{CIP32}, 
validators still need access to the entire state corresponding to such hashes, 
which still requires a significant amount of network communications.
In summary, smart contracts in the UTXO model currently have poor scalability when the contract state becomes large.

\paragraph{Distributing the contract state}
In this paper we propose a framework to improve the performance and scalability of contracts in the UTXO model.
The key idea is to \emph{distribute the contract state over multiple UTXOs}. 
As a basic example, if the contract state contains two variables, we store their values in distinct UTXOs: in this way, contract actions accessing different variables can be parallelized. 
More in general, we distribute arbitrary contract states, including those comprising unbounded data structures.
This is done in a fine-grained way: \eg, key-value maps are distributed point-wise \wrt their finite support.
While distributing the contract state has a clear advantage in parallelization, maintaining the \emph{contract balance} in such distributed approach is problematic. Of course, even if the state is distributed, we still could store the contract balance in a single UTXO, but this would introduce a single point of synchronization, hindering parallelization. 
Alternatively, we could split the balance across multiple UTXOs, but this would not completely solve the problem.
For instance, consider a contract with a balance of 2 tokens, split in two distinct UTXOs. 
If we want to perform two actions, each withdrawing 1 token from the contract, it may happen that both of them attempt to spend the same UTXO, hence causing a conflict, hindering parallelization again.
This problem can manifest itself more frequently when a withdrawal requires spending several UTXOs of lesser value; indeed, conflicts --- which arise when two actions attempt to spend the same output --- become more likely when the actions need to consume many UTXOs.
To address this problem, we introduce a new UTXO design, which separates the contract state (stored in UTXOs) from the contract balance (which is stored in an account-based fashion).
In this way, withdrawing from the contract (\ie from its account) never causes conflicts, provided the balance is sufficient.
We dub our model \emph{hybrid UTXO} (\hutxo), since it combines features of both models.

\paragraph{Securing distributed contracts}
While distributing the contract state and isolating the balance 
enables the parallel validation of transactions, this additional complexity can expose contracts to new kinds of attacks.
Recall that in UTXO-based models, an adversary can always create a new UTXO with arbitrary fields (including the token amount, as long as the adversary spends the needed tokens).
This poses no issue to centralized contracts that involve a single UTXO carrying both the data and the contract balance: indeed, the behaviour of that UTXO is unaffected by the presence of other UTXOs the adversary could create.
By contrast, in distributed contracts, where the contract state is encoded in multiple UTXOs with a small monetary value, an adversary can cheaply forge UTXOs so to corrupt the contract state (see~\Cref{sec:overview}). 

%
To thwart forgery attacks, our \hutxo model proposes a novel combination of techniques:
\begin{itemize}

\item We introduce \emph{contract IDs}: the same ID is shared by all the UTXOs contributing to the contract state and by the associated balance account. 

\item \hutxo validators ensure that new transactions can spawn new UTXOs with pre-existing contract IDs \emph{only} when they are vetted by a script of a previous UTXO already linked to that ID.
In this way, only descendants of a contract-related UTXO can claim to represent a part of the contract state.

\item Finally, we use scripts to ensure that the new UTXOs encode a new state as mandated by the contract logic.

\end{itemize}

\paragraph{A secure compiler to \hutxo}

The last item above highlights a possible issue: since programming contract scripts of UTXO transactions is notoriously error-prone~\cite{Rosetta25fgcs}, developers must be extremely careful not to introduce vulnerabilities in the transaction scripts.
To address this issue, we propose a high-level contract language, dubbed \rulelang (for \hutxo Rule Format), and a compiler from \rulelang to \hutxo transactions. 
\rulelang abstracts from the \hutxo transaction structure, modelling contracts as sets of rules reading and updating the contract state. 
A contract rule can verify a precondition on the contract state, update the state by performing a sequence of variable/map assignments, 
and transfer currency from/to users. 
%
The \rulelang compiler guarantees that the scripts in the \hutxo transactions resulting from compilation enable \emph{exactly} the state/balance updates specified by the \rulelang rules.
Overall, \rulelang relieves developers from worrying about \hutxo-level attacks, so that they can instead focus on the (high-level) contract logic.

\paragraph{Contributions}
In summary, our main contributions are:
\begin{itemize}

\item A new hybrid blockchain model, \hutxo, which extends the UTXO model with Cardano-like contract states, contract IDs, and contract balance accounts.
The transaction validation conditions of \hutxo support the secure distribution of  contract states\iftoggle{arxiv}{ (\Cref{sec:blockchain})}{}. 

\item A high-level contract language, \rulelang, with a secure compiler to \hutxo\iftoggle{arxiv}{ transactions (\Cref{sec:smart-contracts,sec:compilation})}{}.

\item A parallel algorithm for validating transactions blocks in the \hutxo model\iftoggle{arxiv}{ (\Cref{sec:parallelization})}{}.

\item A prototype implementation of a \hutxo validator, featuring both sequential and parallel validation, and an evaluation on a benchmark of contracts inspired by real use cases like crowdfund, registry and multisig wallet\iftoggle{arxiv}{ (\Cref{sec:experiments})}{}. 
Our results show that the parallel validator almost always achieves a speedup close to the number of available threads. 
Furthermore, the distribution of the contract state results in a time and space speedup that grows linearly with the size of the state. This is coherent with our expectations, since in the centralized-state contract, the larger the state, the more time it takes to duplicate it in the redeeming transaction.
Notably, the size of transactions in the distributed-state contract is constant, making it possible to execute contracts with huge dynamic state even in blockchains where the size of transactions is strictly limited (such as, for instance, in Cardano).

\item To foster the reproducibility of our experiments, we have made our simulator and benchmarks publicly available as a Docker image on Zenodo.%
\footnote{\url{https://doi.org/10.5281/zenodo.13889967}}

\end{itemize}

\section{Overview}
\label{sec:overview}

We now present our framework, leveraging on examples to illustrate its main ideas. 
\iftoggle{arxiv}{}{Because of space constraints, we defer the full technical treatment to the appendices.}

\paragraph{Background: the eUTXO model}

In the UTXO model, transaction \emph{outputs} include an amount of assets and a \textit{spending condition}, \ie a script stating how they can be redeemed by another transaction. 
Transaction \emph{inputs}, in turn, refer to unspent outputs of previous transactions, and provide the needed data (the \textit{redeemer}) to unlock the spending conditions of the referenced output. 
Each unspent output (UTXO) can only be consumed once, as a whole, by exactly one input. 
The blockchain state is given by the set of UTXOs. 
While in the basic UTXO model \emph{\`a la} Bitcoin spending conditions are quite limited~\cite{bitcointxm}, in the \emph{extended} UTXO model (eUTXO~\cite{Chakravarty20wtsc}) they can be arbitrary scripts (in a $\txf{validator}$ field). Additionally, transactions outputs can carry arbitrary data (in a $\txf{datum}$ field),
and transactions include a list of their $\txf{signers}$.

\newcommand{\txMiniA}{
  \scalebox{0.6}{
    \hspace{-5pt}
      \begin{tabular}[t]{|l|}
      \hline
      \\[-9pt]
      \multicolumn{1}{|c|}{$\txT[1]$} \\[1pt]
      \hline
      $\cdots$ \\[1pt]
      \hline
      $\txf{out}[0]$ : \\[1pt]
      \hspace{5pt} $\txf{validator} = \pmvA\;\code{in}\;\rtx.\txf{signers}$ \\[1pt]
      \hspace{5pt} $\txf{value} = 1:\tokT$ \\[1pt]
      \hline
    \end{tabular}}
}

\newcommand{\txMiniB}{
  \scalebox{0.6}{
    \hspace{-5pt}    
      \begin{tabular}[t]{|l|}
      \hline
      \\[-9pt]
      \multicolumn{1}{|c|}{$\txT[2]$} \\[1pt]
      \hline
      $\txf{in}[0]${\,}: \\[1pt]
      \hspace{5pt} $\txf{outRef} = \txT[1].\txf{out}[0]{}$    
      \\[1pt]
      \hline
      $\txf{signers} = [\pmvA]$ \\[1pt]
      \hline
      $\txf{out}[0]${\,}: \\[1pt]
      \hspace{5pt} $\txf{validator} =
      \begin{array}{l}\code{owner}\;\code{in}\;\rtx.\txf{signers} \; \code{and} \\
                      \rtx.\txOut[0]{}.\txf{validator} == \txf{validator} \; \code{and} \\ 
                      \rtx.\txf{out}[0].\txf{value} == \txf{value}
                      \end{array}$
                     \\[15pt]
      \hspace{5pt} $\txf{datum} = \{ \code{owner}:\pmvA \}$ \\[3pt]
      \hspace{5pt} $\txf{value} = 1:\tokT$ \\[1pt]
      \hline
    \end{tabular}}
}

\tikzstyle{tx} = [rectangle, minimum width=0cm, minimum height=1cm,text centered, draw=black, fill=gray!10, draw=white, inner sep=0, outer sep=0]

\iftoggle{arxiv}{%
For example, the transaction $\txT[1]$ has a single unspent output holding one token, noted $1:\tokT$. Its script requires the redeeming transaction ($\rtx$) to carry $\pmvA$'s signature:
\begin{center}
  \begin{tikzpicture}[node distance=2cm]
    \node (tx1) [tx] {\txMiniA};
  \end{tikzpicture}
\end{center}

To spend $\txT[1]$'s single output (noted $\txT[1].\txf{out}[0]$), a redeeming transaction must refer to $\txT[1]$ from its inputs and validate the script by having $\pmvA$ as (one of) its signers. This is accomplished by the transaction $\txT[2]$ below: 
\begin{center}
  \begin{tikzpicture}[node distance=2cm]
    \node (tx2) [tx] {\txMiniB};
  \end{tikzpicture}
\end{center}

$\txT[2]$ is signed by $\pmvA$ and its script checks that
\begin{inlinelist}
\item the redeeming transaction is signed by the user stored in the \code{datum} field of the current transaction;
\item the script and the value in the redeeming transaction are the same as in $\txT[2]$.
\end{inlinelist}
Note that, although any transaction redeeming $\txT[2]$ must preserve its script and value, it can change the \code{owner}. 
In a sense, the script implements a non-fungible token (NFT): to change the ownership of the NFT, the current owner must spend the output with a new transaction (signed by herself) that specifies the new owner.
}
{}

\paragraph{A crowdfund contract in eUTXO}

To illustrate our technique, we consider a crowdfund contract with the following workflow. First, donors can deposit any amount of tokens in the contract. Then, after a first deadline, the contract owner can withdraw all the deposited tokens, provided that the total amount is greater than the required goal.
After a second deadline, the donors can take their donations back, if they are still within the contract.

This contract can be implemented in the eUTXO model by storing the whole map of donations in the $\txf{datum}$ field of a single transaction output $\txo$.
Using this approach, making a donation amounts to spending the output $\txo$ and creating a new one $\txoi$ whose $\txf{datum}$ has the same map as $\txo$ except for the donor entry, which is increased by the donation amount. After the first deadline, the contract owner has a chance to withdraw the funds, if there are enough, by spending the last output.
After the second deadline, if the funds are still in the contract, each donor can take their part back, according to the value stored in the map. Doing so requires spending the last output $\txo$ and creating a new one $\txoi$ that has the same $\txf{datum}$ map except that the entry for the donor has been removed.

This contract implementation is, in fact, \emph{centralized}: the whole contract state is stored in a single output. This design choice has three main downsides:
\begin{itemize}
    \item Each donation transaction contains a copy of the old contract state with small changes. With $N$ distinct donors, the state map becomes larger and larger, requiring $O(N^2)$ total blockchain space. This is overly inefficient.
    \item When multiple donors attempt to deposit at the same time, they all have to spend the same output $\txo$ --- leading to \emph{UTXO congestion}~\cite{cardano-utxocongestion}. Since the eUTXO model prevents double spending, all these operations are in conflict and race against each other. Consequently, only one donation will succeed, forcing the other donors to retry with a new transaction.
    \item This makes state updates inherently \emph{sequential}, as they rely on a chain of transactions redeeming each others' outputs. This hinders the parallel validation of transactions.
\end{itemize}

\paragraph{Distributing the contract state}
\label{page:state-updates}

The key idea to overcome these issues is to distribute the contract state across multiple UTXOs. For example, in our crowdfund contract, the state is a map associating participant addresses $p_i$ to their donated amount $a_i \neq 0$. 
We represent it as follows:
\[
[p_1 \mapsto a_1,\, \ldots,\, p_n \mapsto a_n]
\]
The participants not occurring in the above map representation are implicitly associated with the default value \emph{zero}.
To distribute the map, we exploit the standard total ordering of addresses.
For simplicity, assume that $p_{\sf min} < p_1 < \cdots < p_n < p_{\sf max}$, where $p_{\sf min}$ and $p_{\sf max}$ correspond to the minimum and maximum addresses, respectively.
Consequently, we can now denote the state map using the following alternative representation:
\begin{multline*}
(p_{\sf min},p_1)
[p_1 \mapsto a_1]
(p_1,p_2)
[p_2 \mapsto a_2]
(p_2,p_3)
\; \ldots
\\
\; \ldots
(p_{n-1},p_n)
[p_n \mapsto a_n]
(p_n,p_{\sf max})
\end{multline*}
In the new representation, an item $[p_i \mapsto a_i]$ denotes that participant $p_i$ is associated with $a_i \neq 0$.
An item $(p_j, p_k)$ denotes that all the participants in that \emph{open interval} are associated to the default value zero.
To distribute the state, we store each item in its own output. For instance, item $[p_2 \mapsto a_2]$ is stored in the $\txf{datum}$ field of an output $\txo_{p_2}$, while item $(p_2, p_3)$ is stored in the $\txf{datum}$ field of another output $\txo_{p_2,p_3}$.
%

This state distribution makes operating on the contract more complex. 
If participant $p_i$ has already donated $a_i$ tokens, they can donate additional $b_i$ tokens by consuming the output $\txo_{p_i}$ (representing $[p_i \mapsto a_i]$)
and creating a new output $\txo_{p_i}'$ (representing $[p_i \mapsto a_i+b_i]$).
If, instead, $p_i$ has not donated yet, they can donate $b_i$ tokens by \emph{splitting} an open interval. This requires $p_i$ to find the open interval $(p_j,p_k) \ni p_i$ represented by output $\txo_{p_j,p_k}$, consume it, and create \emph{three} new outputs $\txoi_{p_j,p_i}\txoi_{p_i}\txoi_{p_i,p_k}$ to represent the items $(p_j,p_i)[p_i \mapsto b_i](p_i,p_k)$.
Similarly, to take their donation back after the second deadline, $p_i$ must \emph{merge} three items into a single interval item. This requires $p_i$ to find the three outputs $\txo_{p_j,p_i}\txo_{p_i}\txo_{p_i,p_k}$ representing the items $(p_j,p_i)[p_i \mapsto a_i](p_i,p_k)$, spend them, and create a new output $\txo_{p_j,p_k}$
for the new open interval $(p_j,p_k)$.
This effectively updates the map so that $p_i$ is now associated to the default value zero.

This distributed implementation of the contract solves all the issues of the centralized one discussed before, in that:
\begin{itemize}
    \item The $\txf{datum}$ fields now contain a bounded amount of data (one item, $O(1)$), and each operation consumes and creates at most three outputs. Hence, performing $N$ donations only requires $O(N)$ space instead of $O(N^2)$ as in the centralized implementation.
    \item When multiple participants attempt to donate at the same time, only those operations falling in the same interval cause conflicts. After a few donations, the contract state becomes fragmented in several intervals, gradually reducing the conflict probability.
    \item Fewer conflicts result in more transactions with disjoint outputs, creating greater opportunities for parallel validation.
\end{itemize}

\paragraph{Handling more complex states}

In general, the state of a contract does not involve a single map as shown in the previous example, but can comprise multiple maps and variables.
However, the general case can be reduced to the specific single-map case, by a simple \emph{flattening} of the state. For instance, in order to represent maps $\ruleinline{m1}, \ruleinline{m2}$ and variables $\ruleinline{x1},\ruleinline{x2}$, we can use
\[
\begin{array}{l@{\,}l}
  [ &
  \ruleinline{hash("map_m1[12]")} \mapsto v_1,\, \ldots,\, 
  \ruleinline{hash("map_m1[45]")} \mapsto v_2,
  \\ &
  \ruleinline{hash("map_m2[54]")} \mapsto v_3,\, \ldots,\,
  \ruleinline{hash("map_m2[89]")} \mapsto v_4,
  \\ &
  \ruleinline{hash("var_x1")} \mapsto v_5,\, 
  \ruleinline{hash("var_x2")} \mapsto v_6 \,
  ]    
\end{array}
\]
where \ruleinline{hash} is a collision-resistant hash. 
Using hashes as done above ensures the global map keys have constant size. It also helps in spreading the keys in an almost-uniform way over the key space.

\paragraph{Handling tokens}

In the previous discussion of the distributed approach we did not mention how to store the contract balance. While in the centralized approach there is only a  natural choice --- storing all the tokens in the single contract output --- in the distributed contract there is no straightforward solution.

One option is to continue storing the whole balance in a single ``balance output'', even in the distributed case. This, however, would re-introduce conflicts between any pair of operations affecting the contract balance, since they would need to spend the same balance output.
Since this would negate many of the benefits of the distributed approach, we must look for other options.

Another option is to use multiple ``balance outputs'', spreading the balance over multiple UTXOs. This would reduce conflicts since operations spending disjoint outputs can now be executed in parallel.
Still, this solution is not ideal. For instance, when transferring 100 tokens from a contract, one has to consume one or more balance outputs to reach such amount.
Depending on the actual fragmentation of the tokens over balance outputs, it might be impossible to reach the exact amount of 100 tokens, forcing the operation to consume more and then put the exceeding tokens back in the contract by creating a new balance output.
This can increase conflicts, as more tokens need to be consumed than the operation actually requires.
Another drawback is that one must consume a variable amount of balance outputs, making the contract logic more complex.
Finally, since each transaction attempts to consume many balance outputs, it becomes more unlikely for different transactions in the mempool to have disjoint sets of inputs, so increasing the probability of conflicts.
We stress that a single shared input would be enough to cause a conflict between two transactions.

Our hybrid accounting model \hutxo avoids these issues by distributing the contract data across several UTXOs, while keeping the contract balance into a \emph{contract account}, outside of the UTXOs.
By storing the contract balance in a separate account, we effectively make the contract tokens indistinguishable from each other: when transferring  100 tokens from the contract we no longer have to specify \emph{which} 100 tokens (or more) to consume, as we must do in the regular UTXO model.
Using our approach, we will never have to consume more than 100 tokens and  can always run a token transfer in parallel with others, provided there is enough balance in the contract.

Crucially, \hutxo contract accounts only contain tokens, unlike stateful account-based blockchains \emph{\`a la} Ethereum, where contract accounts also store the contract state and code.
We remark that in \hutxo, tokens can be kept both in contract accounts and in UTXOs, while contract data is only stored in the $\txf{datum}$ fields of UTXOs, as in the eUTXO model.
When a UTXO contains distributed contract data, we will not store tokens inside it, preferring instead to deposit them in the contract account.


\paragraph{Transaction validity in \hutxo}

As anticipated in~\Cref{sec:intro}, the distributed approach has several benefits over the centralized one, but there is a drawback: in UTXO-based models, the adversary can freely create outputs with arbitrary $\txf{datum}$ fields, and this makes distributed contracts potentially vulnerable to attacks that are not present in the centralized approach.
%
%
%

To illustrate the attack, consider the crowdfund contract in a state where the adversary $p_A$ has donated $1$ token. 
This state is rendered by items
\mbox{$\cdots(p_j,p_A) [p_A \mapsto 1] (p_A,p_k)\cdots$}
and outputs
$\cdots \txo_{p_j,p_A} \txo_{p_A} \txo_{p_A,p_k} \cdots$
Now, the adversary might forge an output $\hat\txo_{p_A}$ whose $\txf{datum}$ field denotes the item $[p_A \mapsto 100]$.
Note that $\hat\txo_{p_A}$ (like $\txo_{p_A}$) does not hold tokens but only data, so the forgery costs the adversary nothing (other than transaction fees).
By exploiting this forgery, once the deadlines have passed the adversary can perform a refund operation by spending two contract outputs and their own forged output, as in
$\txo_{p_j,p_A}\hat\txo_{p_A}\txo_{p_A,p_k}$, and withdraw in this way $100$ tokens from the contract balance.
This is a clear attack, as it allows the adversary to receive more tokens than they donated.

In the eUTXO model it is hard to protect the distributed crowdfund contract from this attack, since transaction scripts cannot distinguish the forged outputs from the legit ones.
Our \hutxo model protects contracts from this vulnerability by exploiting \emph{contract IDs}: we assign to the distributed \todo{crowdfund?}crowdfund contract a unique ID, which is then used to mark legit UTXOs, \ie those storing valid data for that contract.
When considering a transaction for inclusion in the blockchain, \hutxo validators enforce conditions that prevent the adversary from forging UTXOs with the same ID.
Intuitively, validators only accept a transaction $\txT$ if it falls in one of the following cases%
\iftoggle{arxiv}
{}%
{ (see~\Cref{sec:blockchain} for the actual rules)}%
:
\begin{itemize}

    \item $\txT$ deploys a new contract, creating outputs with a \emph{fresh} contract ID.
    In this way, the adversary can still create an output
    $\hat\txo_{p_A}$ with a forged $\txf{datum}$ field, but only with a distinct contract ID. This prevents the attack, since when redeeming $\txo_{p_j,p_A}\hat\txo_{p_A}\txo_{p_A,p_k}$ the scripts of the two legit outputs can detect that $\hat\txo_{p_A}$ does not belong to the same contract, and reject the transaction.    

    \item $\txT$ performs an operation on an existing contract, spawning new outputs with that contract ID, but only when some input of $\txT$ refers to a pre-existing UTXO with the same contract ID. 
    In this way, the adversary is allowed to create an output
    $\hat\txo_{p_A}$ having the same contract ID of the distributed crowdfund contract. However, its $\txf{datum}$ field must be vetted by the script of the pre-existing contract UTXO. 
    Such contract script can be designed so to accept those $\txf{datum}$s that respect the contract logic. Doing so thwarts the attack, since the adversary can increase the amount of donated tokens in the donations map only if they are actually paying them.

    \item $\txT$ performs an operation not involving contract IDs in any way (\eg, simple token transfers).
    In this way, the adversary can create an output
    $\hat\txo_{p_A}$ with any $\txf{datum}$ field, but only with no associated contract ID.
    As in the first case, such $\hat\txo_{p_A}$ cannot interfere with the contract, preventing the attack.
\end{itemize}

\iftoggle{arxiv}
{The actual validation rules are detailed in~\Cref{sec:blockchain}.}
{}

\paragraph{\rulelang: a language for distributed \hutxo contracts}

%
Distributed contracts have multiple outputs, whose scripts have to precisely interact so to make the intended contract behaviour emerge.
Manually developing these scripts is overwhelmingly error-prone: indeed, distributing the state adds a significant degree of complexity on top of the centralized logic, which is already rather complex to implement in eUTXO~\cite{Rosetta25fgcs}.
Our rule-based language \rulelang simplifies contract development in the distributed approach.
\rulelang rules define the behaviour of the \emph{whole} contract, abstracting from how the contract logic and state are scattered across multiple blockchain-level outputs.
The \rulelang \emph{compiler} reliably generates the (multiple) underlying scripts.

We illustrate \rulelang for our crowdfund contract in~\Cref{fig:crowdfunding-hurf}, deferring its technical treatment to~\Cref{sec:hurf}.
%
The contract starts by listing all its state variables and maps. Map $\ruleinline{m}$ tracks the amount of tokens $\ruleinline{m[a]}$ that have been donated by each donor $\ruleinline{a}$.
Variable $\ruleinline{owner}$ stores the public key of the address that will receive the funds if the $\ruleinline{goal}$ is reached.
%
%
Variable $\ruleinline{t_wd}$ is the deadline after which the owner can withdraw the funds, while $\ruleinline{t_rf}$ is the deadline after which the donors can take their donations back, if not already claimed.

The \emph{rules} $\ruleinline{donate}$, $\ruleinline{withdraw}$, and $\ruleinline{refund}$ specify which operations can be performed on the contract.
Rule $\ruleinline{donate}$ can be invoked by choosing its two parameters $\ruleinline{x}$ and $\ruleinline{a}$, representing the donated amount and the donor address, respectively.
The $\ruleinline{receive(x:T)}$ is a rule \emph{precondition}, checking that $\ruleinline{x}$ tokens have been indeed transferred to the contract upon invoking the rule.
The command $\ruleinline{m[a] = m[a]+x}$ is the rule \emph{effect}: it updates the contract state, tracking the donated amount in the map $\ruleinline{m}$ (note that in the initial state \rulelang maps are implicitly initialized so that $\ruleinline{m[a]}$ is zero for all $\ruleinline{a}$).

Rule $\ruleinline{withdraw}$ takes as a parameter the amount of tokens $\ruleinline{x}$ that the owner wants to withdraw from the contract.
\iftoggle{arxiv}{%
The rule precondition $\ruleinline{require(...)}$ performs several checks. First, it ensures that the owner has authorized this operation with their signature.
Then, it checks that we are within the time window in which the owner can withdraw:
after $\ruleinline{t_wd}$ but before
$\ruleinline{t_rf}$.
Finally, it checks $\ruleinline{x >= goal}$, disallowing the owner to withdraw too few tokens: this effectively prevents the owner from withdrawing any funds when the goal has not been reached.
The rule effect $\ruleinline{owner.send(x:T)}$ transfers $\ruleinline{x}$ tokens to the owner, provided that there are indeed at least $\ruleinline{x}$ tokens within the contract.
}{
The rule precondition $\ruleinline{require(...)}$ ensures that the owner has authorized this operation with their signature, that we are within the right time window, and that the owner is withdrawing enough tokens (so the goal has been reached).
The rule effect $\ruleinline{owner.send(x:T)}$ transfers the tokens to the owner.
}

\iftoggle{arxiv}{%
The last rule $\ruleinline{refund}$ takes an address $\ruleinline{a}$ as parameter, and refunds to $\ruleinline{a}$ its donation if not already claimed.
The rule precondition ensures that the $\ruleinline{t_rf}$ deadline has been passed.
The rule effect transfers $\ruleinline{m[a]}$ tokens back to $\ruleinline{a}$, while resetting the value of $\ruleinline{m[a]}$ to zero.
}{
The last rule $\ruleinline{refund}$ refunds a donation if not already claimed.
}

More in general, \rulelang rule preconditions can 
\begin{inlinelist}
\item check boolean conditions on the parameters and the current contract state, 
\item require signatures by participants, 
\item enforce temporal constraints, and 
\item ensure that funds are transferred to the contract balance.
\end{inlinelist}
Contract effects can modify the contract state through assignments $\ruleinline{y = ...}$ and $\ruleinline{m[...] = ...}$, and they can also transfer tokens out of the contract $\ruleinline{recipient.send(...)}$.
Note that a $\ruleinline{send}$ effect can fail if there are not enough funds in the contract balance, so it implicitly adds such a precondition.
\rulelang effects are executed \emph{simultaneously}, not in sequence (this is why we separate them using the symbol $\ruleinline{|}$ instead of a semicolon).
\Eg, consider rule $\ruleinline{refund}$:
in the effect $\ruleinline{a.send(m[a]:T)}$
the value of $\ruleinline{m[a]}$ is the one relative to the contract state in which the rule was invoked, and not the one affected by the other effect $\ruleinline{m[a] = 0}$.
Note that executing $\ruleinline{m[a] = 0}$ and $\ruleinline{a.send(m[a]:T)}$ in sequence would cause zero tokens to be sent.

\begin{figure}[t]
\begin{lstlisting}[
    language=rulecode, 
    morekeywords={Crowdfund,donate,withdraw,refund},
    caption={\rulelang specification of the crowdfund contract.},
    captionpos=b,
    label=fig:crowdfunding-hurf
]
contract Crowdfund {
  map m; // donors map (from addresses to integers)
  var owner = "pub_key_owner";
  var goal = 1000;
  var t_wd = 10; // withdraw timeout
  var t_rf = 20; // refund timeout
	
  donate(x,a) { // user a donates x tokens
    receive(x:T); // x:T are sent to Crowdfund
    m[a]=m[a]+x; // donors map updated accordingly
  }
  withdraw(x) { // owner gets the entire balance 
                // if the goal is reached 
    require(signedBy(owner)
      && validFrom>=t_wd
      && validTo<t_rf
      && x>=goal);   // goal reached
    owner.send(x:T); // transfers x:T to the owner
  }
  refund(a) { // refunds to a its donation 
              // if not already claimed
    require(validFrom>=t_rf); // deadline expired
    m[a]=0 |        // resets m[a] and (in parallel)
    a.send(m[a]:T); // transfers m[a] tokens to a
  }
}
\end{lstlisting}
\end{figure}


\begin{figure*}[t!]
    \centering
    \begin{tikzpicture}[
        fontscale/.style = {font=\relsize{#1}},
        txstate/.style = {draw=red!60, fill=red!5, rectangle, very thick, minimum height=3em, minimum width=2em, fontscale=-1},
        txdeps/.style={rounded corners, draw=green!60, fill=green!5, very thick, minimum size=10mm, fontscale=-1},
        txfees/.style={rounded corners, draw=green!60, fill=green!5, very thick, minimum size=10mm, fontscale=-1},
        txsigs/.style={rounded corners, draw=blue!60, fill=blue!5, very thick, minimum size=10mm},
        balance/.style={rounded corners, draw=yellow!60, fill=yellow!60, very thick, minimum size=10mm, fontscale=-1},
        rline/.style={draw=blue!60,dashed,->,thick},
        wline/.style={draw=blue!60,solid,->,thick},
        node distance=2cm,
        scale=0.625,
      ]
      \draw [draw=gray,thick,rounded corners] (-4,1.15) rectangle (12.5,-5.30);
      \node [rotate=90] at (-3.6,-2) {\textbf{Transaction}};   
      \node at (-2,0) {\textbf{Inputs}};
      \node at (6,-2) (center) {}; 
      \node[txstate] (logic0) {Logic};
      \draw[rline] (logic0.south) -- (0,-2);
      \node[txfees] at (2,0) (fees0) {Fees};
      \draw[wline] (fees0.south) -- (2,-2);
      \node[txdeps] at (4.2,0) (deps0) {Deposits};
      \node[txdeps] at (4.35,-0.15) (deps1) {Deposits};
      \node[txdeps] at (4.50,-0.30) (deps2) {Deposits};    \draw[wline] (deps2.south) -- (4.50,-2);
      \node[txstate] at (7.2,0) (sread0) {\begin{tabular}{c} State item \\ (read) \end{tabular}};
      \node[txstate] at (7.35,-0.15) (sread1) {\begin{tabular}{c} State item \\ (read) \end{tabular}};
      \node[txstate] at (7.50,-0.30) (sread2) {\begin{tabular}{c} State item \\ (read) \end{tabular}};
      \draw[rline] (sread2.south) -- (7.50,-2);
      \node[txstate] at (10.5,0) (swrite0) {\begin{tabular}{c} State item \\ (write) \end{tabular}};
      \node[txstate] at (10.65,-0.15) (swrite1) {\begin{tabular}{c} State item \\ (write) \end{tabular}};
      \node[txstate] at (10.80,-0.30) (swrite2) {\begin{tabular}{c} State item \\ (write) \end{tabular}};
      \draw[wline] (swrite2.south) -- (10.80,-2);
      \node[txsigs] at (-2,-2) (signers) {Signers};
      \draw[draw=blue!60, dotted, very thick] (-1,-2) -- (12,-2);
      \node at (-2,-4.30) {\textbf{Outputs}};
      \node[txdeps] at (3,-4) (outdeps0) {Deposits};
      \node[txdeps] at (3.15,-4.15) (outdeps1) {Deposits};
      \node[txdeps] at (3.30,-4.30) (outdeps2) {Deposits};
      \draw[wline] (3,-2) -- (outdeps0.north);      
      \node[txstate] at (9,-4) (newstate0) {\begin{tabular}{c} New item state  \end{tabular}};
      \node[txstate] at (9.15,-4.15) (newstate1) {\begin{tabular}{c} New state item \end{tabular}};
      \node[txstate] at (9.30,-4.30) (newstate2) {\begin{tabular}{c} New state item \end{tabular}};
      \draw[wline] (9,-2) -- (newstate0.north);
      \node[balance, chamfered rectangle] at (14,0) (balance0) {\begin{tabular}{c} Old \\ contract \\ balance \end{tabular}};
      \node[balance, chamfered rectangle] at (14,-4) (balance1) {\begin{tabular}{c} New \\ contract \\ balance \end{tabular}};
    \end{tikzpicture}
    \caption{A transaction invoking a \rulelang rule. The inputs and outputs marked with the contract ID are depicted as red squares. Dashed lines represent non-consumed inputs. 
    The contract balance is implicitly updated according to the difference in value between the consumed inputs and the outputs (after fees).
    %
    }
    \label{fig:compiled-tx}
\end{figure*}

\paragraph{The \rulelang compiler}

Our compiler translates \rulelang contracts into a set of \hutxo outputs, split in \emph{logic outputs} and \emph{state outputs}.
For each \rulelang rule, it generates a logic output, whose script ensures that the rule precondition holds and that the effect is correctly applied.
To this purpose, the script checks that any transaction $\txT$ having the logic output in its inputs has the form in~\Cref{fig:compiled-tx}, and satisfies 
the following conditions:
\begin{itemize}

\item $\txT$'s inputs include all the state outputs related to the parts of the state read or written by the \rulelang rule.
This allows the script to evaluate all the \rulelang expressions occurring in the rule,
hence to check the $\ruleinline{require}$ preconditions.

\item $\txT$ includes inputs providing the funds to satisfy the $\ruleinline{receive}$ preconditions. 

\item In order to perform the effect, $\txT$ spends the state outputs representing the state values which are being overwritten, while generating (in the outputs of $\txT$) new state outputs for the new state values.
As discussed before, this might involve splitting and/or merging open intervals as needed.

\item $\txT$ includes in its outputs the ``standard deposits'' needed to transfer tokens to participants as specified by $\ruleinline{send}$ effects.

\item $\txT$ includes an additional input for the transactions fees; further inputs and outputs in $\txT$ are forbidden.

\end{itemize}

Finally, the compiler generates the state outputs for the initial state, following the distributed approach discussed previously.
Their script requires that any transaction $\txT$ having the state output among its inputs must also include in its inputs one logic output.

We remark that our \rulelang compiler is secure, in the sense that the \hutxo contract it generates allows exactly the state updates and tokens transfers specified by the \rulelang rules.
Indeed, if any transaction $\txT$ attempts to update the state by creating a contract state output or to exchange tokens with the contract, the \hutxo validity conditions force $\txT$ either to use a distinct contract ID (in which case, it does not affect the current contract), or to include at least one input related to the current contract ID.
Such an input can only refer to state outputs or logic outputs.
The script for state outputs requires the presence of an input referring to a logic output, so in all cases, a logic output must be present.
The script of logic outputs forces the \rulelang semantics to be respected, and also forbids the creation of \emph{new} logic outputs, effectively preventing the creation of new contract rules.
Since the logic output script forbids $\txT$ to consume its output, this causes the set of logic outputs for a contract to remain the same, forever.
Overall, this forces any valid transaction $\txT$ affecting the contract to have the structure shown in~\Cref{fig:compiled-tx}.

The actual compiler is substantially more complex than the previous description, since it has to deal with the concrete structure of the \hutxo transactions, their validity conditions, and the distribution of the contract state. Because of space constraints, we relegate the full compiler specification to~\Cref{sec:hurf-compiler}.

\section{The \hutxo blockchain model}
\label{sec:blockchain}

In this section we present \hutxo (standing for \emph{hybrid} UTXO), a UTXO-based blockchain model extended with contract state and contract balance accounts.
This model is similar to Cardano's eUTXO model~\cite{cardanoeutxo,Chakravarty20wtsc,Knispel24fmbc}, in that a transaction output contains a $\txf{datum}$ field which can be used to store contract data. 
As in the eUTXO model, 
\iftoggle{arxiv}{\emph{covenants}~\cite{Moser16bw,Oconnor17bw,BLZ20isola}}
{\emph{covenants}~\cite{Moser16bw,Oconnor17bw}}
are used to ensure that that the $\txf{datum}$ is updated according to the smart contract behaviour.
We further extend the eUTXO model by adding a $\txf{ctrId}$ field to the transactions, which represents the ID of the contract which is being affected by the transaction. If no contract is being involved, $\txf{ctrId}$ is set to zero.
Our transaction outputs are then augmented with a boolean $\txf{inContract}$ field, marking which outputs belong to the contract specified by $\txf{ctrId}$.

Finally, we model the balance of contracts as a map form contract IDs to currency. This map is not stored inside the UTXOs nor inside the transactions. Instead, it is a part of the implicit blockchain state, as it happens in the account-based model followed by \eg Ethereum. This makes our blockchain model a hybrid UTXO-account one.

In our model, we will use data of the following primitive types:
\[
\type{Bool}, \type{\mathbb{N}}, \type{PubK}, \type{Hash}, \type{TokenId}, \type{Script}, \type{Data}
\]
The meaning of most of these data types is straightforward. We only mention that $\type{Script}$ is the type of the redeeming scripts, while $\type{Data}$ is a generic type that can hold arbitrary data (concretely represented as a bitstring).

\begin{figure}[t]
\small
\begin{align*}
  \iftoggle{itasec25}{
  & \type{TxId} \eqdef \type{Hash} \qquad \type{CtrId} \eqdef \type{Hash}
  \qquad \type{Wallet} \eqdef \type{TokenId} \rightarrow \type{\mathbb{N}}
  \\
  }{
  & \type{TxId} \eqdef \type{Hash}
  \\
  & \type{CtrId} \eqdef \type{Hash}
  \\
  & \type{Wallet} \eqdef \type{TokenId} \rightarrow \type{\mathbb{N}}
  \\
  }
& \type{OutputRef} \eqdef (\txf{txId}: \type{TxId}, \txf{index}: \type{\mathbb{N}})
\\
& \type{TimeInterval} \eqdef (\txf{from}:\type{\mathbb{N}}, \txf{to}:\type{\mathbb{N}})
\\
& \type{Output} \eqdef (\txf{value}: \type{Value}, \ \txf{validator}: \type{Script},
\\
& \qquad \qquad \quad \txf{datum}: \type{Data}, \txf{inContract} : \type{Bool})
\\
& \type{Input} \eqdef ( \txf{outRef}: \type{OutputRef}, \txf{redeemer}: \type{Data}, \txf{\txf{spent}}: \type{Bool})
\\
& \type{Tx} \eqdef (
    \txf{in}: \type{List[Input]}, 
    \txf{out}: \type{List[Output]}, 
    \txf{signers}: \type{List[PubK]},
\\
& \hspace{28pt} 
    \txf{validityTime}: \type{TimeInterval},
    \txf{fee}: \type{\mathbb{N}}, 
    \txf{ctrId}: \type{CtrId})
\\
& \type{Accounts} \eqdef \type{CtrId} \rightarrow \type{Value}
\\
& \type{Ledger} \eqdef ( \txf{txs}: \type{List[Tx]}, \txf{accounts} : \type{Accounts}, \txf{time} : \type{\mathbb{N}})
\end{align*}
\iftoggle{itasec25}{\vspace{-5mm}}{}
\caption{Types for the \hutxo model.}
\label{fig:hutxo:types}
\iftoggle{itasec25}{\vspace{-5mm}}{}
\end{figure}

Our model is formalized in~\Cref{fig:hutxo:types}. 
There, we describe the format of $\type{Input}$s, $\type{Output}$s, and transactions ($\type{Tx}$).
%
Below, we provide some intuition about the fields of a transaction $\txT: \type{Tx}$. 
\begin{itemize}

\item Outputs store tokens (denoted by field $\txf{value}: \type{Value}$) and data (denoted by field $\txf{datum} : \type{Data}$).
Each output has a $\txf{validator}$ script, \ie a program that allows it to specify the condition under which it can be either spent or checked by another transaction.
The $\txf{inContract}$ flag determines whether the output and its data are part of a contract or not.

\item Every input references a previous unspent output.
The $\txf{spent}$ flag determines whether the output is \emph{spent} (\ie, $\txT$ grabs its tokens) or to \emph{checked} (\ie, $\txT$ just reads its fields, including the $\txf{datum}$,  leaving it unspent%
\footnote{The ability of reading output data without spending the output is similar to Cardano's \emph{checked inputs}~\cite{CIP31}.}). 
The output script must be satisfied regardless of the $\txf{spent}$ flag. 
An input also contains a $\txf{redeemer}$: a piece of data that can be checked by the referred output script.

\item The $\txf{validityTime}$ field specifies time constraints for the validation of the transaction. 

\item The $\txf{fee}$ field specifies the amount of native cryptocurrency that the transaction pays to validators.

\item The $\txf{ctrId}$ field uniquely identifies a contract in the blockchain. 
More specifically, an output $\txo$ with $\txo.\txf{inContract} = \true$ in $\txT$ is considered part of the contract specified by $\txT.\txf{ctrId}$. Indeed, the current contract state is collectively represented by all such unspent outputs, distributed across all the transactions with that $\txf{ctrId}$.

\end{itemize}

The $\type{Ledger}$ state comprises the list of the transactions that have been appended so far ($\txf{txs}$), the contract balances ($\txf{accounts}$), and the $\txf{time}$, abstractly represented as a natural.
The $\txf{accounts}$ field is a mapping that associates to each contract ID the amount of tokens it owns. 
Since contracts have their balance represented in the $\txf{accounts}$ map, there is no need to also store their tokens in the UTXOs. For this reason, the outputs having $\txf{inContract} = \true$ usually carry a $\txf{value}$ of zero tokens. Note that such value-less outputs can still be useful, since they can hold a part of the contract state in their $\txf{datum}$, as well as a $\txf{validator}$ script to implement the contract behaviour.

\paragraph{Validity of \hutxo transactions}
Below we list the conditions needed for the validity of a transaction $\txT : \type{Tx}$ in a given $(\bcB, \accA,t) : \type{Ledger}$.
As an auxiliary notion, we say that an output is \emph{unspent} in $\bcB$ 
if there is no transaction $\txT$ in $\bcB$ that has an input referring to that output with $\txf{spent}=\true$.
Then, we say that a transaction $\txT$ is valid when:
\begin{enumerate}

\item \label{it:validity-1}
All inputs of $\txT$ refer to unspent outputs in $\bcB$.

\item \label{it:validity-2}
The inputs of $\txT$ with $\txf{spent}=\true$ refer to distinct outputs in $\bcB$. 
    Note instead that the inputs of $\txT$ with $\txf{spent}=\false$ may refer to the same output of any other input.

\item \label{it:validity-3}
At least one input of $\txT$ has $\txf{spent}=\true$.

\item \label{it:validity-4}
The current time $t$ falls within the $\txf{validityTime}$ interval of $\txT$.

\item \label{it:validity-5}
For each input of $\txT$ (including those with $\txf{spent}=\false$), if it refers to the output $\txo$ then  evaluating $\txo.\txf{validator}$ yields $\true$.

\item \label{validation:ctrId-preservation} 
For each input of $\txT$, if it refers to some output $\txo$ in $\txTi$  with $\txo.\txf{inContract}=\true$, then $\txTi.\txf{ctrId} = \txT.\txf{ctrId}$.

\item \label{validation:ctrId-creation} 
If all inputs of $\txT$ refer to outputs with $\txf{inContract}=\false$, then:
\begin{enumerate}
\item If all outputs of $\txT$ have $\txf{inContract}=\false$, then $\txT.\txf{ctrId} = 0$.
\item Otherwise, if some output of $\txT$ has $\txf{inContract}=\true$, then $\txT.\txf{ctrId}$ is equal to the hash of the $\txf{outRef}$ of the first input of $\txT$ with $\txf{spent}=\true$.
\end{enumerate} 

\item \label{validation:no-contract} 
If $\txT.\txf{ctrId} = 0$, let $v_{in}$ be the sum of all the $\txo.\txf{value}$ where $\txo$ is referred by an input of $\txT$ with $\txf{spent}=\true$. Moreover, let $v_{\it out}$ be equal to the sum of all the $\txo'.\txf{value}$ of outputs $\txo'$ of $\txT$, and $v_{\txf{fee}}$ be a value of $\txT.\txf{fee}$ units of the native cryptocurrency.
We require that $v_{in}  \geq v_{out} + v_{\txf{fee}}$.

\item \label{it:validity-9}
If $\txT.\txf{ctrId} \neq 0$, then
let $bal = \accA[\txT.\txf{ctrId}]$ be the current contract balance, and let $v_{\it in}, v_{\it out}, v_{\it fee}$ as in 
\Cref{validation:no-contract}.
We require that $v_{\it in} + {\it bal} \geq v_{\it out} + v_{\it fee}$.

\end{enumerate}

\iftoggle{itasec25}{}{
We now discuss the above validity conditions.
\Cref{it:validity-1} requires that the inputs have not already been spent by a previous transaction.
\Cref{it:validity-2} prevents $\txT$ from double spending the same output.
\Cref{it:validity-3} requires $\txT$ to spend at least one previous output. In this way, $\txT$ can be appended at most once to the ledger.
\Cref{it:validity-4} ensures that the time constraints are respected.
\Cref{it:validity-5} requires that $\txT$ satisfies each script $\txo.\txf{validator}$ of each output $\txo$ referred to in the inputs of $\txT$. 
\Cref{validation:ctrId-preservation,validation:ctrId-creation} control the creation of new contract outputs, so that they can not be forged. More specifically, \Cref{validation:ctrId-preservation} 
restricts the outputs that can be created inside an existing contract $\txf{ctrId}$, by requiring that some input refers to a previous output $\txo$ in the same contract. This makes it possible for $\txo.\txf{validator}$ to impose its own requirements on the new outputs, rejecting output forgeries by allowing only those allowed by the contract logic.
By contrast, \Cref{validation:ctrId-creation} handles the case where a fresh contract is deployed: this happens when there are no contract inputs but some contract outputs. Here $\txT.\txf{ctrId}$ is mandated to be a fresh contract ID, so that one can not maliciously reuse a previous ID and forge outputs inside such contract.
\Cref{validation:no-contract,it:validity-9} ensure that the currency provided by the inputs is enough to cover for all the outputs and the fees. If some contract is involved by the transaction, then the contract balance can also contribute to pay outputs and fees.
}

\paragraph{Updating the ledger}
Appending a valid transaction $\txT$ makes the ledger state $(\bcB, \accA, t)$ evolve into a new state $(\bcB \txT, \accAi, t')$.
The accounts map is unchanged if $\txT$ is not a contract action (\ie, $\txT.\txf{ctrId} = 0$), while it is updated at point $\ctrId = \txT.\txf{ctrId}$ otherwise:
\iftoggle{arxiv}{%
\begin{align}
\label{eq:account-update}
\accAi[\ctrId] =& 
\\
&\begin{cases}
    \accA[\ctrId] + v_{\it in} - v_{\it out} - v_{\it fee} 
    & \text{if } \ctrId = \txT.\txf{ctrId} \neq 0 
    \\
    \accA(\ctrId)
    & \text{otherwise}
\end{cases}
\nonumber
\end{align}
}
{
\begin{equation}
\label{eq:account-update}
\accAi[\ctrId] \; = \; 
\begin{cases}
    \accA[\ctrId] + v_{\it in} - v_{\it out} - v_{\it fee} 
    & \text{if } \ctrId = \txT.\txf{ctrId} \neq 0 
    \\
    \accA(\ctrId)
    & \text{otherwise}
\end{cases}
\end{equation}
}
where $ v_{\it in}$, $v_{\it out}$, $v_{\it fee} $ are defined as in \Cref{validation:no-contract} of the previous paragraph.
The time update is non-deterministic, with the only constraint that $t' \geq t$. We do not specify further how time is handled in the model: it could be either tied to real time or to the block height of the ledger. Our model is independent from this choice.

\paragraph{Scripts}
For our purposes the actual script language is immaterial, so hereafter we do not explicitly provide its syntax and semantics. 
Given a transaction $\txT$ referring to an output $\txo$ in its inputs, we just assume that the script $\txo.\txf{validator}$ can read:
\begin{enumerate}[(a)]
\item all the fields of $\txT$;
\item all the fields of the \emph{sibling} outputs, \ie those referred to by the inputs of $\txT$ (including $\txo$ itself). 
\end{enumerate}
In particular, by item (a) it follows that the script can access the $\txf{redeemer}$ field in the input of $\txT$ referring to $\txo$. 
This is analogous to the way witnesses are used in Bitcoin.
Moreover, the script can access both $\txo$ (by item (b)) and each output $\txo'$ of $\txT$ (by item (a)). This makes it possible, for instance, to enforce a covenant by checking that $\txo'.\txf{validator} = \txo.\txf{validator}$ and that $\txo'.\txf{datum}$ is related to  $\txo.\txf{datum}$ according to the specific covenant logic one wants to implement.
Furthermore, by item (b), $\txo.\txf{validator}$ can allow $\txT$ to spend $\txo$ only if $\txT$ simultaneously spends a sibling $\txo''$ of a wanted form. More specifically, $\txo''$ can be required to be a contract output by checking $\txo''.\txf{inContract}$.



\section{The \rulelang contract language}
\label{sec:smart-contracts}
\label{sec:hurf}

Although the transactions validation mechanism outlined in \Cref{sec:blockchain} allows one to define programmatic exchanges of tokens, it does not provide a sharp notion of smart contract. This is because the above-mentioned programmatic exchanges can  result from the interaction of multiple different scripts distributed among several transactions: in this sense, the smart contract could be seen as an emerging high-level behaviour of the low-level behaviour of these transactions.  

In this section we provide a higher-level model of smart contracts, \rulelang (\hutxo Rule Format), where the programmatic exchange of tokens is specified by a set of rules.
We will show in~\Cref{sec:compilation} how to compile and execute \rulelang contracts on \hutxo blockchains.
In this way, \rulelang provides a lower bound to the expressiveness of \hutxo contracts, as any \rulelang contract can be effectively implemented in \hutxo.  

\paragraph*{Contracts as rules}

Syntactically, a \rulelang contract comprises:
\begin{itemize}
\item a declaration of the variables and maps that contribute to its \emph{state} ($\cstate{}$), possibly including their initial values;
\item a set of \emph{rules} ($\vec{\ruleR}$) that specify how the contract state $\cstate{}$ and \emph{balance} $\bal$ can be updated. 
\end{itemize}
\Cref{fig:bank-simple-rules} shows a simple example of contract.
Users execute contracts by repeatedly choosing and firing contract rules.
Each rule $\ruleR$ consists of three parts: 
\begin{itemize}
\item a \emph{signature} providing the rule name and the sequence of formal parameters to be instantiated by users upon firing the rule;
\item a set of \emph{preconditions} that must be met in order for the rule to be fired;
\item an \emph{effect} that can modify the contract state $\cstate{}$ and transfer tokens to users, affecting the contract balance $\bal$ accordingly.
\end{itemize}

To formalize the contract state, we first define the \emph{base values} $\type{BVal}$ as
\[
	\type{BVal} \eqdef \type{Bool} \mid \type{\mathbb{Z}} \mid \type{String}
\]
%
Then, we make state $\cstate{}$ associate each variable \ruleinline{x} to a base value $\cstate{\ruleinline{x}}:\type{BVal}$,
and each each map \ruleinline{m} to a function from tuples of base values to a single base value, \ie:
\[
	\cstate{\ruleinline{m}} : \type{BVal}^* \rightarrow \type{BVal}
\]
To denote the function $\cstate{\ruleinline{m}}$
we will write
\[
	\cstate{\ruleinline{m}} = [(l^1_1, \cdots, l^1_n) \mapsto v_1] \cdots [ (l^k_1, \cdots, l^k_n) \mapsto v_k ]
\] 
for the function that takes the value $v_i$ at the point $(l_1^i, \cdots, l_n^i)$, and the \emph{default value} $0$ at any other point.

Accessing the state of a map requires one to specify the map name \ruleinline{m} and the point $(l_1,\ldots,l_n)$ where it is accessed. We will refer to the pair $(\ruleinline{m}, (l_1\ldots,l_n))$ as a \emph{map location}.

The contract balance $\bal$ is a $\type{Wallet}$ (\ie, a map from token types to non-negative integers) that represents the amount of tokens owned by the contract.


\paragraph{Expressions}

Expressions \ruleinline{e} appearing in preconditions and effects have the following form:
\begin{itemize}
\item boolean, integer, and string constants;
\item \ruleinline{e1 op e2}, where \ruleinline{op} is an arithmetic or boolean operator;
\item \ruleinline{not e}, to negate a boolean;
\item \ruleinline{if e0 then e1 else e2}, conditional;
\item \ruleinline{m[e1,...,ek]}, accessing the map \ruleinline{m} at point \ruleinline{(e1,...,ek)};
\item \ruleinline{hash(e1,...,ek)}, hashing a tuple of values;
\item \ruleinline{e1 @ e2}, concatenating two strings;
\item \ruleinline{len(e)}, the length of a string;
\item \ruleinline{substr(e, e1, e2)}, taking the substring from character in position \ruleinline{e1} to the one in position \ruleinline{e2} of a given string \ruleinline{e};
\item \ruleinline{toStr(e1,...,ek)}, converting a tuple of values to a string;
\item \ruleinline{signedBy(e)}, checking whether user \ruleinline{e} authorizes the rule invocation;
\item \ruleinline{validFrom}, \ruleinline{validTo}, the validity temporal bounds for the rule invocation;
\item rule parameters and state variables.
\end{itemize}

Note that the language of expressions has no operators to iterate over maps.
This ensures that the evaluation time of expressions does not depend on the size of maps. 
W.l.o.g.\ there are no expressions to obtain the contract balance: if needed, the developer can record in contract variables the balances of the different tokens stored by the contract. 
\todo{forward pointer}
However, doing so may affect the parallelizability of contracts, so these variables must be dealt with carefully.

\paragraph{Preconditions}
A rule precondition is a list of requirements of the following forms:
\begin{itemize}
\item zero or more \ruleinline{receive(e:T)}, enabling the rule only if some user sends exactly \ruleinline{e} tokens of type \ruleinline{T} to the contract account.
\item one \ruleinline{require(e)}, enabling the rule only if \ruleinline{e} evaluates to \ruleinline{true}.
\end{itemize}

\paragraph{Effects}
The effect of a rule allows to simultaneously assign values to contract variables and maps, while also transferring assets from the contract balance to a given user.

The effect, denoted as \ruleinline{S1 | ... | SN},
simultaneously executes statements \ruleinline{Si} of the following forms:
\begin{itemize}

\item \ruleinline{x = e}, assigning a value to variable \ruleinline{x};

\item \ruleinline{m[e1,...,ek] = e'}, updating the map \ruleinline{m} at point \ruleinline{(e1,...,ek)};

\item \ruleinline{e.send(e':T)}, sending \ruleinline{e'} tokens of type \ruleinline{T} from the contract balance to a user (whose public key is) \ruleinline{e}.
If the contract balance is insufficient, the rule cannot be fired.

\end{itemize}

We stress that the execution of statements is simultaneous: all the expressions in the effect are first evaluated in the \emph{old} state, which is then updated.
To ensure that such update is well-defined, we postulate that in each rule, each variable and map location is assigned at most once.
More specifically, we postulate that effects satisfy the following:
\begin{itemize}


\item if the effect comprises \ruleinline{x = e} and \ruleinline{x = e'}, then \ruleinline{e} and \ruleinline{e'} must be equal;

\item if the effect comprises \ruleinline{m[e1,...,ek] = e} and \ruleinline{m[f1,...,fk] = e'} with \ruleinline{e} different from \ruleinline{e'},
then the rule \ruleinline{require} precondition must check that
\begin{lstlisting}[language=rulecode]
not (e1==f1 && ... && ek==fk)
\end{lstlisting}

\end{itemize} 
Hereafter, for simplicity we will not explicitly include this check in our rules.

\paragraph*{Example: bank}

We specify in \Cref{fig:bank-simple-rules} a bare-bone bank contract.
The contract state consists of a single map \ruleinline{m} that associates the users' public keys to the amount of tokens of type \code{T} they have deposited.
The contract has two rules: \ruleinline{deposit} allows users to transfer tokens to the contract, and \ruleinline{withdraw} allows them to get their funds back.
\begin{figure}
\begin{lstlisting}[language=rulecode, morekeywords={Bank,deposit,withdraw}]
contract Bank {
  map m;
	
  deposit(x, a) { 
    receive(x:T);   // precondition
    m[a] = m[a]+x;  // effect
  }
  withdraw(x, a) {
    require(m[a]>=x && signedBy(a));
    m[a] = m[a]-x | a.send(x:T);
  }
}
\end{lstlisting}
\caption{A simple bank contract.}
\label{fig:bank-simple-rules}
\end{figure}

\paragraph{Contract behaviour}

To describe the behaviour of a contract, besides its state and balance we must also consider the tokens held by users.
These are needed to provide inputs to the contract (to fulfill its \ruleinline{receive} conditions), and to transfer tokens from the contract to users (through the \ruleinline{send} statement).
We denote by $\confDep[x]{\pmvA}{\bal}$ the fact that a user $\pmvA$ holds a \emph{deposit} $\bal$ of tokens. The name $x$ is used to discriminate between different deposits of the same user. For instance, in the same system configuration we can have both $\confDep[x]{\pmvA}{\walenum{\waltok{1}{\tokT}}}$ and $\confDep[y]{\pmvA}{\walenum{\waltok{2}{\tokT}}}$ for two distinct deposits of $\waltok{1}{\tokT}$ and $\waltok{2}{\tokT}$ both owned by $\pmvA$.
Invoking a rule spends deposits, burning their name, while executing \ruleinline{send} creates new deposits with fresh names.

As usual, several contracts might be simultaneously active, possibly instantiating the same set of rules multiple times.
To account for that, we model a \emph{contract instance} as
$\confContr[{\ctrId}]{\vec{\ruleR}}{\cstate{}, \bal}$
where $\vec{\ruleR}$ is the set of contract rules, 
and $\ctrId[i]$ is a unique identifier of the instance.

We then define the system \emph{configuration} as:
\begin{align*}
\confG \; = \; 
& \confContr[{\ctrId[1]}]{\vec{\ruleR}_1}{\cstate[1]{}, \bal[1]} 
\mid \cdots \mid 
\confContr[{\ctrId[n]}]{\vec{\ruleR}_n}{\cstate[n]{}, \bal[n]} 
\mid 
\\
& \confDep[x_1]{\pmvA[1]}{\bali[1]} 
\mid \cdots \mid 
\confDep[x_k]{\pmvA[k]}{\bali[k]} \mid t
\end{align*}
where 
$\confContr[{\ctrId[i]}]{\vec{\ruleR}_i}{\cstate[i]{}, \bal[i]}$
are the contract instances, 
$\confDep[x_i]{\pmvA[i]}{\bali[i]}$ are the deposits, 
and $t$ is the current time.
In the initial configuration there are no contract instances, and the timestamp is zero.

Configurations advance through \emph{actions}. An action can perform one of the following: 
\begin{itemize}
\item deploy a new contract instance, spending some value $\bal + \txf{fee}$ from users' deposits to add $\confContr[\ctrId]{\ruleR}{\cstate{}, \bal}$ to the configuration, where $\ctrId$ is a fresh identifier;
\item fire a contract rule, affecting the state and balance of a contract instance, while possibly consuming and/or creating  deposits;
\item allow users to redistribute the tokens in their deposits;
\item make the time pass.
\end{itemize}

We detail below the actions of the second kind, called contract actions.
Contract actions must specify all the data needed to uniquely determine its behavior. 
In particular, each contract action must specify:
\begin{itemize}

\item the unique identifier $\ctrId$ of the affected instance;

\item the name of the rule in $\vec{\ruleR}$ that is being invoked, as well as its actual parameters;

\item the signers who authorize the action. There can be any number of signers (possibly zero); 

\item the names of each deposit that is spent to satisfy each \ruleinline{receive} precondition in the invoked rule; 

\item an additional deposit name, specifying a fee paid to fire the action. Including this fee also ensures that actions can not be replayed;

\item the validity interval, constraining the time at which the rule can be performed.

\end{itemize}

We remark a few differences between the data included in our actions and those occurring in Ethereum transactions.
First, our actions allow multiple signers, while in Ethereum each transaction is signed by a single externally-owned account.
Second, our actions can spend multiple deposits and transfer tokens from different users to the contract. In Ethereum, the tokens transferred to the contract can only come from the sender's account.
Third, our actions include an explicit validity interval, while
Ethereum transactions do not. This is because Ethereum contracts can access the current block number, and then use this information to implement the wanted time constraints.
The above-mentioned differences between our model and Ethereum are due to our underlying compilation target, which is a \hutxo-based blockchain instead of an account-based one. 

When an action invokes a rule, if its preconditions are satisfied, the configuration is updated accordingly. The deposits spent by the action are removed from the configuration, while new deposits are added to implement the \ruleinline{send} effects. The contract balance $\bal$ is updated according to the \ruleinline{receive} preconditions and \ruleinline{send} effects. The contract state $\cstate{}$ is updated according to the assignments to variables and maps specified in the effect.

\section{Compiling \rulelang to \hutxo}
\label{sec:compilation}
\label{sec:hurf-compiler}

In this section we show how to compile \rulelang contracts into the \hutxo model. 

Recall that a contract with rules $\vec{\ruleR}$ can be instantiated multiple times. Each instance 
$\confContr[\ctrId]{\vec{\ruleR}}{\cstate{},\bal}$
is associated to a unique contract identifier $\ctrId$.
At the \hutxo level, the rules $\vec{\ruleR}$ and the current state $\cstate{}$ of the contract $\ctrId$ are represented by those UTXOs whose $\txf{inContract}$ flag is true and whose enclosing transaction has $\txf{ctrId} = \ctrId$.
We will refer to such outputs as the \emph{contract outputs} for $\ctrId$.

We first show how to represent the contract state and balance on the ledger and how to update them. Then, we show how to compile contract rules into output scripts.

\subsection{Representing the contract state and balance}
\label{subsec:state-encoding}

We represent the state $\cstate{}$ and balance $\bal$ of the contract $\ctrId$ as follows:
\begin{itemize}
    \item The balance $\bal$ is simply stored as a $\type{Wallet}$ within the $\txf{accounts}$ map of the $\type{Ledger}$. Formally, we have
    \[
        \txf{accounts}(\ctrId) = \bal
    \]
    \item The state $\cstate{}$ is instead stored across multiple unspent contract outputs for $\ctrId$.
    To achieve this, we proceed in two steps: first, we \emph{flatten} $\cstate{}$ into a single map $\cstateOut{}$ from hashes to values. Then, we \emph{distribute} $\cstateOut{}$ across multiple contract outputs.
\end{itemize}

\paragraph{Flattening $\cstate{}$ to $\cstateOut{}$}
The state $\cstate{}$ associates variables \ruleinline{x} to values $\cstate{\ruleinline{x}} = v$, and maps \ruleinline{m} to functions $\cstate{\ruleinline{m}} = [(l_1,\cdots,l_n) \mapsto v] \cdots$.
We flatten this structure, creating a new state map $\cstateOut{}$ that associates the hash of each variable name $\ruleinline{x}$ and the hash of each map location
$(\ruleinline{m}, (l_1, \cdots, l_n))$ to their respective values.
More formally, we represent each variable association
\[
	\cstate{\ruleinline{x}} = v
\]
with the association
\begin{equation}
	\label{eq:sigma-var-definition}
	\cstateOut{\ruleinline{hash("var\_x")}} =v
\end{equation}
Similarly, we represent each map association
\[
	\cstate{\ruleinline{m}} =  [(l^1_1, \cdots, l^1_n) \mapsto v_1] \cdots [ (l^k_1, \cdots, l^k_n) \mapsto v_k ]
\]
with the associations
\begin{align} 
	\nonumber
	\cstateOut{\ruleinline{hash("map\_m["@}\ruleinline{toStr(} l^1_1 \cdots l^1_n \ruleinline{)@"]")}} &= v_1
	\\
	\label{eq:sigma-map-definition}	
	\ldots \qquad   \qquad \ldots  \qquad \qquad  \ldots \qquad&  
	\\
	\nonumber
    \cstateOut{\ruleinline{hash("map\_m["@}\ruleinline{toStr(} l^k_1 \cdots l^k_n \ruleinline{)@"]")}} &= v_k
\end{align}

We let the map $\cstateOut{}$ associate all the other hashes to the \emph{default value} $0$.
This representation $\cstateOut{}$ of the state $\cstate{}$ is well defined, provided that there are no hash collisions. Since such collisions are extremely unlikely we simply assume that they will never happen. 
We assume that the codomain of the hash function \ruleinline{hash} is the open interval $(\minHash, \maxHash)$. We will denote such values in hexadecimal format: $\minHash = \ruleinline{00...}$ and $\maxHash = \ruleinline{ff...}$.

We provide an example in~\Cref{table:state-outputs},
showing how a state $\cstate{}$ is flattened to $\cstateOut{}$.

\paragraph{Encoding $\cstateOut{}$ as outputs}
To represent the map $\cstateOut{}$ as contract outputs we first notice that $\cstateOut{}$ takes the default value $0$ everywhere, except for some hashes which we denote by $h_1 \cdots h_n$.
For simplicity, let those $n$ hashes be ordered, so that $h_1 < h_2 < \cdots < h_n$. 
We can then visualize $\cstateOut{}$ as follows: 
\begin{equation}
    \label{eq:Sigma-state}
	\cstateOut{h} = 
		\begin{cases}
			0 & h \in (\minHash, h_1) 
			\\
			v_1 & h= h_1
			\\
			\cdots & 
			\\
			0 & h \in (h_{i-1}, h_{i}) 
			\\
			v_i & h= h_i
			\\ 
			0 & h \in (h_{i}, h_{i+1}) 
			\\
			\cdots & 
			\\
			v_n & h= h_n
			\\
			0 & h \in (h_{n},\maxHash)  
		\end{cases}
\end{equation}
We distribute $\cstateOut{}$ on the ledger by creating an output for each of the cases shown above, amounting to $2n+1$ outputs. This is exemplified in~\Cref{fig:state-function-hash}.

Among the generated outputs, $n$ are used to certify that $\cstateOut{h_i}$ has a specific non-default value $v_i$. 
The remaining $n+1$ outputs are used to certify that for any $h$ in the open interval $(h_i,h_{i+1})$, $\cstateOut{h}$ takes the default value $0$. 

We define \emph{state outputs} as follows:
\begin{itemize}
\item \todo{$\txo_h(v)$} $\txo_h$ is the output certifying that $\cstateOut{}$ takes a non-default value $v$ in point $h$. We set the $\txf{datum}$ field of $\txo_h$ to $[\ruleinline{"non-default"}, h, v]$. 
\item $\txo_{h',h''}$ is the output certifying that $\cstateOut{}$ takes the default value in the open interval $(h',h'')$. We set the $\txf{datum}$ field of $\txo_{h',h''}$ to $[\ruleinline{"default"}, h', h'']$.
\end{itemize}
In both cases, we make these outputs have $value = 0$ and $\txf{inContract} = true$.
Finally, we let $\txf{script} = \txscript_{state}$ as described in \Cref{subsec:contract-logic}.


\begin{table*}
\footnotesize
\begin{minipage}{0.2\textwidth}
	\centering
	
	\begin{tabular}{c | c}
		Name & Value
		\\
		\hline
		x  & 3
		\\
		y & 0
		\\
		m& $[1\rightarrow 10][6\rightarrow 100]$ 
	\end{tabular}
	\caption*{\small State $\cstate{}$ of a contract.}
\end{minipage}
\begin{minipage}{0.35\textwidth}
	\centering
		\begin{tabular}{c | c | c | c}
		Name & Key & Key hash $h$ & $\cstateOut{h}$
		\\
		\hline
		x  &  var\_x & ed5ba7\dots & 3
		\\
		y  & var\_y & bd1138\dots & 0
		\\
		m[1]& map\_m[1] & ea00c6\dots & 10
		\\
		m[6]  & map\_m[6] & 9c90ac\dots & 100
	\end{tabular}
	\caption*{\small Hash keys and mapped values, defining $\cstateOut{}$.}
\end{minipage}
\begin{minipage}{0.385\textwidth}
	\centering
		\begin{tabular}{c||c c|c c}
		 & Default &  & Non-default & 
		\\
		\hline	
		Key &  from hash & to hash & Key hash & value 
		\\
		\hline
		--- &  0 &  9c90a\dots &&
		\\
		map\_m[6] & & &9c90a\dots & 100
		\\
		--- & 9c90a\dots &ea00c\dots &&
		\\
		map\_m[1] & & &ea00c\dots & 10
		\\
		---  & ea00c\dots &ed5ba\dots  &&
		\\
		var\_x & & &ed5ba\dots & 3
		\\
		---  &  ed5ba\dots & fffff\dots	&&
	\end{tabular}
	\caption*{\small Outputs representing the contract. Since $y$ takes the value 0, it is already represented in the default ranges.}
\end{minipage}
\caption{From contract state to the outputs.}
\label{table:state-outputs}
\end{table*}

\begin{figure}
\includegraphics[width=\columnwidth]{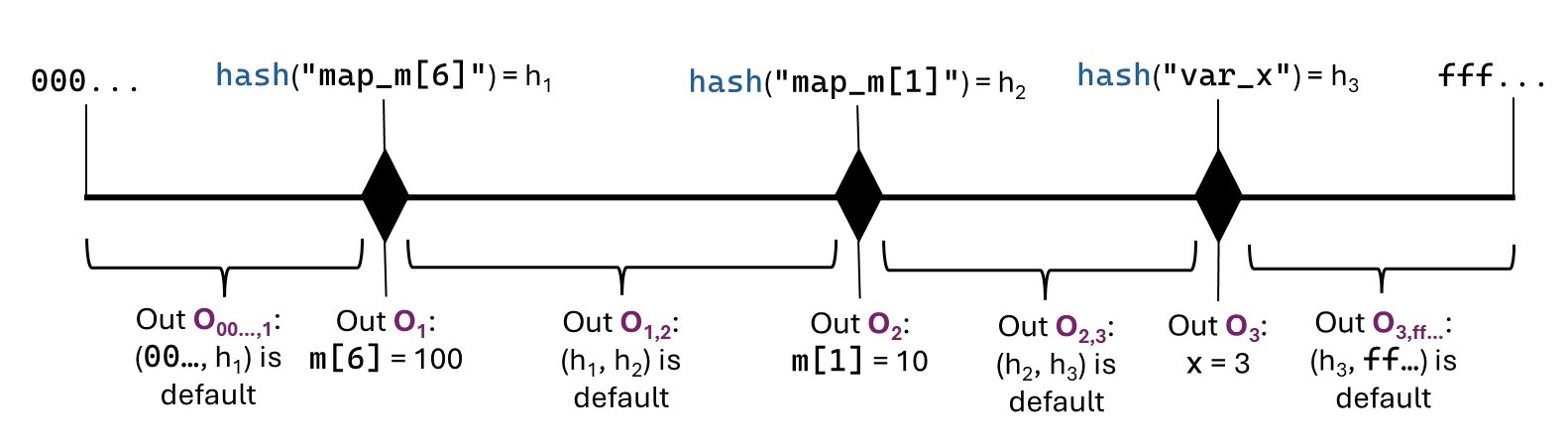}
\caption{Visualizing $\cstateOut{}$.}
\label{fig:state-function-hash}
\end{figure}

\subsection{Reading and updating the state}
\label{subsec:state-updates}

To implement the contract behavior, we will use suitable transaction scripts to verify that the contract evolves according to the contract rules. In order to do so, such scripts need to read and update the contract state. 
This is realized by making the redeeming transaction include inputs referring to those state outputs that represent the part of the state that is being accessed. More in detail, reading from the state is done by inputs having $\txf{spent} = \false$, while updating the state is done by inputs having $\txf{spent} = \true$. In the case of updates, the redeeming transaction must also include new state outputs to represent the updated state.


\paragraph{Reading the state}
In order to read the value of a contract variable \ruleinline{x} or of a map access \ruleinline{m[v,...]}, we first compute the hash $h$ of the relevant key as in \Cref{eq:sigma-var-definition,eq:sigma-map-definition}, and then craft a transaction including among its inputs the state output that certifies the value of $\cstateOut{h}$, \ie either $\txo_h$ or some $\txo_{h_0,h_1}$ with $h\in (h_0,h_1)$.  
This allows the scripts to access the $\txf{datum}$ field, and read the value of $\cstateOut{h}$. Note that this reading operation must not spend the output, so we set $\txf{spent}=\false$ in the input.

\paragraph{Updating $\cstateOut{}$: intuition}

To explain how the state can be updated, we first consider the case where the update only involves a single point $\cstateOut{h}$, and then generalize this to the general case where multiple points are updated.

In order to update $\cstateOut{}$ on a point a transaction needs to spend the old state outputs and to create new ones.
This is handled in four possible ways, depending on whether the old value $\cstateOut{h}=v$ is the default or not, and on whether the new value $\cstateOuti{h} = v'$ is the default or not: 
\begin{itemize}

\item If $\cstateOut{h}=v\neq 0$ and we want to update it to $\cstateOuti{h}=v'\neq 0$, then the transaction needs to consume the output $\txo_h$ and create a new output $\txo'_h$ containing the new value $v'$.

\item If $\cstateOut{h}=v\neq 0$ and we want to update it to $\cstateOuti{h}=0$, then the transaction needs to consume three outputs $\txo_{h_0,h}$, $\txo_h$, and $\txo_{h,h_1}$ and merge them into a single range of default values, creating an output $\txo'_{h_0,h_1}$.

\item If $\cstateOut{h}=0$ and we want to update it to $\cstateOuti{h}=v'\neq 0$, then the transaction needs to consume the single output $\txo_{h_0,h_1}$ (where $h_0<h<h_1$), and produce three outputs: $\txo'_{h_0,h}$, $\txo'_h$, and $\txo'_{h,h_1}$, splitting the default range into two and adding a new non-default output $\txo'_h$ containing the new value $v'$.

\item Finally, if $\cstateOut{h}=0$ and we want to update it to $\cstateOuti{h}=0$ (this can happen due to the effect of a rule), the transaction does not need to consume nor create any output. However a non-spending input is still needed to ensure that the old value is actually the default: this input is $\txo_{h_0,h_1}$, where $h_0<h<h_1$.
\end{itemize}

\paragraph{Problems with updating the state on multiple points}
If we need to update $\cstateOut{}$ on multiple points, we might be tempted to handle each update independently, applying the technique presented above on each point. However, this does not always work. 

For instance, let $\cstateOut{}$ be such that $\cstateOut{h'} = 1$ and $\cstateOut{h''} = 2$, with $h'<h''$ and $ \cstateOut{h} = 0$ for any other $h$.
If we try to update both $h'$ and $h''$ to 0 within a single transaction, we would need to redeem output $\txo_{h',h''}$ twice (once to update $\cstateOut{h'}$ and once again to update $\cstateOut{h''}$).
This is already problematic. 
Furthermore, the first update would create the output $\txo_{\minHash,h''}$ while the second one would create the output $\txo_{h', \maxHash}$. However, these intervals overlap, so the new outputs would fail to correctly represent the updated state $\cstateOut'{}$. 
Indeed, a correct transaction should instead produce a single output $\txo_{\minHash,\maxHash}$.

This is not the only case in which these simultaneous updates are problematic: consider the opposite case, in which $\cstateOut{}$ is 0 everywhere, and we want a single transaction to change both  $\cstateOut{h'}$ and  $\cstateOut{h''}$ to 1. 
Following the technique described above, this transaction would produce both the triple of state outputs $\txo_{\minHash,h'}$, $\txo_{h'}$, $\txo_{h',\maxHash}$, and another triple $\txo_{\minHash,h''}$, $\txo_{h''}$, $\txo_{h'',\maxHash}$. 
Once again, this does not correctly represent the updated state.
Indeed, the output $\txo_{\minHash,h''}$ would wrongly claim that $\cstateOuti{h'}=0$ (and similarly $\txo_{h',\maxHash}$ would claim $\cstateOuti{h''}=0$).

\paragraph{Updating $\cstateOut{}$ in the correct way}
Here we detail an algorithm that, given a list of updates $\vec{u} = (h_1 \mapsto v_1) \cdots (h_n \mapsto v_n)$ and an initial state $\cstateOut{}$, constructs a list of inputs $\vec{in}$ and a list of outputs $\vec{out}$ that a transaction would need to include in order to perform the updates in $\vec{u}$ to the state.
Intuitively, this algorithm suitably splits and/or joins the intervals affected by the updates.

We assume that the list of updates $\vec{u} = (h_1\mapsto v_1) \cdots (h_n \mapsto v_n)$ given as input to our algorithm is sorted, so that if $i < j$ then $h_i < h_j$. 
Note that, by requiring a strict inequality, we effectively forbid multiple updates at the same point (consistently with the fact that we want to perform all of them simultaneously).

We start by constructing the list of inputs $\vec{in}$ affected by the updates $\vec{u}$. 
Given an initial state $\cstateOut{}$, the algorithm repeats the following steps to add inputs to the initially empty list $\vec{in}$. 
\begin{enumerate}

\item If $\vec{u}$ is empty, terminate.

\item If $\vec{u} = (h \mapsto v) \vec{u'}$, with $\cstateOut{h} \neq 0$ and $v \neq 0$, then append $\txo_{h}$ to $\vec{in}$ and remove $(h\mapsto v)$ from $\vec{u}$.

\item  If $\vec{u} = (h \mapsto 0) \vec{u'}$, with $\cstateOut{h} \neq 0 $, then find $h',h''$ for which there exist outputs $\txo_{h',h},\txo_{h},\txo_{h,h''}$.
Append $\txo_{h',h}$ to $\vec{in}$  unless it already occurs in it. Then append $\txo_h$, and $\txo_{h,h''}$ to $\vec{in}$. Finally, remove $(h\mapsto 0)$ from $\vec{u}$. 

\item If  $\vec{u} = (h \mapsto v) \vec{u'}$, with $\cstateOut{h} = 0$ then find  $h',h''$ for which there exists an output $\txo_{h',h''}$ with $h' < h < h''$.
Append $\txo_{h,h'}$ to $\vec{in}$ unless it already occurs in it. Then remove $(h \mapsto v)$ from $\vec{u}$.

\end{enumerate}

Constructing the outputs list $\vec{out}$ is done by the partial function $\outMap$ defined in \Cref{fig:output-list}. 
When called as $\outMap(\vec{in}, \vec{u})$ the function checks whether the list of inputs $\vec{in}$ can support the state updates $\vec{u}$, and in such case it returns the list of outputs $\vec{out}$ to be included in the transaction to drive the wanted state update.
Intuitively, this function iterates in parallel over the lists $\vec{in}$ and $\vec{u}$, finding the points where the input intervals need to be split or merged, producing suitable outputs in the process.

For instance, let $h_1 < h_2 < h_3 < h_4$, and consider the following state $\cstateOut{}$:
\[
	\cstateOut{h} = \begin{cases}
			1 & \text{if } h = h_1
			\\
			3 & \text{if } h = h_3
			\\
			4 & \text{if } h = h_4
			\\
			0 & \text{otherwise}
	\end{cases}
\]
Let $\vec{u} = (h_2 \mapsto 2)(h_3 \mapsto 0)$ be the wanted state updates. 
Assume that $\cstateOut{}$ is currently represented by the UTXOs $\txo_{\minHash,h_1}$, $\txo_{h_1}$, $\txo_{h_1,h_3}$, $\txo_{h_3}$, $\txo_{h_3,h_4}$, $\txo_{h_4}$, $\txo_{h_4,\maxHash}$.
We generate a transaction to update the state.
Running the input generation algorithm we obtain the list of inputs $\vec{in} = \txo_{h_1,h_3} \txo_{h_3} \txo_{h_3,h_4}$. Running the output generation algorithm as $\outMap(\vec{in}, \vec{u})$ we obtain $\vec{out} = \txo'_{h_1,h_2} \txo'_{h_2} \txo'_{h_2,h_4}$.
Indeed, the algorithm splits $\txo_{h_1,h_3}$ to account for $h_2$ being changed from the default value to 2, and merges $\txo_{h_3}$ into the output $\txo'_{h_2,h_4}$ representing a new default interval.

\begin{figure}
    \centering
    \includegraphics[scale=0.4]{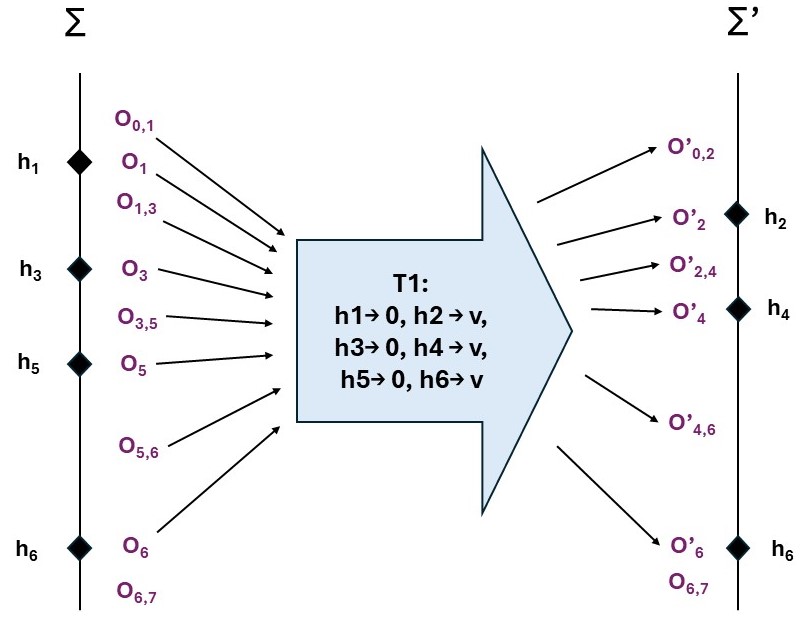}
    \caption{The state can be modified drastically with a single transaction.}
    \label{fig:big-tx-example}
\end{figure}

\begin{figure}
    \centering
    \includegraphics[scale=0.4]{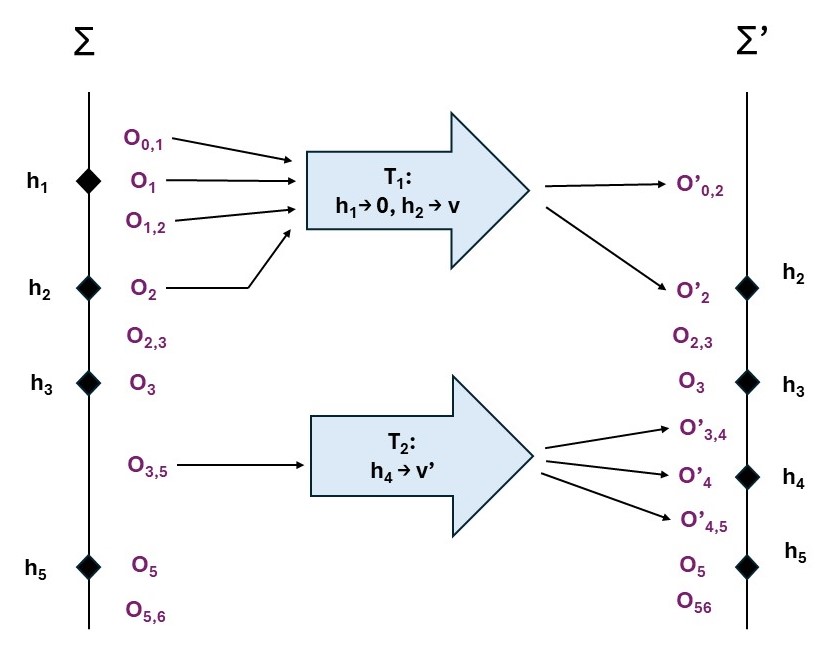}
    \caption{As long as they do not consume the same part of the state, transactions can be parallelized.}
    \label{fig:parallel-tx-example}
\end{figure}

\paragraph{Example: multiple state updates} 
We show in~\Cref{fig:big-tx-example} the effect of a transaction $\txT[1]$ that updates the state at six distinct points, consuming eight state outputs while spawning six new ones. 
The inputs and outputs in the figure are those computed by the input and output generation algorithms for the updates 
\[
\vec{u} = (h_1 \mapsto 0, h_2 \mapsto v, h_3\mapsto 0, h_4\mapsto v, h_5\mapsto 0, h_6\mapsto v)
\]
with $v\neq 0$.
Taken individually, the update part $(h_1 \mapsto 0)$ would require the state outputs $\txo_{0,1},\txo_{1},\txo_{1,3}$ to be merged, but the update part $(h_2 \mapsto v)$ would instead require $\txo_{1,3}$ to be split. The net effect of $(h_1 \mapsto 0, h_2 \mapsto v)$ is that the new outputs include $\txo'_{0,2},\txo'_{2}$. A similar interaction is caused by $(h_3\mapsto 0, h_4\mapsto v)$. By contrast, the update part $(h_5\mapsto 0, h_6\mapsto v)$ does not require a split, since $\txo_6$ already associates $h_6$ to a non-default value.

We remark that the updates $\vec{u}$ modify the state simultaneously, therefore we bundle them in a single transaction. 
Indeed, regardless of the size of $\vec{u}$, the inputs and outputs generation algorithms always craft a single transaction that performs them all.

\paragraph{Example: parallel updates}
\Cref{fig:parallel-tx-example} shows how two transactions updating the state at different points can be validated in parallel. There, $\txT[1]$ updates the state at points $h_1,h_2$
while $\txT[2]$ updates the state at point $h_4$.
The inputs of the two transactions do not overlap, so a parallel validation is possible.

Note that updating the state at different points is not, in general, sufficient to guarantee that two transactions are parallelizable. Indeed, had $\txT[1]$ performed the update $(h_3 \mapsto 0)$, we would still have disjoint update points ($h_3$ vs. $h_4$) but both transactions would attempt to consume $\txo_{3,5}$ ($\txT[1]$ to merge it with $\txo_{2,3}$,  $\txT[2]$ to split it).
This makes the two transaction conflict: only one of them can be appended to the current blockchain, and their validation can not be made in parallel.
After one of them is put on the blockchain, the other update can still be performed, but using a different transaction involving the new UTXOs.

\begin{figure}
\begin{equation*}
		\outMapi(\emptyseq, \emptyseq) = \emptyseq
\end{equation*}
\begin{equation*}
	\outMapi(\txo_{h',h''}, \emptyseq) = \txo_{h',h''}
\end{equation*}
\begin{align*}
	\outMapi( &\txo_{h',h''} \vec{in},(h \mapsto v) \vec{u}) =
	\\ 
	=& \begin{cases}
		\txo_{h',h''} \outMap( \vec{in}, (h\mapsto v)\vec{u})  &\text{if } h>h''
		\\
		\outMap(\txo_{h',h''} \vec{in},(h \mapsto v) \vec{u}) &\text{otherwise}
	\end{cases} \notag
\end{align*}
\begin{equation*}
		\outMap(\emptyseq, \emptyseq) = \emptyseq
\end{equation*}
\begin{align*}
	& \ \outMap(\txo_{h',h''} \vec{in}, (h\mapsto v)\vec{u}) =
	\\
	=&\begin{cases}
		\outMapi(\txo_{h',h''}\vec{in},\vec{u}) &\text{if $v = 0$, $h \in (h',h'')$ }
		\\
		\txo_{h',h} \txo_{h} \outMapi( \txo_{h,h''} \vec{in} , \vec{u}) &\text{if $v\neq 0$, $h \in (h',h'')$ }
	\end{cases} \notag
\end{align*}
\begin{equation*}
\outMap( \txo_h \vec{in}  , (h\mapsto v \neq 0)\vec{u} )
= \txo'_h \outMap( \vec{in}  , \vec{u} )
\end{equation*}
\begin{equation*}
 \outMap( \txo_{h',h} \txo_h \txo_{h,h''} \vec{in} , (h\mapsto0)\vec{u} )
= \outMapi( \txo_{h',h''} \vec{in}  , \vec{u} )
\end{equation*}
\caption{Definition of the partial function $\outMap$. In those cases not explicitly handled by the equations above, the operation fails.
The definition uses the auxiliary function $\outMapi$.
}
\label{fig:output-list}
\end{figure}

\subsection{Contract logic}
\label{subsec:contract-logic}
We now show how to implement the contracts of \Cref{sec:smart-contracts} on the blockchain.
In \Cref{subsec:state-encoding,subsec:state-updates} we saw how to represent, read and update the state. Here, starting from a set of rules $\vec{\ruleR}$, we create a set of contract outputs that constrain the state updates to follow the rules.

To deploy a contract, we append a transaction $\txT[init]$ with a fresh contract identifier, as required by~\cref{validation:ctrId-creation} of the transaction validity conditions (see~\cpageref{validation:ctrId-creation}).
The transaction $\txT[init]$ contains a contract output $\txo^{\ruleR}$ for each contract rule $\ruleR$,
as well as a set of state outputs to represent the
initial contract state, as explained in~\Cref{subsec:state-encoding}.

Once the contract is deployed, one can execute a rule $\ruleR$ by appending a transaction $\txT$ referring to the output $\txo^{\ruleR}$ as a non-spending input.
This causes the script $\txo^{\ruleR}.\txf{script}$ to be executed.
We define such script so to verify that $\txT$ correctly updates the current contract state as mandated by rule $\ruleR$. This is done by inspecting all the inputs and the outputs of $\txT$.
In \Cref{fig:compiled-tx} we show the structure of $\txT$.

We now describe $\txo^{\ruleR}$ in more detail. We set its fields as follows: $\txf{value} = 0$, $\txf{datum} = \ruleinline{"logic"}$, $\txf{inContract} = \true$. Further, $\txo^{\ruleR}.\txf{script}$ verifies that:
\begin{itemize}

\item $\txo^{\ruleR}$ is included as the first input in the redeeming transaction. This ensures that only one rule (\ie, $\ruleR$) is executed by the redeeming transaction $\txT$.

\item The first input of $\txT$ specifies $\txf{spent}=\false$. Overall, this ensures that $\txo^{\ruleR}$ can never be consumed.

\item The second input of $\txT$ spends a non-contract output and its value is equal to $\txT.\txf{fee}$. This allows the input to pay transaction fees without affecting the contract state and balance. 

\item If $k$ is the cumulative number of variables and map accesses that are read by rule $\ruleR$, then the script checks that the next $k$ inputs specify $\txf{spent}=\false$ and refer to contract outputs with a $\ruleinline{"default"}$ or $\ruleinline{"non-default"}$ datum, reading the values in the process. The script checks that the hash-keys in the input datums are the expected ones (so that we do not read some other part of the state). Moreover, the rule parameters are read from the $\txf{redeemer}$ field of the first input of $\txT$. After this step, the script is able to evaluate all the expressions occurring in $\ruleR$.

\item If $n$ is the number of \ruleinline{receive} preconditions of $\ruleR$, then the script checks that the next $n$ inputs spend non-contract outputs contributing the exact values specified in those \ruleinline{receive} preconditions.

\item The expression in the $\ruleinline{require}$ precondition of $\ruleR$ evaluates to $\true$.



\item If $m$ is the number of \ruleinline{send} statements in $\ruleR$, then the script checks that the first $m$ outputs in $\txT$ set $\txf{inContract} = \false$ and pay the intended amount of tokens to the intended recipients, as specified by the rule.

\item Finally, the script checks that the remaining inputs and outputs perform the state update mandated by the rule. To this purpose, the script first computes the list of the remaining inputs $\vec{in}$ and the list of state updates $\vec{u}$. The script checks that the inputs in $\vec{in}$ have $\txf{spent}=\true$. Then it computes $\vec{out} = \outMap(\vec{in}, \vec{u})$ as per \Cref{fig:output-list}. If that fails, then $\txT$ is rejected. Otherwise, the script verifies that the remaining outputs of $\txT$ are equal to $\vec{out}$. In this way, this step ensures that there are no extraneous inputs or outputs \wrt those needed to implement the intended state update.

\end{itemize}

By construction, when the ``logic'' output $\txo^{\ruleR}$ is referred as an input in $\txT$, the contract state must be updated according to rule $\ruleR$. We prevent the contract state to be modified in any other way from those specified by the rules. This is done by making the script $\txscript_{state}$ of all state outputs to require the presence of some ``logic'' output $\txo^{\ruleR}$. More precisely, $\txscript_{state}$ 
checks that in the redeeming transaction $\txT$ the first input refers to a contract output whose $\txf{datum}$ is $\ruleinline{"logic"}$.

\paragraph{Example}
Consider a contract with variables \ruleinline{a,w,y,z} and map \ruleinline{m}, having the following rule:
\begin{lstlisting}[language=rulecode,morekeywords={example}]
example(x) {
   receive(z:T0);
   require(z>10);
   w=m[x]-y | m[z]=7+m[1] | a.send(1:T1);
}
\end{lstlisting}
Consider the execution of the rule in the state $\cstate{}$ where: 
\begin{gather*}
    \cstate{\ruleinline{y}} = 3 \quad 
    \cstate{\ruleinline{w}} = 2 \quad 
    \cstate{\ruleinline{z}} = 15 \quad
	\cstate{\ruleinline{a}} = \ruleinline{"pubkey_a"}
	\\
	 \cstate{\ruleinline{m}} = [14 \mapsto 1][27 \mapsto 3]
\end{gather*}
and let the contract balance be $\bal = \walenum{\waltok{100}{\tokT[1]}}$.

The flattened state map $\cstateOut{}$ takes as input the hashes of variable names and map location. These are generated as in \Cref{eq:sigma-var-definition,eq:sigma-map-definition}. Assume that they are ordered as follows:
\begin{align*}
	\minHash &< h_y < h_w < h_z < h_a <
	\\ 
	&< h_{m[1]} < h_{m[14]} < h_{m[15]} < h_{m[27]} < 
    \maxHash
\end{align*}

We now craft a transaction $\txT$ that executes the rule with parameter $\ruleinline{x} = 27$.
The first input of $\txT$ must refer to the output $\txo^{\ruleinlinesmall{example}}$ that encodes the rule logic. Moreover, this input must have $\txf{redeemer}$ set to $27$, specifying the value of $\ruleinline{x}$.
The second input of $\txT$ must spend a previous output contributing the fees.
Inputs from the third to the seventh access (without spending them) the state outputs relative to the values read by the contract.
More precisely, we have: $\txo_{h_y}$ for \ruleinline{y}, $\txo_{h_z}$ for \ruleinline{z}, $\txo_{h_a}$ for \ruleinline{a}, $\txo_{h_a, h_{m[14]}}$ for \ruleinline{m[1]} (since $h_a <  h_{m[1]} < h_{m[14]}$), and $\txo_{h_{m[27]}}$ for \ruleinline{m[x]}, in that order.
The values stored in the outputs referenced from these inputs are then used throughout the script to evaluate the expressions.
The eight input must spend an output contributing $\waltok{15}{\tokT[0]}$ to satisfy the \ruleinline{receive(z:T0)} precondition.
Finally, inputs from the ninth to the twelfth spend the outputs relative to the state updates. Note that the updates are $\vec{u} = (h_w \mapsto 0),(h_{m[15]} \mapsto 7)$ since in the state $\cstate{}$ the expression \ruleinline{m[x]-y} evaluates to $3-3=0$, while \ruleinline{z} evaluates to $15$ and \ruleinline{7+m[1]} evaluates to $7+0=7$.
To achieve the wanted update $\vec{u}$, we include in $\txT$ the inputs $\vec{in}$ specified by the input generation algorithm of~\Cref{subsec:state-updates}. These include
$\txo_{h_y,h_w}, \txo_{h_w}, \txo_{h_w,h_z}$ which are needed for $h_w \mapsto 0$ and $\txo_{h_{m[14]}, h_{m[27]}}$ which is needed for $h_{m[15]} \mapsto 7$. These inputs have $\txf{spent}$ set to $\true$.

\smallskip
We also generate the outputs for the updates, according to $\outMap(\vec{in},\vec{u})$. Since $h_w \mapsto 0$ causes \ruleinline{w} to change from a non-default value to the default one, the inputs $\txo_{h_y,h_w}, \txo_{h_w}, \txo_{h_w,h_z}$ are merged into a single new output $\txo'_{h_y,h_z}$.
Instead, since $h_{m[15]} \mapsto 7$ causes \ruleinline{m[15]} to change from the default value to a non-default one, the input $\txo_{h_{m[14]}, h_{m[27]}}$ is split into three new outputs $\txo'_{h_{m[14]},h_{m[15]}}$, $\txo'_{h_{m[15]}}$, and $txo'_{h_{m[15]},h_{m[27]}}$.

\smallskip
The outputs of $\txT$ also include an output to implement the \ruleinline{send} statement, transferring $\waltok{1}{\tokT[1]}$ to the public key $\ruleinline{"pubkey_a"}$ (the result of the evaluation of \ruleinline{a}). This output has $\txf{inContract}$ set to false, and the owner of the corresponding private key can freely spend it.

After the transaction is appended to the blockchain, the state of the contract is updated to 
\begin{gather*}
  \cstatei{\ruleinline{y}} = 3 \ \  \cstatei{\ruleinline{w}} = 0 \ \ \cstatei{\ruleinline{z}} = 15 \ \ \cstatei{\ruleinline{a}} = \ruleinline{"pubkey_a"}
\\
	 \cstatei{\ruleinline{m}} = [14 \mapsto 1][15 \mapsto 7][27 \mapsto 3]
\end{gather*}
and the contract balance is updated to:
\[
\bal' = \walenum{ \waltok{15}{\tokT[0]}, \waltok{99}{\tokT[1]}}
\]

\paragraph{Contract deployment}
To deploy a new contract instance $\confContr[\ctrId]{\vec{\ruleR}, \cstate{}}{\bal{}}$ we need to append a suitable transaction $\txT[init]$ to the blockchain. The first input of $\txT[init]$ must spend a previous non-contract output, contributing the fees. We set $\txT[init].\txf{fee}$ to its value.
As specified by the validity conditions (\cref{validation:ctrId-creation} at page~\pageref{validation:ctrId-creation}), the hash of this input also determines a fresh contract identifier $\ctrId$. Accordingly, we set $\txT[init].\txf{ctrId}$ to this value.
We then fill the outputs of $\txT[init]$, adding one contract output $\txo^{\ruleR}$ for each rule $\ruleR\in\vec{\ruleR}$, as well as the sequence of state outputs $\txo_{\minHash,h_0},\txo_{h_0},\txo_{h_0,h_1},\txo_{h_1},\ldots,\txo_{h_n,\maxHash}$ according to the initial state $\cstate{}$ of the contract.
These outputs have zero $\txf{value}$, and have as their scripts the ones described before.

If the required initial balance $\bal{}$ is non-zero, we also include in $\txT[init]$ one or more (non-contract) spent input,
whose overall value amounts to $\bal$. Since all the outputs of $\txT[init]$ have zero value, the currency received from the inputs is transferred to the contract balance.
The $\txf{redeemer}$ fields of all the inputs is set so to satisfy the scripts of the outputs they are spending. The $\txf{spent}$ field of all the inputs is set to $\true$.
Finally, $\txT[init]$ is signed by all the users whose authorization is required to spend the inputs.

\subsection{Discussion}
\label{subsec:discussion}

We now discuss several aspects of our approach, as well as a few possible variants.

\paragraph{Security of the compiler}
The security of our approach hinges on precisely controlling the form of the contract outputs, \ie those UTXOs having $\txf{inContract} = \true$ and the contract $\txf{ctrId} = \gamma$ in their transaction.
Indeed, it is crucial that such UTXOs correctly represent the current contract state as well as the logic operations corresponding to contract rules. More specifically, the contract outputs must either have $\txf{datum} = \ruleinline{"logic"}$ and check the correct application of a contract rule, or have 
a $\txf{datum}$ of one the forms $\ruleinline{"default"}$/$\ruleinline{"non-default"}$ and contribute to represent the contract state as in~\Cref{eq:Sigma-state}.

Breaking this invariant would allow a malicious user to disrupt the execution of the contract. Adding a new $\ruleinline{"logic"}$ contract output allows updating the contract state without following the contract rules. Removing a $\ruleinline{"logic"}$ contract output would instead prevent the execution of a contract rule. Moreover, adding a spurious state output ($\ruleinline{"default"}$/$\ruleinline{"non-default"}$) could cause confusion on the value of a contract variable or map location: reading the state could then use one of the two different values on the blockchain, potentially leading to vulnerabilities. Finally, removing a state output in an uncontrolled way can prevent reading and updating that part of the state, effectively preventing the execution of the rules which attempt that.

This invariant on the contract outputs is enforced at multiple levels. First, our blockchain model prevents users from freely creating new UTXOs in an existing contract.
To see why, assume we have an existing contract with $\txf{ctrId}= \gamma$. The validation conditions for new transactions
(see \cpageref{validation:ctrId-creation,validation:ctrId-preservation}) allow new contract outputs to be created only in two cases:
\begin{itemize}
    \item \Cref{validation:ctrId-preservation} allows to create new contract outputs $\txo'$, but only when at least one input refers to a previous contract output $\txo$ having the same $\txf{ctrId}$. Hence, one can create contract outputs $\txo'$ with $\txf{ctrId} = \ctrId$ only when the previous script $\txo.\txf{script}$ accepts it.

    \item \Cref{validation:ctrId-creation}
    allows to create new contract outputs $\txo'$, but only when there are no input referring to a contract output $\txo$, and when the $\txf{ctrId}$ of $\txo'$ is generated from a spent input. This makes $\txf{ctrId}$ a fresh contract identifier because of the no-double-spending property. Hence, this $\txf{ctrId}$ must be different from $\ctrId$, making the contract output $\txo'$ belong to a new contract.
\end{itemize}

Consequently, the creation of new contract outputs inside $\txf{ctrId}= \ctrId$ can only be done with the permission of one of the previous output scripts within the contract. This allows to enforce our invariant at the script level, by making the contract-related scripts verify the newly created contract outputs. Indeed, no new $\ruleinline{"logic"}$ outputs are accepted, and the new $\ruleinline{"default"}$/$\ruleinline{"non-default"}$ outputs are only accepted when they are part of a state update mandated by some contract rule.

\paragraph{Simplicity of the scripts}
As discussed in~\Cref{subsec:contract-logic} our scripts have to perform several checks on the fields of the redeeming transaction $\txT$ and the UTXOs referred to by $\txT$.
We remark that such checks do not involve any other parts of the blockchain, so they can be performed within the minimalistic script model described in~\Cref{sec:blockchain}.

We note that the mentioned checks would also be feasible in a loop-free script language. Indeed, the most complex check our scripts need to perform is the one on the outputs defined in~\Cref{fig:output-list}. While that algorithm is presented as a recursive function, the number of recursive calls it performs is at most $2n$
where $n$ is the number of the assignments in the contract rule effect, which is statically known.
It is therefore feasible to unfold the recursion up to that level and obtain an equivalent loop-free script.

\paragraph{Improving the compiler}
The transactions produced by our compiler can be optimized and made slightly more efficient in a few scenarios. For instance, when updating the state according to $\vec{u} = (h \mapsto v)$ it is possible that the new value $v$ is actually the same as the old value. Our compiler always consumes the old state output $\txo$ and creates a new one $\txo'$, even the value does not change. In such case, it would suffice reading $\txo$ using a transaction input with $\txf{spent} = \false$ and check that the old value is indeed $v$. 
Doing so would be more efficient since it would generate smaller transactions (one fewer output), possibly reducing the fees.
Further, it would increase the opportunities of parallelization by avoiding conflicts with other transactions reading $\txo$ (see~\Cref{sec:parallelization}).

Our compiler can also generate multiple inputs for the same part of the state, as it happens when a contract variable is both read and updated. This could be optimized by omitting the redundant inputs, at the cost of some additional complexity in the scripts.

\paragraph{Closing contracts}
Our contracts do not feature any means of termination. Once deployed, a contract lives forever: its logic outputs and state outputs are never fully removed from the UTXO set. At best, the number of state outputs can decrease when setting some part of the state to the default value. We could extend our contract model and  compiler to allow a complete shutdown of the contract. To achieve that, a special $\ruleinline{"logic"}$ output could verify that a suitable programmed precondition for the shutdown is satisfied, and instantly consume all the $\ruleinline{"logic"}$ outputs, effectively forbidding any new firing of contract rules. At the same time, a new $\ruleinline{"logic"}$ output $\txo_{del}$ is spawned to start the consumption of all the existing state outputs, following the order given by the hashes stored in their $\txf{datum}$ fields, from $\minHash$ to $\maxHash$. This can be realized by making $\txo_{del}$ keep the hash $h$ of the next state output to consume in $\txo_{del}.\txf{datum}$. The only way to spend $\txo_{del}$ would then be to consume the outputs $\txo_{del},\txo_{h,h'},\txo_{h'}$ (for any $h'$) and produce a new output $\txo_{del}'$ with a new hash $h'$ in its $\txf{datum}$. In this way, the deletion process can continue until $\maxHash$ is found, at which point no new output needs to be produced, completing the contract shutdown.

\paragraph{Avoiding spam transactions}
Our logic and state outputs have their $\txf{value}$ field set to $0$, so that creating a new state output only costs the transaction $\txf{fee}$. This might allow contracts to generate an excessively large amount of UTXOs, polluting the UTXO set that validator nodes must handle. To incentivize a more conservative use of state storage, fees could be increased for state outputs. For instance, the fee could be made proportional to the transaction size as in Cardano~\cite{CardanoFeeEstimator}.

Alternatively, a small ``collateral'' value could be deposited in the state outputs' $\txf{value}$. Such collateral could then be refunded when the state output is consumed (including when the contract is shut down). This would be analogous to the ``rent'' mechanism used by Solana~\cite{solana-white-paper}.

\section{Parallel validation of transactions}
\label{sec:parallelization}

\paragraph{Conflicts}
In order to parallelize the validation of transactions, it is paramount to be able to understand the nature of conflicts that can hinder or prevent parallel execution.
Such conflicts can be observed at the contract (\rulelang) level or at the transaction (\hutxo) level.

A first kind of contract-level conflict is caused by rule preconditions becoming false. For example, consider a user who wants to call rule $\ruleinline{f}$, another users who wants to call rule $\ruleinline{g}$, and the effect of $\ruleinline{f}$ makes the precondition of $\ruleinline{g}$ false, and vice versa.
There is no way to perform both $\ruleinline{f}$ and $\ruleinline{g}$, whether in parallel or sequentially: only one of them can be performed. We name this a \emph{hard conflict}. This class of conflicts also includes cases where $\ruleinline{f}$ and $\ruleinline{g}$ attempt to withdraw tokens from the contract balance, but there are not enough for both operations.

A different kind of contract-level conflict is caused by data dependencies: for example, when $\ruleinline{f}$ modifies some parts of the state that are then accessed by $\ruleinline{g}$. This is not a problem if validation is sequential, but hinders parallel validation, since we can not run $\ruleinline{g}$ until $\ruleinline{f}$ has modified the state. We name this a \emph{soft conflict} since the operations can still be executed, just not in parallel.

Contract-level conflicts directly induce transaction-level conflicts. However, at the transaction level there also arise conflicts that do not have a contract-level counterpart.

The simplest case of a transaction-level conflict is a double spending attempt, \ie two transactions trying to spend the same input. 
This is again a hard conflict, since, due to the validation rules, only one of the two transactions can be performed.

Note that double spending can also arise when compiling two rule invocations having a soft conflict at the contract level, if the compilation is performed relatively to the \emph{same state outputs}.
For instance, consider a contract with rule $\ruleinline{f(v) \{ x = v; \}}$, which is being invoked as $\ruleinline{f(1)}$ and $\ruleinline{f(2)}$.
Assume that in the current blockchain state, the value for variable $\ruleinline{x}$ is represented by output $\txo_x$.
Compiling the two invocations would lead to two transactions both spending $\txo_x$ and generating $\txo_x'$ and $\txo_x''$ for the new values $1$ and $2$, respectively.
This causes a double spending attempt on $\txo_x$, hence a hard conflict at the transaction level, even if it is only a soft conflict at the contract level.
To avoid the hard conflict, we must take state updates into account when compiling the two rule invocations.
We first compile $\ruleinline{f(1)}$ and obtain a transaction that spends $\txo_x$ and produces $\txo_x'$ for the new value $1$. Then, we compile $\ruleinline{f(2)}$ according to this new state, producing a transaction that spends $\txo_x'$ and produces $\txo_x''$ for the new value $2$.
In this way, the two transactions form a chain because of their dependency.
Now we only have a soft conflict at the transaction level, since no double spending is attempted.
Both transactions can be added to the blockchain, even if some parallelism is lost.

Finally, we remark that two operations that have no conflicts at the contract-level may still cause a conflict at the transaction-level due to a \say{implicit} data dependency. 
Consider a contract with rules:
\begin{lstlisting}[language=rulecode, morekeywords={set_x,set_y}]
set_x(v) { x = v; }
set_y(v) { y = v; }
\end{lstlisting}
and assume that the current state associates to each variable and map the default value.
Two rule invocations $\ruleinline{set\_x(1)}$ and $\ruleinline{set\_y(2)}$ apparently are not in conflict, since they affect distinct variables.
However, after compiling these invocations to transactions, both variables $\ruleinline{x}$ and $\ruleinline{y}$ are represented by the \emph{same} state output $\txo_{\minHash,\maxHash}$, hence a double spending hard conflict appears.
Again, as remarked above, this can be turned into a soft conflict by taking the updated state into account: we first generate a transaction that consumes 
$\txo_{\minHash,\maxHash}$ and produces
$\txo_{\minHash,x}\txo_x\txo_{x,\maxHash}$
to perform $\ruleinline{set\_x(1)}$, and then generate another transaction that 
consumes $\txo_{x,\maxHash}$ and produces
$\txo_{x,y}\txo_y\txo_{y,\maxHash}$
to perform $\ruleinline{set\_y(2)}$.
In this way, we only cause a soft conflict.

\todo{capire se conviene spostare} Since \hutxo --- as well as eUTXO --- does not feature contract-to-contract calls, so avoiding conflicts between the states of different contracts. Note in passing that, even if one opted to extend our model so to allow contract-to-contract calls, one would still be able to detect conflicts easily, since the part of the state which is being affected must still be explicitly mentioned in the transaction inputs.


\paragraph{Parallel validation}

Our validator takes as input a sequence of transactions, constructs its longest conflict-free prefix, and sends it to a pool of worker threads for validation.
%
Then, it commits to the blockchain the transactions that passed validation, so updating the blockchain state, and restarts by processing the next prefix.

Detecting conflicts in our \hutxo model can be done efficiently.
Note that we can focus on transaction-level conflicts, since each contract-level conflict is also a conflict at the level of transactions. 
%
Finding double spending attempts is straightforward since each transaction explicitly mentions the outputs it is trying to consume or to read (through inputs with $\txf{spent} = \true$ or $\false$, respectively).

Dependency conflicts (either hard or soft) can also be detected in a similar fashion, by examining the explicit inputs and outputs of transactions.

For balance-related conflicts, we adopt a conservative execution approach.
We first compute the tokens that each transaction is going to consume from the contract balance, as specified in~\Cref{eq:account-update}. Then, if such tokens are still available in the contract balance, we make a thread validate the transaction while temporarily marking those tokens as frozen, making them unavailable for other transactions. If the transaction is ultimately accepted, the tokens are consumed, while on rejection they are thawed. We call our approach as conservative since we only validate transactions when we are sure there is enough contract balance for them.

Our parallel validation algorithm is efficient. Beyond the standard execution of scripts and verification of digital signatures, it requires only a modest amount of additional bookkeeping for the affected UTXOs and the contract balances.
If we use a balanced search tree to store those, we only pay an additional logarithmic cost for each input and output, and each involved contract.
This can be further improved using hash tables, obtaining a constant overhead.

\section{Experimental evaluation}
\label{sec:experiments}

We implement our approach as a prototype tool named \toolname, featuring:
\begin{itemize}

\item a blockchain node simulator for our \hutxo model, comprising two alternative transaction validators: a \emph{sequential} one that only uses one thread, and a \emph{parallel} one that can exploit a thread pool;

\item routines to generate the transactions needed to deploy and execute \rulelang contracts, coherently with the \rulelang compiler; 

\item a benchmark of use cases, to be discussed below.

\end{itemize}

To make our experimental evaluation realistic, we use the same cryptographic primitives as Cardano, \ie 
BLAKE2b-512 for hashing, and Ed25519 for signatures. 
For simplicity, our simulator diverges from Cardano in some aspects, which however only affect the results of the experiments in a marginal way. 
First, \toolname only focusses on the execution of contracts, and does not include those functionalities of validators that are orthogonal to that aim (\eg, the consensus protocol).
Furthermore, we simplify the generation of transactions in our experiments:
\begin{itemize}
\item First, we use sequential numbers instead of hashes as transaction IDs, since it makes it easier to generate chains of transactions that refer to each other.
\iftoggle{arxiv}{
The actual cryptographic hash function is still used for all other purposes, including reading and updating the contract state as detailed in~\Cref{sec:compilation}, and in particular in~\Cref{eq:sigma-var-definition,eq:sigma-map-definition} which define the expected hashes that occur in state outputs.
}{
The actual cryptographic hash function is still used for all other purposes, including the flattening of the state
mentioned in~\Cref{sec:overview}.
}
\item Second, we do not actually sign the transactions: instead, when a signature on a transaction would need to be checked, \toolname verifies a fixed signature on a fixed message. We ensure that this step is not optimized away when compiling our tool, so not to invalidate the performance measurements.
\end{itemize}

Our simulator was developed in Rust, and amounts to ${\sim}{3500}$ Lines of Code and $125\mathrm{KB}$.
The experiments code instead amounts to $2200$ LoC and $73\mathrm{KB}$.

\subsection{Experimental setup}
\label{sec:experimental-setup}

For each benchmark we setup and execute a series of experiments, each comprising the following actions:
\begin{enumerate}
\item We fix the number $N_{\it threads}$ of worker threads available for the node, beyond a main thread.
\item We generate a sequence $\vec{T}$ of transactions to be validated, representing the interactions of the users with the contracts considered in the experiment.
This sequence does not include \emph{hard conflicts}, but does include some \emph{soft conflicts} in the form of transaction dependencies, \ie transactions whose inputs refer to the outputs of previous transactions in the sequence.
\todo{se si riesce, scrivere in positivo, dicendo che si potrebbe simulare anche il miner, ma non lo facciamo perché il computational overhead della scelta delle tx da includere nel blocco è ortogonale alla loro validazione}
This choice is coherent with simulating a \emph{validator} that is verifying proposed blocks whose transactions have a very high chance to be free from hard conflicts. By contrast, we do not simulate \emph{miners} that build blocks selecting transaction from the mempool, which instead contains arbitrary transactions from the users, often involving hard conflicts.

\item We start a timer, and proceed to validate the transactions in the sequence $\vec{T}$. More precisely:
\begin{itemize}
\item If $N_{\it threads} = 0$ we run the \emph{sequential} validator, which processes the transactions in $\vec{T}$ one by one.
\item Otherwise, a pool of $N_{\it threads}$ worker threads is created. The main thread repeatedly scans the sequence $\vec{T}$, searching for the longest prefix of conflict-free transactions (as detailed in~\Cref{sec:parallelization}), and sends them to the worker threads for validation. When the worker threads complete a batch of transactions, the main thread commits the valid transactions in the batch on the ledger.
\end{itemize}
\item Finally, after the whole sequence $\vec{T}$ has been processed, we stop the timer and report its value.
\end{enumerate}

All the measures are the average over 10 runs on a 3GHz 64-bit Intel Xeon Gold 6136 CPU and a GNU/Linux OS (x86\_64-linux) with 8 cores and 64 GB of RAM.

\begin{figure*}
\vspace{-5pt}
\begin{subfigure} 
    \centering
    \begin{minipage}{0.495\textwidth}
        \centering
\centering
\scriptsize
\begin{tikzpicture}
\begin{axis}[
width  = 1\linewidth,
height = \iftoggle{arxiv}{8cm}{6cm},
legend style = {at={(0.05,0.85)},anchor=west},
legend style={font=\tiny},
xlabel={Num. Users},
scaled x ticks=base 10:0,
xtick={1000,5000,10000,15000,20000},
xticklabel style = {rotate=0,anchor=north},
\iftoggle{arxiv}{}{xlabel style={yshift=0.2cm}},
ylabel={Time (s)},
scaled y ticks=base 10:0,
ytick scale label code/.code={},
yticklabel={\pgfmathprintnumber{\tick}},
ylabel absolute, 
ylabel style={yshift=-0.4cm},
cycle list name=exotic,
thick
]
\pgfplotstableread[col sep=comma]{results/cf-time.csv}\data
\addplot table[x index = {0}, y index = {1}]{\data};
\addplot table[x index = {0}, y index = {2}]{\data};
\iftoggle{arxiv}
{\legend{centralized (sequential),distributed (sequential)}}
{\legend{Time centr/seq,Time distr/seq}}
\end{axis}
\begin{axis}[
axis y line*=right,
axis x line=none,
width  = 1\linewidth,
height = \iftoggle{arxiv}{8cm}{6cm},
ylabel={Space (GB)},
scaled y ticks=base 10:0,
ytick scale label code/.code={},
ylabel absolute, 
y label style={at={(axis description cs:1.25,0.5)}},
legend style = {at={(0.05,0.65)},anchor=west},
legend style={font=\tiny},
cycle list name=exotic,
thick,
dotted
]
\pgfplotstableread[col sep=comma]{results/cf-space.csv}\data
\addplot table[x index = {0}, y index = {1}]{\data};
\addplot table[x index = {0}, y index = {2}]{\data};
\iftoggle{arxiv}
{\legend{centralized (sequential),distributed (sequential)}}
{\legend{Space centr/seq,Space distr/seq}}
\end{axis}
\end{tikzpicture}
    \end{minipage}
    \begin{minipage}{0.495\textwidth}
        \centering

\centering
\scriptsize
\begin{tikzpicture}
\begin{axis}[
width  = 1\linewidth,
height = \iftoggle{arxiv}{8cm}{6cm},
\iftoggle{arxiv}
{legend style = {at={(0.95,0.65)},anchor=east}}
{legend style = {at={(0.95,0.65)},anchor=east}},
legend style={font=\tiny},
xlabel={Num. Users},
scaled x ticks=base 10:0,
xtick={1000,5000,10000,15000,20000},
xticklabel style = {rotate=0,anchor=north},
\iftoggle{arxiv}{}{xlabel style={yshift=0.2cm}},
ylabel={Speedup: distr/par \emph{vs.} distr/seq},
scaled y ticks=base 10:0,
ytick scale label code/.code={},
yticklabel={\pgfmathprintnumber{\tick}},
ylabel absolute, 
ylabel style={yshift=-0.4cm},
ytick={1,2,3,4,5,6,7},
cycle list name=exotic,
thick
]
\pgfplotstableread[col sep=comma]{results/cf-speedup.csv}\data
\addplot table[select coords between index={0}{7}, x index = {0}, y index = {1}]{\data};
\addplot table[select coords between index={0}{7}, x index = {0}, y index = {3}]{\data};
\addplot table[select coords between index={0}{7}, x index = {0}, y index = {4}]{\data};
\addplot table[select coords between index={0}{7}, x index = {0}, y index = {8}]{\data};
\legend{Sequential,1+2 threads,1+3 threads,1+7 threads}
\end{axis}
\end{tikzpicture}
    \end{minipage}
    \vspace{-10pt}
    \caption{Time, space, and speedup for the Crowdfund experiments.}
    \label{fig:cf}
\end{subfigure}
\vspace{10pt}
\begin{subfigure} 
    \centering
    \begin{minipage}{0.495\textwidth}
        \centering

\centering
\scriptsize
\begin{tikzpicture}
\begin{axis}[
width  = 1\linewidth,
height = \iftoggle{arxiv}{8cm}{6cm},
legend pos=south east,
legend style={font=\tiny},
xlabel={Conflict percentage},
xtick={0,10,20,30,40,50,60,70,80,90,100},
xticklabel style = {rotate=0,anchor=north},
\iftoggle{arxiv}{}{xlabel style={yshift=0.2cm}},
ylabel={Time (s)},
scaled y ticks=base 10:-3,
ytick scale label code/.code={},
yticklabel={\pgfmathprintnumber{\tick}},
ylabel absolute, 
ylabel style={yshift=-0.4cm},
cycle list name=exotic,
thick
]
\pgfplotstableread[col sep=comma]{results/map.csv}\data
\addplot table[x index = {0}, y index = {1}]{\data};
\addplot table[x index = {0}, y index = {2}]{\data};
\addplot table[x index = {0}, y index = {3}]{\data};
\addplot table[x index = {0}, y index = {4}]{\data};
\legend{Sequential,1+2 threads,1+3 threads,1+7 threads}
\draw[dotted,lightgray] (60,0) -- (54,3200);  
\end{axis}
\end{tikzpicture}
    \end{minipage}
    \begin{minipage}{0.495\textwidth}
        \centering
\centering
\scriptsize
\begin{tikzpicture}
\begin{axis}[
width  = 1\linewidth,
height = \iftoggle{arxiv}{8cm}{6cm},
legend pos=north east,
legend style={font=\tiny},
xlabel={Conflict percentage},
xtick={0,10,20,30,40,50,60,70,80,90,100},
xticklabel style = {rotate=0,anchor=north},
\iftoggle{arxiv}{}{xlabel style={yshift=0.2cm}},
ylabel={Speedup: distr/par \emph{vs.} distr/seq},
scaled y ticks=base 10:0,
ytick={1,2,3,4,5,6},
ytick scale label code/.code={},
yticklabel={\pgfmathprintnumber{\tick}},
ylabel absolute, 
ylabel style={yshift=-0.4cm},
cycle list name=exotic,
thick
]
\pgfplotstableread[col sep=comma]{results/map.csv}\data
\addplot table[x index = {0}, y index = {5}]{\data};
\addplot table[x index = {0}, y index = {6}]{\data};
\addplot table[x index = {0}, y index = {7}]{\data};
\addplot table[x index = {0}, y index = {8}]{\data};
\legend{Sequential,1+2 threads,1+3 threads,1+7 threads};
\end{axis}
\end{tikzpicture}
    \end{minipage}
    \vspace{-10pt}
    \caption{Timings and speedup for the Map experiments.}
    \label{fig:map}
\end{subfigure}
\iftoggle{arxiv}{\end{figure*}\begin{figure*}}
{\vspace{10pt}}
\begin{subfigure}
    \centering
    \begin{minipage}{0.495\textwidth}
        \centering
\centering
\scriptsize
\begin{tikzpicture}
\begin{axis}[
width  = 1\linewidth,
height = \iftoggle{arxiv}{8cm}{6cm},
legend pos=north east,
legend style={font=\tiny},
xlabel={$N_{\it threads}$},
xtick={0,1,2,3,4,5,6,7},
xticklabel style = {rotate=0,anchor=north},
\iftoggle{arxiv}{}{xlabel style={yshift=0.2cm}},
ylabel={Time (s)},
scaled y ticks=base 10:0,
ytick scale label code/.code={},
yticklabel={\pgfmathprintnumber{\tick}},
ylabel absolute, 
ylabel style={yshift=-0.4cm},
cycle list name=exotic,
thick
]
\pgfplotstableread[col sep=comma]{results/multisig-time.csv}\data
\addplot table[x index = {0}, y index = {1}]{\data};
\addplot table[x index = {0}, y index = {2}]{\data};
\addplot table[x index = {0}, y index = {3}]{\data};
\addplot table[x index = {0}, y index = {4}]{\data};
\legend{N=2,N=4,N=6,N=10}
\end{axis}
\end{tikzpicture}
    \end{minipage}
    \begin{minipage}{0.495\textwidth}
        \centering
\centering
\scriptsize
\begin{tikzpicture}
\begin{axis}[
width  = 1\linewidth,
height = \iftoggle{arxiv}{8cm}{6cm},
legend pos=north east,
legend style={font=\tiny},
xlabel={$N_{\it threads}$},
xtick={1,2,3,4,5,6,7},
xticklabel style = {rotate=0,anchor=north},
\iftoggle{arxiv}{}{xlabel style={yshift=0.2cm}},
ylabel={Speedup: distr/par \emph{vs.} distr/seq},
scaled y ticks=base 10:0,
ytick scale label code/.code={},
yticklabel={\pgfmathprintnumber{\tick}},
ylabel absolute, 
ylabel style={yshift=-0.4cm},
cycle list name=exotic,
thick
]
\pgfplotstableread[col sep=comma]{results/multisig-speedup.csv}\data
\addplot[blue,error bars/.cd, y dir=both, y explicit] table[y error plus expr=\thisrowno{2}-\thisrowno{1}, y error minus expr=\thisrowno{1}-\thisrowno{3}]{\data};
\end{axis}
\end{tikzpicture}
    \end{minipage}
    \vspace{-10pt}
    \caption{Timings and speedup for the Multisig experiments.}
    \label{fig:multisig}
\end{subfigure}
\vspace{10pt}
\begin{subfigure}
    \centering
    \begin{minipage}{0.495\textwidth}
        \centering
        \centering
\scriptsize
\begin{tikzpicture}
\begin{axis}[
width  = 1\linewidth,
height = \iftoggle{arxiv}{8cm}{6cm},
legend pos=north west,
legend style={font=\tiny},
xlabel={Num. Users},
xtick={1000,20000,50000},
scaled x ticks=base 10:0,
xticklabel style = {rotate=0,anchor=north},
\iftoggle{arxiv}{}{xlabel style={yshift=0.2cm}},
ylabel={Time (s)},
scaled y ticks=base 10:0,
ytick scale label code/.code={},
yticklabel={\pgfmathprintnumber{\tick}},
ylabel absolute, 
ylabel style={yshift=-0.4cm},
cycle list name=exotic,
thick
]
\pgfplotstableread[col sep=comma]{results/registry-time.csv}\data
\addplot table[x index = {0}, y index = {1}]{\data};
\addplot table[x index = {0}, y index = {3}]{\data};
\addplot table[x index = {0}, y index = {4}]{\data};
\addplot table[x index = {0}, y index = {8}]{\data};
\legend{Sequential,1+2 threads,1+3 threads,1+7 threads}
\end{axis}
\end{tikzpicture}
    \end{minipage}
    \begin{minipage}{0.495\textwidth}
        \centering
        \centering
\scriptsize
\begin{tikzpicture}
\begin{axis}[
width  = 1\linewidth,
height = \iftoggle{arxiv}{8cm}{6cm},
\iftoggle{arxiv}
{legend style = {at={(0.95,0.5)},anchor=east}}
{legend style = {at={(0.95,0.65)},anchor=east}},
legend style={font=\tiny},
xlabel={Num. Users},
xtick={1000,20000,50000},
scaled x ticks=base 10:0,
xticklabel style = {rotate=0,anchor=north},
\iftoggle{arxiv}{}{xlabel style={yshift=0.2cm}},
ylabel={Speedup: distr/par \emph{vs.} distr/seq},
scaled y ticks=base 10:0,
ytick scale label code/.code={},
yticklabel={\pgfmathprintnumber{\tick}},
ylabel absolute, 
ylabel style={yshift=-0.4cm},
ytick={1,2,3,4,5,6,7},
cycle list name=exotic,
thick
]
\pgfplotstableread[col sep=comma]{results/registry-time.csv}\data
\addplot table[x index = {0}, y index = {9}]{\data};
\addplot table[x index = {0}, y index = {11}]{\data};
\addplot table[x index = {0}, y index = {12}]{\data};
\addplot table[x index = {0}, y index = {16}]{\data};
\legend{Sequential,1+2 threads,1+3 threads,1+7 threads}
\end{axis}
\end{tikzpicture}
    \end{minipage}
    \vspace{-10pt}
    \caption{Timings and speedup for the Registry experiments.}
    \label{fig:registry}
\end{subfigure}
\end{figure*}


\begin{table*}[h]
\centering
\begin{filecontents*}{results/cf.csv}
Validator,c250,c500,c1000,c2500,c5000,c10000,c15000,c20000,d250,d500,d1000,d2500,d5000,d10000,d15000,d20000,d50000,d1000000
Sequential,0.040,0.092,0.244,1.026,3.522,16.172,41.945,86.758,0.047,0.094,0.191,0.491,0.963,1.943,2.929,3.908,10.827,200.446
1+1 threads,0.057,0.110,0.268,0.977,3.075,11.283,23.991,41.257,0.057,0.102,0.201,0.506,1.007,2.011,3.021,4.061,10.008,200.656
1+2 threads,0.067,0.117,0.280,1.017,3.221,11.702,26.067,45.100,0.037,0.059,0.143,0.265,0.525,1.033,1.538,2.051,5.088,101.515
1+3 threads,0.078,0.116,0.281,1.044,3.283,13.653,29.776,67.941,0.029,0.045,0.081,0.183,0.359,0.698,1.054,1.389,3.441,67.976
1+4 threads,0.087,0.147,0.281,1.116,3.515,13.236,30.089,61.785,0.026,0.038,0.064,0.144,0.276,0.539,0.801,1.063,2.609,51.101
1+5 threads,0.069,0.118,0.311,1.136,3.523,14.744,38.905,77.655,0.024,0.038,0.056,0.120,0.228,0.461,0.654,0.863,2.104,41.046
1+6 threads,0.057,0.129,0.321,1.109,3.540,15.176,39.192,77.371,0.026,0.041,0.052,0.103,0.195,0.375,0.556,0.741,1.786,34.335
1+7 threads,0.056,0.138,0.320,1.106,3.656,17.385,44.246,81.322,0.028,0.041,0.052,0.093,0.172,0.332,0.488,0.638,1.544,29.528
\end{filecontents*}
\csvreader[
    tabular = ccccccccccccc,
    table head = \toprule \textbf{Validator} & \textbf{C(250)} & \textbf{D(250)} & \textbf{C(500)} & \textbf{D(500)} & \textbf{C(1K)} & \textbf{D(1K)} & \textbf{C(10K)} & \textbf{D(10K)} & \textbf{C(20K)} & \textbf{D(20K)} & \textbf{D(50K)} & \textbf{D(1M)} \\\midrule,
    table foot = \bottomrule,
]{results/cf.csv}
{
Validator=\validator,c250=\cCCV,c500=\cD,c1000=\cM,c2500=\cMMD,c5000=\cVK,c10000=\cXK,c15000=\cXVK,c20000=\cXXK,d250=\dCCV,d500=\dD,d1000=\dM,d2500=\dMMD,d5000=\dVK,d10000=\dXK,d15000=\dXVK,d20000=\dXXK,d50000=\dDK,d1000000=\dMK}{%
\validator
& \cCCV
& \dCCV
& \cD
& \dD
& \cM
& \dM
& \cXK
& \dXK
& \cXXK
& \dXXK
& \dDK
& \dMK
}%
\caption{Timings for the Crowdfund experiments (C=centralized, D=distributed).}
\label{tab:cf}
\end{table*}

\subsection{Results}

The common goal of our experiments is to measure the speedup obtained by
distributing the contract state and by parallelizing the validation of transactions.
To this purpose, we consider a benchmark of contracts with different features.


\paragraph{Crowdfund}

We experiment with the crowdfund contract of~\Cref{sec:overview}.
The interest in this contract is that it includes methods that simultaneously update maps and transfer tokens, but still in a way that  leads to an efficient parallelization.
This was expected, because donors operate on disjoint parts of the state, and token transfers from the contract are always successful, so reducing conflicts.
We craft experiments for the centralized and distributed versions of the contract, for each value of $N_{\it threads}$ (so covering both sequential and parallel validators), and for varying numbers of donors.
In the transaction sequence $\vec{T}$ all the donors make their donations, and then claim a refund after the deadline.
%

\Cref{fig:cf} (left) shows the time (solid lines) and space (dotted lines) used by the sequential validator to process $\vec{T}$, for the centralized and the distributed versions of the contract.
In the figures, we denote by ``$1+n$ threads'' the runs of the parallel validator with $1$ main thread and $N_{\it threads}=n$ workers.
We see that as the number of donors grows, the centralized contract becomes less efficient than the distributed one. 
This was expected, since replicating the full state in each output takes more time and memory. Since each donation adds an entry in the map, the overall memory consumed is quadratic \wrt the number of donors, as confirmed by the figure. 
%
By contrast, the distributed contract scales linearly, making it possible to run it with a much larger number of users (up to 1M donors in our experiments, see~\Cref{tab:cf}). 
Furthermore, the size of transactions in the centralized contract grows linearly, while in the distributed one it is constant. Hence, in eUTXO blockchains the contract would only be usable for a small number of donors. \Eg, in Cardano transactions have a 16KB size cap, and so the contract cannot handle more than a few hundred donors.
\Cref{fig:cf} (right) shows, for the distributed version of the contract, the time speedup of the parallel validator over the sequential one for varying values of $N_{\it threads}$.
We observe that the speedup grows in the number of users, being only bounded by the number of available threads.
This increase in the speedup is coherent with the fact that the percentage of (soft) conflicts, which only depends on the number of users, decreases as this number increases. 
For instance, for 250 users we have $18.23\%$ of conflicts, and this value  drops to $1.64\%$ for 20K users.
%
These conflicts are mostly due to the splitting and merging of map intervals (as discussed at page~\pageref{page:state-updates}), which occur during the $\ruleinline{donate}$ and $\ruleinline{refund}$ operations.
%
%
%



\paragraph{Map}

We experiment with a basic map contract featuring only a single rule:
\begin{lstlisting}[
    language=rulecode,
    %caption={The Map contract in \rulelang.},
    %captionpos=b,
    %label=fig:map-code,
    morekeywords={ConflictMap,inc}
]
contract ConflictMap {
  map (int => int) m;
  inc(i) { m[i] = m[i] + 1; }
}
\end{lstlisting}

The purpose of this experiment is to measure the impact of \emph{soft conflicts} due to data dependencies on performance.
Invoking the $\ruleinline{inc}$ rule with distinct values of $\ruleinline{i}$ updates distinct parts of the state, allowing the parallel validator to fully exploit its threads,
while invoking the rule multiple times with the same $\ruleinline{i}$ causes a data dependency conflict, preventing the validator to parallelize the dependent invocations.
Given any probability value $p$, we generate the input transaction sequence $\vec{T}$ so that the probability of causing such a conflict is $p$, and observe the impact on performance.
%

\Cref{fig:map} displays the results of our experiment with different validators and values of $p$.
We observe that the parallel validator is generally faster than the sequential one when using at least $2$ worker threads.
The sequential validator becomes faster only when $p > 60\%$ (dotted vertical line in~\Cref{fig:map}), \ie when the conflict probability is very high.
This corresponds to the highly unrealistic scenario where more than half of the transactions update the same map location. 
%
%
When $p$ assumes more realistic values, we observe that the speedup gets fairly close to the number of workers, scaling linearly with them.




\begin{figure}
\begin{lstlisting}[
    language=rulecode,
    morekeywords={MultiSig_N,authorize,deposit,withdraw},
    caption={The Multisig wallet contract in \rulelang.},
    captionpos=b,
    label={fig:multisig-code}
]
contract MultiSig_N {
  const int K = N / 2;
  address owner = "owner_public_key";
  map (address => bool) auth;

  authorize(a) {
    require signedBy(owner);
    auth[a] = true;
  }
  
  deposit(x) { receive(x:T); }

  withdraw(a1,...,aK,b,x) {
    require(a1 < a2 && ... && aK-1 < aK
      && signedBy(a1) && ... && signedBy(aK)
      && auth[a1] && ... && auth[aK]);
    b.send(x:T);
  }
}
\end{lstlisting}
\end{figure}

\paragraph{Multisig wallet}
We consider a wallet contract with three rules: $\ruleinline{authorize}$, $\ruleinline{deposit}$, and $\ruleinline{withdraw}$
(see~\Cref{fig:multisig-code}).
The owner authorizes $N$ privileged users, half of which must agree to each withdraw operation by signing it. Instead, any user can deposit funds.
We measure the performance of the contract for $N=2,4,6,10$, so to observe the impact of multiple signature verification.
In each experiment, the sequence of transactions $\vec{T}$ authorizes $N$ users, and then performs $100K$ random operations. 
When the wallet holds no tokens, we perform a $\ruleinline{deposit}$ operation, otherwise we flip an unbiased coin and perform either 
a $\ruleinline{deposit}$ or a $\ruleinline{withdraw}$ (signed by $N/2$ users).
Validating a transaction also involves verifying an additional signature to pay the transaction fees.

The results are displayed in~\Cref{fig:multisig}.
As expected, the average running time roughly scales with the average number of verified signatures. Assuming that the number of deposits is close to the number of withdrawals, the number of verified signatures for $N$ participants is $S(N) = 100K \cdot (0.5 \cdot 2 + 0.5 \cdot (N/2+1)) = 25K \cdot N + 150K$.
Observing the timings in~\Cref{fig:multisig} (left) we see that the time for each $N$ is roughly proportional to $S(N)$, confirming that the cost of signature verification is the main factor affecting performance.
\Cref{fig:multisig} (right) shows the speedups of the parallel validator over the sequential one for a varying number of threads. We observe that the speedup does not depend on the value of $N$ but only on the number of threads, confirming that contracts that make heavy use of signature verification can fully exploit parallel validation.
Overall, we measured \mbox{${\sim}{1.5}\%$} of soft conflicts due to data dependencies. This is not affected in any significant way by the number of users $N$, nor by the number of threads.


\paragraph{Registry}


\begin{figure}
\begin{lstlisting}[
    language=rulecode,
    morekeywords={Registry,register,claim,own},
    caption={The registry contract in \rulelang.},
    captionpos=b,
    label={fig:registry-code}
]
contract Registry {
  const int register_accuracy = 10;
  const int claim_time = 100;
  map (hash => int) reserve;
  map (hash => int) priority;
  map (hash => address) owner;

  register(hh,t) {
    require(reserve[hh] == 0 
      && validFrom == t 
      && validTo == t+register_accuracy);
    reserve[hh] = t;
  } 
  claim(h,a) {
    require(reserve[hash(h,a)] != 0
      && (priority[h] == 0 || 
          priority[h] > reserve[hash(h,a)]));
    priority[h] = reserve[hash(h,a)];
  }
  own(h,a) {
    require(validFrom == reserve[hash(h,a)] + register_accuracy + claim_time
      && priority[h] == reserve[hash(h,a)]
      && owner[h] == 0);
    owner[h] = a;
  }
}
\end{lstlisting}
\end{figure}

We consider a notarization / registry contract that allows users to register a document and claim its ownership (see~\Cref{fig:registry-code}).
The contract implements an anti-spoof mechanism based on temporal constraints to prevent adversaries from intercepting registration requests and claim the documents as their own.
Users start by calling a rule to record in a map the \emph{hash of the hash} of their document, and a current timestamp.
This is needed to prevent frontrunning attacks where an adversary ``steals'' the actual document hash, and registers it before the legit user.
After that, the user calls another rule to reveal the actual hash, and store it in another map.
This rule allows the legit user (\ie, the one who was the first to register the double hash) to overwrite any adversaries' claims. 
Finally, the user who has claimed a hash can permanently record its ownership.
The corresponding rule uses time constraints to ensure that the legit user can reveal the hash, as per the second rule.
The interest in this contract is in the combination of maps and hash functions; hashes are computed in the second and third rule to obtain the double hash.
%
%
In each experiment, the input transaction sequence $\vec{T}$ starts with
$N$ distinct users invoking $\ruleinline{register}$.
Users then perform a $\ruleinline{claim}$ on their documents, wait for some deadlines to expire, and finally invoke $\ruleinline{own}$ to finalize their ownership claims.
\Cref{fig:registry} displays the timings and speedup for the distributed version of the contract.
We observe that the parallel validator is significantly faster than the sequential one. 
Using $1+2$ threads, the speedup is always greater than $1$ even for a small number of users (\eg, $1.47$ for $N = 250$). 
Using $1+7$ threads, the speedup is always greater than $1$, and it tends asymptotically to the bound $7$ as the number of users grows.
This is due to a low number of soft conflicts ($< 2\%$).

\section{Limitations}

We discuss a few limitations and shortcomings of our approach.

\paragraph{Blockchain model}

The \hutxo model originally combines several mechanisms in the realm of UTXO-based blockchains, such as contract IDs and
the segregation of the contract balance from the UTXOs.
These mechanisms can not be easily implemented on top of existing platforms like Cardano, but require adapting the underlying validation layer of the blockchain itself.
This is the reason why we performed our experiments in a node simulator,
instead of directly on-top of an existing blockchain.
%
For the sake of simplicity, in the \hutxo model we allowed transaction outputs carrying no cryptocurrency, but only data.
Using such an approach in the real world could attract \emph{spam} transactions, making it possible to flood the blockchain with a large amount of unspent outputs, which can be problematic for the blockchain nodes.
This can however be prevented by adding some anti-spam mechanism to the model. For instance, one could require a minimum amount of currency on each output, as a sort of collateral. This would cause participants to pay a little extra when a smart contract has to store new data, and receive that amount back when the data is finally deleted.
Alternatively, preventing spam could be integrated with
the fee mechanism, by making the transaction fees depend on the
number of new UTXOs spawned by the transaction.
Similar mechanisms are already being used in several blockchains (\eg Cardano).

\paragraph{Contract language}

Our \rulelang contract language is simple but not as expressive and full-featured as Solidity. In particular, \rulelang lacks loops, allowing contract developers to only define rules that run in constant time and space.
Consequently, some Solidity contracts can not be easily translated into \rulelang. For instance, a Solidity function can scan an array and sum its elements, if enough gas is provided for the whole computation.
In \rulelang, implementing this behaviour requires using \emph{multiple} rule invocations:
more precisely, the array could be represented as a \rulelang map, and scanning all its elements requires repeatedly invoking \rulelang rules.
Designing such rules is cumbersome, compared to Solidity, and special care must be taken to ensure atomicity.
Indeed, an adversary must be prevented to interfere with the computation by invoking other rules in the middle of it.
Remarkably, this limitation does not affect many real-world smart contracts. Even in Solidity, designers optimize gas consumption, and for that reason usually avoid scanning a dynamic data structure. Best practices suggest crafting smart contracts so that functions only require a small amount of gas.

\paragraph{Parallelization}
To take advantage of the inherent parallelism in the \hutxo model and \rulelang, it is important to have many updates to different parts of the smart contract state.
While this property is frequently enjoyed by contracts, not all of them are amenable to parallelization.
For instance, consider an Automated Market Maker contract that allows participants to perform swaps between two types of tokens
(\eg, as in \code{UniswapV2Pair}).
The state of such contract contains two integer values representing the reserves of each token type. Any token swap affects both values, also implicitly updating the exchange rate.
When several participants operate the AMM simultaneously, their transactions can not be parallelized, since they all read and write on the same part of the state.
Also note that the mere fact that a contract can be expressed in \rulelang does not guarantee that it can be efficiently parallelized.

As another shortcoming, our \hutxo model does not exploit all the theoretical opportunities for parallelization.
For instance, when performing multiple \emph{commutative} operations on one part of the contract state, it could be beneficial to store the affected data inside the \emph{account} instead of UTXO.
In that way, the operations no longer involve transactions trying to spend the same input, so they are no longer in conflict.
Out \hutxo model exploits this idea only for the account balance, but it could be extended to other parts of the state, on a case-by-case basis.

\paragraph{Experiments}

Our experiments involve synthetic data. Since our technique requires a custom blockchain model (\hutxo) and smart contracts specifically designed for it (\eg, using \rulelang), we do not have access to actual real-world transaction data to experiment with.
Ideally, one could try to replicate the data from an existing blockchain, adapting it to fit the \hutxo model. This task, however, is overwhelmingly complex: it would require selecting a large number of blocks, inspecting each transaction, finding the original source code of \emph{all} the involved contracts, manually translating them into \rulelang (assuming this is actually possible), and constructing analogous \hutxo blocks.

\section{Related work}
\label{sec:related}

Although the problem of parallelizing transaction execution 
is common to all kinds of blockchains, the solutions are
radically different depending on whether the blockchain
is UTXO-based or account-based.
More specifically:
\begin{itemize}

\item in UTXO blockchains, each output may be spent only once: this implies that, when validating a set of possibly conflicting transactions, different orderings will lead to different transactions being included in a block. 
Therefore, in order to effectively parallelize transactions one needs to decrease the conflict probability, \ie the probability that their spent inputs are not disjoint. 
In our approach, we achieved this by distributing the contract state across multiple UTXOs, and by moving the contract balance to an account, rather than keeping it into UTXOs (see~\Cref{sec:overview}).

\item in account-based blockchains, two transactions directed to the same contract can be executed in any order, but different orders may possibly result in different states (even when both transactions are included in the block).
Since validation must be deterministic, parallelizing transactions requires to precisely estimate the data dependencies between them, in order to detect if they can be executed in any order, resulting in the same state.
This can be done with different techniques, 
\eg optimistic execution~\cite{Dickerson17podc,Anjana19pdp},
static analysis~\cite{BGM21lmcs},
or by annotating transactions with data dependencies~\cite{solana-transactions}.


\end{itemize}

Furthermore, account-based blockchains can be partitioned into \emph{stateful} when a contract account carries the whole state of the contract besides the contract code, and \emph{stateless} when contract accounts carry no state.
Ethereum, Algorand, and Tezos are notable instances of the stateful account-based model, while Solana, Aptos and SUI are instances of the stateless one.
The stateful of stateless nature of a blockchain deeply affects the techniques to parallelize transactions and the obtained speedup.
We discuss below approaches to parallelize transactions in both models, relating them to ours.

\paragraph{Stateful account-based blockchains}

One the first works studying transaction parallelization in the account-based model is \cite{Dickerson17podc}, which defines a technique for the parallel formation and validation of blocks in Ethereum.
Block proposers use speculative execution techniques to find a possible schedule for the parallel execution of the transactions in a block.
The obtained schedule is then included in the block, so to allow validators to replicate the same parallel execution when validating the block. 
The speedup obtained on the proposed (synthetic) benchmark is 1.33x for block formation and 1.69x for block validation, using 3 threads.
A main difference between this approach and ours is that in the UTXO model no explicit schedule for the parallel validation is needed, since transaction inputs implicitly provide the dependencies among transactions.
Therefore, our technique does not require to include the schedule as meta-data in blocks, unlike~\cite{Dickerson17podc}. 
Another main difference is that, in the account-based model, the state is not explicitly stored within transactions and is computed during the validation process: for this reason, it is impossible to validate a transaction until all its dependencies are validated.
Instead, in our \hutxo model, transaction outputs store (parts of) the contract state: consequently, transactions depending on such outputs can have their scripts checked even when the validation of their dependencies is not yet completed (or even started).
Dependencies only matter in \hutxo when checking that the contract balance is non negative after each transaction. 
This check requires to compute the resulting contract balance after the execution of the dependencies: note that such computation is inexpensive, requiring only a few additions per transaction.
This in principle allows a theoretical speedup in \hutxo greater than that achievable in account-based models.
This is confirmed by our experiments in~\Cref{sec:experiments}, where we have achieved a speedup close to the number of available threads. 


The Groundhog model~\cite{Ramseyer24groundhog}  proposes a parallelization technique that, rather than
searching for a particular ordering of transactions, relies on smart contracts to perform commutative operations. This leads to the same result in any order, so they can be easily parallelized.
Since limiting contracts to only use commutative operations would make Groundhog rather inexpressive, a reserve-commit mechanism is exploited to also allow a few selected primitives that might not always commute.
Such primitives are realized by first checking a set of constraints (during the reserve phase) on the parallel threads, and then applying their effects (during commit), which are now guaranteed to commute.
Similarly to our model, some of the burden is placed on the contract code, which must strive to invoke the primitives without violating the constraints, so to allow for parallelization. 
Our model is arguably simpler to understand: \rulelang contracts use familiar programming constructs, and ensuring parallelism requires to avoid repeated accesses to the same part of the state.
By contrast, in Groundhog potentially non-commutative contract operations must be implemented as commutative operations subject to constraints, which is not as natural.

A few works~\cite{BGM21lmcs,Pirlea21pldi} study the use of static analysis to approximate the portion of the contract state affected by the execution of a transaction, and from this infer a parallel scheduling of transactions.
Compared to these works, parallelizing transactions in our \hutxo model does not rely on static analysis, since the parts of the contract state affected by their execution is already explicit in the inputs of transactions.

The work~\cite{BGM21lmcs} considers an abstract blockchain model, in order to achieve general results that are applicable both to UTXO-based and account-based blockchains.
Assuming a static analysis of the parts of the contract state that are read or written by transactions, they construct an \emph{occurrence net}~\cite{Best87tcs} that describes the possible concurrent executions of transactions, and prove that these  executions are semantically equivalent to the sequential ones.
Interpreting these occurrence nets as transaction schedules, they show how miners and validators can parallelize the execution of transactions, considering Bitcoin and Ethereum as case studies. 
Our work refines the results in~\cite{BGM21lmcs} regarding the parallelization of Bitcoin transactions, developing a technique that makes it possible to parallelize the execution of more expressive contracts, like the ones in Cardano~\cite{Rosetta25fgcs}.

The work~\cite{Pirlea21pldi} develops a static analysis of smart contracts (written in the Scilla contract language ~\cite{Sergey19pacmpl}) that allows to distribute transactions across multiple \emph{shards}, \ie partitions of the whole set of validators~\cite{Luu16sharding}.
Notably, while most works on sharding focus on simple asset-transfer transactions, \cite{Pirlea21pldi} was the first one to enable the parallel sharded execution of transactions targeting the \emph{same} contract.
To this purpose, the analysis in~\cite{Pirlea21pldi} soundly approximates the effects of a transaction in terms of which part of the state the transaction may write, and whether the writes that affect the same part of state commute.
This information is then used to partition the contract state, so that each shard can independently execute (in parallel to the other shards) the transactions that only affect parts of the state assigned to the shard.
Similarly to our work, also \cite{Pirlea21pldi} enables the fine-grained distribution of state, but in the account-based model, while our technique addresses the UTXO model.
Working on the account-based model allows~\cite{Pirlea21pldi} to perform further optimizations, \eg executing in parallel commutative operations on the same part of the state.
This is not possible in UTXO (as well as in our \hutxo), since each state update requires spending transaction outputs.
To make this optimization possible in \hutxo, we could move to an account-based part of the state for which commutative operations are frequent, similarly to what we have done for the contract balance.
Another key difference is that the sharding technique in~\cite{Pirlea21pldi} requires to statically assign each part of the contract state to shard, making that shard the unique executor of transactions that write that parts of the state.
This could give rise in an unbalanced workload between different shards, in case some parts of the state are more used than others.
Instead, in \hutxo model does not require such static assignment: each transaction can be executed by any worker thread. 
This avoids the above-mentioned problem of workload unbalance.



\paragraph{Stateless account-based blockchains}

In the stateless model, the contract state is scattered across multiple accounts, which must be supplied as parameters along with method invocations.
Therefore, to detect when two transactions targeting the same contract can be executed concurrently it suffices to check if their arguments are disjoint.
Based on this, various parallelization techniques are natively implemented in Solana~\cite{solana-white-paper,solana-parallelization}, SUI~\cite{sui-parallelization} and APTOS~\cite{aptos-parallelization,Gelashvili23ppopp}
and in other custom chains~\cite{Cheng21pvldb}.

A main difference between our approach and the parallelization techniques on stateless account-based blockchains is that
our parallelization is more fine-grained.
To illustrate, assume that the contract state includes a map. 
In stateless blockchains, this map will be usually stored in a \emph{single} contract account: therefore, if two transactions try to write the map (even at \emph{distinct} keys), they cannot be parallelized.
Instead, our fine-grained distribution of the state scatters the map across \emph{multiple} UTXOs: this allows two write operations at distinct keys to be parallelized, with high probability.

\paragraph{UTXO-based blockchains}

To the best of our knowledge, this is the first study to enable the parallel validation of transactions in the UTXO model, also considering transactions targeting the same contract. 
Other papers addressing the scalability problem in the UTXO model devise mechanisms to execute transactions across multiple shards~\cite{Al-Bassam18ndss,Fletcher-Smith23cnis}.
While these works support cross-shard contract transactions, they cannot parallelize transactions targeting the same contract --- so inefficiently processing high-throughput contracts.
In particular, each of the contracts in~\Cref{sec:experiments} would be executed sequentially by the same shard. 
The work~\cite{Muller23dltrp} considers the problem of conflicting transactions, devising an extended UTXO model that performs optimistic updates and tracks the dependencies causing the conflicts. Blockchain consistency is achieved through a conflict resolution mechanism.
By contrast, our approach is conservative: transactions are checked for conflicts before they are sent to worker threads, as discussed in~\Cref{sec:experimental-setup}. Also, writing different parts of the contract state does not create conflicts.

\paragraph{Implementing \hutxo on top of existing UTXO blockchains}

To the best of our knowledge, it is not possible to implement \hutxo on top of existing UTXO-based blockchains without extending the blockchain model in some way.
The closest target is Cardano, which is based on an extended UTXO model (eUTXO) featuring Turing-expressive scripts and tokens with programmable policies.
A main feature of \hutxo that is not present in Cardano is contract IDs, which are pivotal to partition UTXOs according to the contract of which they represent part of the state.
It is unclear to us whether the current features of Cardano can be exploited to implement contract IDs: the most promising approach  seems to involve token minting policies, and using custom tokens to represent the fact that an output belongs to a contract. 
Here, a contract ID would be represented as a freshly minted token type, and the minting policy should allow duplicating existing tokens to spawn new outputs within the contract.
This approach is similar to the one adopted in ~\cite{Vinogradova24fc}.
Instead, if contract IDs cannot be implemented on top of  Cardano, then Cardano should be extended either with native support for contract IDs, or with some lower-level feature that makes it possible to support them, such as \emph{neighbourhood covenants}~\cite{BLZ21csf}.

Another main feature of \hutxo which Cardano lacks is contract accounts.
As already discussed in~\Cref{sec:overview}, contract accounts could in principle be replaced with UTXOs: the contract balance would then be spread over a set of UTXOs.
This increases the complexity of token transfers from/to the contract and  leads to more conflicts, making parallelization less likely to happen.
While this solution would be suboptimal, at least it could be realized on top of the already-existing Cardano infrastructure.


\paragraph{Representing the contract state}

To represent the state of a contract we first flatten it into a single map $\cstateOut{}$.
This map is represented by storing the key-value pairs for non-zero values, and open intervals of keys for zero values (the default).
Using key-value pais and open intervals is similar to how DNSSEC~\cite{dnssec} represents its Resource Records (RRs).
DNSSEC uses both ``positive'' RRs to authoritatively resolve a domain name to an address, and ``negative'' RRs to authoritatively assert that no valid domain name exists within a certain interval.
After all the RRs for a domain are signed (possibly off-line), a DNS server is able to answer any query within that domain and certify its positive or negative answer.
Our approach in similar, in that a state output can attest either that $\cstateOut{}$ holds a given non-zero value at one point (positive information), or that $\cstateOut{}$ is zero inside an interval (negative information).
%
\iftoggle{arxiv}{
Unlike DNSSEC RRs, our state outputs are not signed: their authority comes instead from the invariant discussed in~\Cref{subsec:discussion}.}{
Unlike DNSSEC RRs, our state outputs are not signed: their authority comes instead from the blockchain and contract scripts invariants discussed in~\Cref{sec:overview}.
}

\paragraph{Contract IDs}

We exploit contract IDs to ensure contract isolation from maliciously created UTXOs which claim that some state part has a value chosen by the adversary.
Thwarting this attack requires constraining the use of contract IDs, both at the blockchain and contract script level.
A similar approach is found in \textsc{Zexe}~\cite{Bowe20sp} whose aim is the execution of smart contracts over a privacy-preserving UTXO platform (an extension of Zerocash~\cite{Sasson14Zerocash}).
Contract IDs are also used in \text{Zexe} to achieve contract isolation.
\textsc{Zexe} is the foundation of the \textsc{Aleo} blockchain which integrates an account-like public state with a UTXO-based private state.
The goals of \textsc{Zexe} differ from ours, focusing on the offline private execution of smart contracts over UTXO, instead of the parallel validation of transactions.

\paragraph{Languages for UTXO-based contracts}

The work~\cite{BMZ23ieeesp} proposes a high-level language (called \textsc{HeLLUM}) for UTXO-based contracts,  
and a compiler from \textsc{HeLLUM} 
to a clause-based intermediate-level language (called \textsc{Illum}), inspired by process algebras.
Compared to our \rulelang, \textsc{HeLLUM} is closer to standard imperative programming languages: even if it does not feature loops, it supports conditionals and sequential composition of commands.
By contrast, the effects in \rulelang must be written as a single simultaneous assignment.
Since the intermediate code transformations of the  \textsc{HeLLUM} compiler eventually produce a simultaneous assignment, it seems feasible to adapt the same technique so to compile \textsc{HeLLUM} into \rulelang, instead of \textsc{Illum}.
Another key difference between our work and~\cite{BMZ23ieeesp} is that \textsc{HeLLUM}/\textsc{Illum} contracts are executed in a centralized fashion, while \rulelang contracts enable the distribution of state and the parallel execution of transactions.
Other contract languages for UTXO blockchains, like \eg Plutus Tx and Aiken for Cardano, support a low-level programming style, where implementing a contract action amounts to checking, in a transaction redeem script, that the new contract state is a correct and authorized update of the old one. 
This requires the developer to reason at the level of the  transaction fields that encode the contract state; by contrast, our \rulelang allows us to abstract from low-level transactions.


\section{Conclusions}
\label{sec:conclusions}

We have presented a technique to improve the performance of smart contracts in the UTXO model, based on distributing the contract state and parallelizing the validation of transactions.
We discuss below a few possible alternatives\iftoggle{arxiv}{ to our approach}{}.

When dealing with large-state contracts, an alternative technique to reduce their footprint is to store only the \emph{hash} of the contract state in the transactions, while relegating the actual state in a separate data structure (possibly, in a transient way). 
This would require only one output, making the approach centralized.
Validating a transaction would still require validators to access the old and new states from the separate data structure, so that scripts can verify that the state was updated correctly.
In order to save space, old states could, in principle, be discarded from the data structure after several blocks are put on top of the blockchain after the latest state.
The main disadvantage of this approach, compared to ours, is that by removing the old states, the blockchain can no longer be fully revalidated: validators must blindly accept that old state updates were correct because several blocks have been appended after them.

A more sophisticated centralized approach would be to represent the state as a Merkle tree, storing the root hash in the transaction $\txf{datum}$.
Updating a state would then require to include as transaction witnesses/redeemers only those parts of the Merkle tree that are actually accessed by scripts, similarly to Taproot~\cite{BIP341} for Bitcoin.
A downside of this approach is that the transaction scripts are significantly more complex than ours, since they must traverse the state Merkle trees (which could represent dynamic data structures) and check that they are the intended ones.
By contrast, our \rulelang language allows us to seamlessly operate on the contract state, automatically generating the \hutxo scripts. 
Another main downside is that, since the approach is centralized, it prevents parallel updates to different parts of the state.

Smart contracts in the UTXO model suffer from the so-called ``UTXO congestion'' problem~\cite{cardano-utxocongestion}. When multiple users want to perform an operation that affects the same part of the state of the same contract, they all need to sign and broadcast transactions that attempt to spend the same UTXO. Only one of such transactions will be appended to the blockchain, updating the contract state, while all the others are invalid and then rejected. To perform the wanted operation, users must adapt their transactions to the new state output and sign it again. This process must be repeated by each user until they win the race.
To mitigate this issue, we could introduce a controlled form of \emph{transaction malleability}, allowing block proposers to adapt the inputs of the previously signed transactions to the new contract state outputs \emph{without} invalidating the signatures.
This would allow users to sign a transaction to order the execution of a contract operation ``in any state'', as in account-based blockchains.
Users must carefully choose whether to sign a transaction in the regular way or in this ``malleable'' way, since the latter might cause the operation to be executed in an unwanted state, or expose the user to transaction-ordering attacks~\cite{Babel23clockwork,Torres21frontrunner}.
\iftoggle{anonymous}{}{\paragraph*{Acknowledgments}

Work partially supported by project SERICS (PE00000014)
under the MUR National Recovery and Resilience Plan funded by the
European Union -- NextGenerationEU, and by PRIN 2022 PNRR project DeLiCE (F53D23009130001).}
\bibliographystyle{ACM-Reference-Format}
\bibliography{main}
}
{

\iftoggle{anonymous}{}{\paragraph*{Acknowledgments}

Work partially supported by project SERICS (PE00000014)
under the MUR National Recovery and Resilience Plan funded by the
European Union -- NextGenerationEU, and by PRIN 2022 PNRR project DeLiCE (F53D23009130001).}

\bibliographystyle{ACM-Reference-Format}
\bibliography{main}


\begin{thebibliography}{43}


\ifx \showCODEN    \undefined \def \showCODEN     #1{\unskip}     \fi
\ifx \showDOI      \undefined \def \showDOI       #1{#1}\fi
\ifx \showISBNx    \undefined \def \showISBNx     #1{\unskip}     \fi
\ifx \showISBNxiii \undefined \def \showISBNxiii  #1{\unskip}     \fi
\ifx \showISSN     \undefined \def \showISSN      #1{\unskip}     \fi
\ifx \showLCCN     \undefined \def \showLCCN      #1{\unskip}     \fi
\ifx \shownote     \undefined \def \shownote      #1{#1}          \fi
\ifx \showarticletitle \undefined \def \showarticletitle #1{#1}   \fi
\ifx \showURL      \undefined \def \showURL       {\relax}        \fi
\providecommand\bibfield[2]{#2}
\providecommand\bibinfo[2]{#2}
\providecommand\natexlab[1]{#1}
\providecommand\showeprint[2][]{arXiv:#2}

\bibitem[Al{-}Bassam et~al\mbox{.}(2018)]%
        {Al-Bassam18ndss}
\bibfield{author}{\bibinfo{person}{Mustafa Al{-}Bassam},
  \bibinfo{person}{Alberto Sonnino}, \bibinfo{person}{Shehar Bano},
  \bibinfo{person}{Dave Hrycyszyn}, {and} \bibinfo{person}{George Danezis}.}
  \bibinfo{year}{2018}\natexlab{}.
\newblock \showarticletitle{Chainspace: {A} Sharded Smart Contracts Platform}.
  In \bibinfo{booktitle}{\emph{Network and Distributed System Security
  Symposium ({NDSS})}}. \bibinfo{publisher}{The Internet Society}.
\newblock


\bibitem[Androulaki et~al\mbox{.}(2018)]%
        {Androulaki18eurosys}
\bibfield{author}{\bibinfo{person}{Elli Androulaki}, \bibinfo{person}{Artem
  Barger}, \bibinfo{person}{Vita Bortnikov}, \bibinfo{person}{Christian
  Cachin}, \bibinfo{person}{Konstantinos Christidis},
  \bibinfo{person}{Angelo~De Caro}, \bibinfo{person}{David Enyeart},
  \bibinfo{person}{Christopher Ferris}, \bibinfo{person}{Gennady Laventman},
  \bibinfo{person}{Yacov Manevich}, \bibinfo{person}{Srinivasan Muralidharan},
  \bibinfo{person}{Chet Murthy}, \bibinfo{person}{Binh Nguyen},
  \bibinfo{person}{Manish Sethi}, \bibinfo{person}{Gari Singh},
  \bibinfo{person}{Keith Smith}, \bibinfo{person}{Alessandro Sorniotti},
  \bibinfo{person}{Chrysoula Stathakopoulou}, \bibinfo{person}{Marko Vukolic},
  \bibinfo{person}{Sharon~Weed Cocco}, {and} \bibinfo{person}{Jason Yellick}.}
  \bibinfo{year}{2018}\natexlab{}.
\newblock \showarticletitle{{Hyperledger} {Fabric}: a distributed operating
  system for permissioned blockchains}. In
  \bibinfo{booktitle}{\emph{{EuroSys}}}. \bibinfo{publisher}{{ACM}},
  \bibinfo{pages}{30:1--30:15}.
\newblock
\urldef\tempurl%
\url{https://doi.org/10.1145/3190508.3190538}
\showDOI{\tempurl}


\bibitem[Anjana et~al\mbox{.}(2019)]%
        {Anjana19pdp}
\bibfield{author}{\bibinfo{person}{Parwat~Singh Anjana}, \bibinfo{person}{Sweta
  Kumari}, \bibinfo{person}{Sathya Peri}, \bibinfo{person}{Sachin Rathor},
  {and} \bibinfo{person}{Archit Somani}.} \bibinfo{year}{2019}\natexlab{}.
\newblock \showarticletitle{An Efficient Framework for Optimistic Concurrent
  Execution of Smart Contracts}. In \bibinfo{booktitle}{\emph{Euromicro Int.
  Conf. on Parallel, Distributed, and Network-Based Processing ({PDP})}}.
  \bibinfo{pages}{83--92}.
\newblock
\urldef\tempurl%
\url{https://doi.org/10.1109/EMPDP.2019.8671637}
\showDOI{\tempurl}


\bibitem[{APTOS Foundation}(2024)]%
        {aptos-parallelization}
\bibfield{author}{\bibinfo{person}{{APTOS Foundation}}.}
  \bibinfo{year}{2024}\natexlab{}.
\newblock \bibinfo{title}{Network - Blockchain - Execution}.
\newblock
  \bibinfo{howpublished}{\url{https://aptos.dev/en/network/blockchain/execution}}.
\newblock


\bibitem[Atzei et~al\mbox{.}(2018)]%
        {bitcointxm}
\bibfield{author}{\bibinfo{person}{Nicola Atzei}, \bibinfo{person}{Massimo
  Bartoletti}, \bibinfo{person}{Stefano Lande}, {and} \bibinfo{person}{Roberto
  Zunino}.} \bibinfo{year}{2018}\natexlab{}.
\newblock \showarticletitle{A formal model of {Bitcoin} transactions}. In
  \bibinfo{booktitle}{\emph{{Financial} Cryptography}}
  \emph{(\bibinfo{series}{LNCS}, Vol.~\bibinfo{volume}{10957})}.
  \bibinfo{publisher}{Springer}, \bibinfo{pages}{541--560}.
\newblock
\urldef\tempurl%
\url{https://doi.org/10.1007/978-3-662-58387-6}
\showDOI{\tempurl}


\bibitem[Babel et~al\mbox{.}(2023)]%
        {Babel23clockwork}
\bibfield{author}{\bibinfo{person}{K. Babel}, \bibinfo{person}{P. Daian},
  \bibinfo{person}{M. Kelkar}, {and} \bibinfo{person}{A. Juels}.}
  \bibinfo{year}{2023}\natexlab{}.
\newblock \showarticletitle{Clockwork Finance: Automated Analysis of Economic
  Security in Smart Contracts}. In \bibinfo{booktitle}{\emph{IEEE Symposium on
  Security and Privacy}}. \bibinfo{publisher}{IEEE Computer Society},
  \bibinfo{pages}{622--639}.
\newblock
\urldef\tempurl%
\url{https://doi.org/10.1109/SP46215.2023.00036}
\showDOI{\tempurl}


\bibitem[Bartoletti et~al\mbox{.}(2025)]%
        {Rosetta25fgcs}
\bibfield{author}{\bibinfo{person}{Massimo Bartoletti},
  \bibinfo{person}{Lorenzo Benetollo}, \bibinfo{person}{Michele Bugliesi},
  \bibinfo{person}{Silvia Crafa}, \bibinfo{person}{Giacomo~Dal Sasso},
  \bibinfo{person}{Roberto Pettinau}, \bibinfo{person}{Andrea Pinna},
  \bibinfo{person}{Mattia Piras}, \bibinfo{person}{Sabina Rossi},
  \bibinfo{person}{Stefano Salis}, \bibinfo{person}{Alvise Span{\`{o}}},
  \bibinfo{person}{Viacheslav Tkachenko}, \bibinfo{person}{Roberto Tonelli},
  {and} \bibinfo{person}{Roberto Zunino}.} \bibinfo{year}{2025}\natexlab{}.
\newblock \showarticletitle{Smart contract languages: {A} comparative
  analysis}.
\newblock \bibinfo{journal}{\emph{Future Gener. Comput. Syst.}}
  \bibinfo{volume}{164} (\bibinfo{year}{2025}), \bibinfo{pages}{107563}.
\newblock
\urldef\tempurl%
\url{https://doi.org/10.1016/J.FUTURE.2024.107563}
\showDOI{\tempurl}


\bibitem[Bartoletti et~al\mbox{.}(2021a)]%
        {BGM21lmcs}
\bibfield{author}{\bibinfo{person}{Massimo Bartoletti},
  \bibinfo{person}{Letterio Galletta}, {and} \bibinfo{person}{Maurizio
  Murgia}.} \bibinfo{year}{2021}\natexlab{a}.
\newblock \showarticletitle{A theory of transaction parallelism in
  blockchains}.
\newblock \bibinfo{journal}{\emph{Log. Methods Comput. Sci.}}
  \bibinfo{volume}{17}, \bibinfo{number}{4} (\bibinfo{year}{2021}).
\newblock
\urldef\tempurl%
\url{https://doi.org/10.46298/LMCS-17(4:10)2021}
\showDOI{\tempurl}


\bibitem[Bartoletti et~al\mbox{.}(2020)]%
        {BLZ20isola}
\bibfield{author}{\bibinfo{person}{Massimo Bartoletti},
  \bibinfo{person}{Stefano Lande}, {and} \bibinfo{person}{Roberto Zunino}.}
  \bibinfo{year}{2020}\natexlab{}.
\newblock \showarticletitle{Bitcoin Covenants Unchained}. In
  \bibinfo{booktitle}{\emph{{ISoLA}}} \emph{(\bibinfo{series}{LNCS},
  Vol.~\bibinfo{volume}{12478})}. \bibinfo{publisher}{Springer},
  \bibinfo{pages}{25--42}.
\newblock
\urldef\tempurl%
\url{https://doi.org/10.1007/978-3-030-61467-6\_3}
\showDOI{\tempurl}


\bibitem[Bartoletti et~al\mbox{.}(2021b)]%
        {BLZ21csf}
\bibfield{author}{\bibinfo{person}{Massimo Bartoletti},
  \bibinfo{person}{Stefano Lande}, {and} \bibinfo{person}{Roberto Zunino}.}
  \bibinfo{year}{2021}\natexlab{b}.
\newblock \showarticletitle{Computationally sound {Bitcoin} tokens}. In
  \bibinfo{booktitle}{\emph{{IEEE} Computer Security Foundations Symposium
  ({CSF})}}. \bibinfo{pages}{1--15}.
\newblock
\urldef\tempurl%
\url{https://doi.org/10.1109/CSF51468.2021.00022}
\showDOI{\tempurl}


\bibitem[Bartoletti et~al\mbox{.}(2024)]%
        {BMZ23ieeesp}
\bibfield{author}{\bibinfo{person}{Massimo Bartoletti},
  \bibinfo{person}{Riccardo Marchesin}, {and} \bibinfo{person}{Roberto
  Zunino}.} \bibinfo{year}{2024}\natexlab{}.
\newblock \showarticletitle{Secure compilation of rich smart contracts on poor
  {UTXO} blockchains}. In \bibinfo{booktitle}{\emph{{IEEE} {European}
  {Symposium} on {Security} and {Privacy}}}.
\newblock


\bibitem[Ben~Sasson et~al\mbox{.}(2014)]%
        {Sasson14Zerocash}
\bibfield{author}{\bibinfo{person}{Eli Ben~Sasson}, \bibinfo{person}{Alessandro
  Chiesa}, \bibinfo{person}{Christina Garman}, \bibinfo{person}{Matthew Green},
  \bibinfo{person}{Ian Miers}, \bibinfo{person}{Eran Tromer}, {and}
  \bibinfo{person}{Madars Virza}.} \bibinfo{year}{2014}\natexlab{}.
\newblock \showarticletitle{Zerocash: Decentralized Anonymous Payments from
  Bitcoin}. In \bibinfo{booktitle}{\emph{2014 IEEE Symposium on Security and
  Privacy}}. \bibinfo{pages}{459--474}.
\newblock
\urldef\tempurl%
\url{https://doi.org/10.1109/SP.2014.36}
\showDOI{\tempurl}


\bibitem[Best and Devillers(1987)]%
        {Best87tcs}
\bibfield{author}{\bibinfo{person}{Eike Best} {and} \bibinfo{person}{Raymond~R.
  Devillers}.} \bibinfo{year}{1987}\natexlab{}.
\newblock \showarticletitle{Sequential and Concurrent Behaviour in Petri Net
  Theory}.
\newblock \bibinfo{journal}{\emph{Theoretical Computer Science}}
  \bibinfo{volume}{55}, \bibinfo{number}{1} (\bibinfo{year}{1987}),
  \bibinfo{pages}{87--136}.
\newblock
\urldef\tempurl%
\url{https://doi.org/10.1016/0304-3975(87)90090-9}
\showDOI{\tempurl}


\bibitem[Bowe et~al\mbox{.}(2020)]%
        {Bowe20sp}
\bibfield{author}{\bibinfo{person}{Sean Bowe}, \bibinfo{person}{Alessandro
  Chiesa}, \bibinfo{person}{Matthew Green}, \bibinfo{person}{Ian Miers},
  \bibinfo{person}{Pratyush Mishra}, {and} \bibinfo{person}{Howard Wu}.}
  \bibinfo{year}{2020}\natexlab{}.
\newblock \showarticletitle{{ZEXE:} Enabling Decentralized Private
  Computation}. In \bibinfo{booktitle}{\emph{2020 {IEEE} Symposium on Security
  and Privacy}}. \bibinfo{publisher}{{IEEE}}, \bibinfo{pages}{947--964}.
\newblock
\urldef\tempurl%
\url{https://doi.org/10.1109/SP40000.2020.00050}
\showDOI{\tempurl}


\bibitem[Br{\"{u}}njes and Gabbay(2020)]%
        {Brunjes20isola}
\bibfield{author}{\bibinfo{person}{Lars Br{\"{u}}njes} {and}
  \bibinfo{person}{Murdoch~James Gabbay}.} \bibinfo{year}{2020}\natexlab{}.
\newblock \showarticletitle{{UTxO-} vs Account-Based Smart Contract Blockchain
  Programming Paradigms}. In \bibinfo{booktitle}{\emph{ISoLA}}
  \emph{(\bibinfo{series}{LNCS}, Vol.~\bibinfo{volume}{12478})}.
  \bibinfo{publisher}{Springer}, \bibinfo{pages}{73--88}.
\newblock
\urldef\tempurl%
\url{https://doi.org/10.1007/978-3-030-61467-6\_6}
\showDOI{\tempurl}


\bibitem[Cardano(2022)]%
        {cardanoeutxo}
\bibfield{author}{\bibinfo{person}{Cardano}.} \bibinfo{year}{2022}\natexlab{}.
\newblock \bibinfo{title}{{EUTXO} Handbook}.
\newblock
  \bibinfo{howpublished}{\url{https://ucarecdn.com/3da33f2f-73ac-4c9b-844b-f215dcce0628/EUTXOhandbook_for_EC.pdf}}.
\newblock


\bibitem[Chakravarty et~al\mbox{.}(2020)]%
        {Chakravarty20wtsc}
\bibfield{author}{\bibinfo{person}{Manuel M.~T. Chakravarty},
  \bibinfo{person}{James Chapman}, \bibinfo{person}{Kenneth MacKenzie},
  \bibinfo{person}{Orestis Melkonian}, \bibinfo{person}{Michael~Peyton Jones},
  {and} \bibinfo{person}{Philip Wadler}.} \bibinfo{year}{2020}\natexlab{}.
\newblock \showarticletitle{The Extended {UTXO} Model}. In
  \bibinfo{booktitle}{\emph{Financial Cryptography and Data Security
  Workshops}} \emph{(\bibinfo{series}{LNCS}, Vol.~\bibinfo{volume}{12063})}.
  \bibinfo{publisher}{Springer}, \bibinfo{pages}{525--539}.
\newblock
\urldef\tempurl%
\url{https://doi.org/10.1007/978-3-030-54455-3\_37}
\showDOI{\tempurl}


\bibitem[Dickerson et~al\mbox{.}(2017)]%
        {Dickerson17podc}
\bibfield{author}{\bibinfo{person}{Thomas~D. Dickerson}, \bibinfo{person}{Paul
  Gazzillo}, \bibinfo{person}{Maurice Herlihy}, {and} \bibinfo{person}{Eric
  Koskinen}.} \bibinfo{year}{2017}\natexlab{}.
\newblock \showarticletitle{Adding Concurrency to Smart Contracts}. In
  \bibinfo{booktitle}{\emph{{ACM} {Symposium} on {Principles} of {Distributed}
  {Computing} ({PODC})}}. \bibinfo{publisher}{ACM}, \bibinfo{pages}{303--312}.
\newblock
\urldef\tempurl%
\url{https://doi.org/10.1145/3087801.3087835}
\showDOI{\tempurl}


\bibitem[Fletcher-Smith and Sallal(2023)]%
        {Fletcher-Smith23cnis}
\bibfield{author}{\bibinfo{person}{Cayo Fletcher-Smith} {and}
  \bibinfo{person}{Muntadher Sallal}.} \bibinfo{year}{2023}\natexlab{}.
\newblock \showarticletitle{Security Analysis of Blockchain Layer-One Sharding
  Based {Extended-UTxO} Model}. In \bibinfo{booktitle}{\emph{Communications,
  Networking, and Information Systems}}. \bibinfo{publisher}{Springer},
  \bibinfo{pages}{95--123}.
\newblock


\bibitem[Garamv{\"{o}}lgyi et~al\mbox{.}(2022)]%
        {Garamvolgyi22icse}
\bibfield{author}{\bibinfo{person}{P{\'{e}}ter Garamv{\"{o}}lgyi},
  \bibinfo{person}{Yuxi Liu}, \bibinfo{person}{Dong Zhou}, \bibinfo{person}{Fan
  Long}, {and} \bibinfo{person}{Ming Wu}.} \bibinfo{year}{2022}\natexlab{}.
\newblock \showarticletitle{Utilizing Parallelism in Smart Contracts on
  Decentralized Blockchains by Taming Application-Inherent Conflicts}. In
  \bibinfo{booktitle}{\emph{{IEEE/ACM} International Conference on Software
  Engineering ({ICSE})}}. \bibinfo{publisher}{{ACM}},
  \bibinfo{pages}{2315--2326}.
\newblock
\urldef\tempurl%
\url{https://doi.org/10.1145/3510003.3510086}
\showDOI{\tempurl}


\bibitem[Gelashvili et~al\mbox{.}(2023)]%
        {Gelashvili23ppopp}
\bibfield{author}{\bibinfo{person}{Rati Gelashvili}, \bibinfo{person}{Alexander
  Spiegelman}, \bibinfo{person}{Zhuolun Xiang}, \bibinfo{person}{George
  Danezis}, \bibinfo{person}{Zekun Li}, \bibinfo{person}{Dahlia Malkhi},
  \bibinfo{person}{Yu Xia}, {and} \bibinfo{person}{Runtian Zhou}.}
  \bibinfo{year}{2023}\natexlab{}.
\newblock \showarticletitle{{Block-STM}: Scaling Blockchain Execution by
  Turning Ordering Curse to a Performance Blessing}. In
  \bibinfo{booktitle}{\emph{{ACM} {SIGPLAN} Annual Symposium on Principles and
  Practice of Parallel Programming ({PPoPP})}}. \bibinfo{publisher}{{ACM}},
  \bibinfo{pages}{232--244}.
\newblock
\urldef\tempurl%
\url{https://doi.org/10.1145/3572848.3577524}
\showDOI{\tempurl}


\bibitem[Hammond(2022)]%
        {CardanoFeeEstimator}
\bibfield{author}{\bibinfo{person}{Kevin Hammond}.}
  \bibinfo{year}{2022}\natexlab{}.
\newblock \bibinfo{title}{{Plutus} fee estimator: find out the cost of
  transacting on {Cardano}}.
\newblock
  \bibinfo{howpublished}{\url{https://iohk.io/en/blog/posts/2022/01/21/plutus-fee-estimator-find-out-the-cost-of-transacting-on-cardano/}}.
\newblock


\bibitem[Hoffman(2023)]%
        {dnssec}
\bibfield{author}{\bibinfo{person}{Paul~E. Hoffman}.}
  \bibinfo{year}{2023}\natexlab{}.
\newblock \bibinfo{title}{{DNS Security Extensions (DNSSEC)}}.
\newblock \bibinfo{howpublished}{RFC 9364}.
\newblock
\urldef\tempurl%
\url{https://doi.org/10.17487/RFC9364}
\showDOI{\tempurl}


\bibitem[{IOHK}(2022)]%
        {cardano-utxocongestion}
\bibfield{author}{\bibinfo{person}{{IOHK}}.} \bibinfo{year}{2022}\natexlab{}.
\newblock \bibinfo{title}{How to write a scalable {Plutus} app -- {UTXO}
  congestion}.
\newblock
  \bibinfo{howpublished}{\url{https://plutus-apps.readthedocs.io/en/latest/plutus/howtos/writing-a-scalable-app.html\#utxo-congestion}}.
\newblock


\bibitem[Jones(2021a)]%
        {CIP32}
\bibfield{author}{\bibinfo{person}{Michael~Peyton Jones}.}
  \bibinfo{year}{2021}\natexlab{a}.
\newblock \bibinfo{title}{Inline datums}.
\newblock
\newblock
\newblock
\shownote{\url{https://cips.cardano.org/cip/CIP-32}}.


\bibitem[Jones(2021b)]%
        {CIP31}
\bibfield{author}{\bibinfo{person}{Michael~Peyton Jones}.}
  \bibinfo{year}{2021}\natexlab{b}.
\newblock \bibinfo{title}{Reference inputs}.
\newblock
\newblock
\newblock
\shownote{\url{https://cips.cardano.org/cip/CIP-31}}.


\bibitem[Knispel et~al\mbox{.}(2024)]%
        {Knispel24fmbc}
\bibfield{author}{\bibinfo{person}{Andre Knispel}, \bibinfo{person}{James
  Chapman}, \bibinfo{person}{Joosep Jääger}, \bibinfo{person}{Ulf Norell},
  \bibinfo{person}{Orestis Melkonian}, \bibinfo{person}{Alasdair Hill}, {and}
  \bibinfo{person}{William DeMeo}.} \bibinfo{year}{2024}\natexlab{}.
\newblock \showarticletitle{Formal specification of the {Cardano} blockchain
  ledger, mechanized in {Agda}}. In \bibinfo{booktitle}{\emph{Workshop on
  Formal Methods for Blockchains ({FMBC})}}.
\newblock


\bibitem[Luu et~al\mbox{.}(2016)]%
        {Luu16sharding}
\bibfield{author}{\bibinfo{person}{Loi Luu}, \bibinfo{person}{Viswesh
  Narayanan}, \bibinfo{person}{Chaodong Zheng}, \bibinfo{person}{Kunal Baweja},
  \bibinfo{person}{Seth Gilbert}, {and} \bibinfo{person}{Prateek Saxena}.}
  \bibinfo{year}{2016}\natexlab{}.
\newblock \showarticletitle{A Secure Sharding Protocol For Open Blockchains}.
  In \bibinfo{booktitle}{\emph{{ACM} {SIGSAC} Conference on Computer and
  Communications Security}}. \bibinfo{publisher}{{ACM}},
  \bibinfo{pages}{17--30}.
\newblock
\urldef\tempurl%
\url{https://doi.org/10.1145/2976749.2978389}
\showDOI{\tempurl}


\bibitem[M{\"o}ser et~al\mbox{.}(2016)]%
        {Moser16bw}
\bibfield{author}{\bibinfo{person}{Malte M{\"o}ser}, \bibinfo{person}{Ittay
  Eyal}, {and} \bibinfo{person}{Emin~G{\"u}n Sirer}.}
  \bibinfo{year}{2016}\natexlab{}.
\newblock \showarticletitle{{Bitcoin} covenants}. In
  \bibinfo{booktitle}{\emph{Financial Cryptography Workshops}}
  \emph{(\bibinfo{series}{LNCS}, Vol.~\bibinfo{volume}{9604})}.
  \bibinfo{publisher}{Springer}, \bibinfo{pages}{126--141}.
\newblock
\urldef\tempurl%
\url{https://doi.org/10.1007/978-3-662-53357-4\_9}
\showDOI{\tempurl}


\bibitem[M{\"{u}}ller et~al\mbox{.}(2023)]%
        {Muller23dltrp}
\bibfield{author}{\bibinfo{person}{Sebastian M{\"{u}}ller},
  \bibinfo{person}{Andreas Penzkofer}, \bibinfo{person}{Nikita Polyanskii},
  \bibinfo{person}{Jonas Theis}, \bibinfo{person}{William Sanders}, {and}
  \bibinfo{person}{Hans Moog}.} \bibinfo{year}{2023}\natexlab{}.
\newblock \showarticletitle{Reality-based {UTXO} Ledger}.
\newblock \bibinfo{journal}{\emph{Distributed Ledger Technol. Res. Pract.}}
  \bibinfo{volume}{2}, \bibinfo{number}{3} (\bibinfo{year}{2023}),
  \bibinfo{pages}{1--33}.
\newblock
\urldef\tempurl%
\url{https://doi.org/10.1145/3616022}
\showDOI{\tempurl}


\bibitem[O’Connor and Piekarska(2017)]%
        {Oconnor17bw}
\bibfield{author}{\bibinfo{person}{Russell O’Connor} {and}
  \bibinfo{person}{Marta Piekarska}.} \bibinfo{year}{2017}\natexlab{}.
\newblock \showarticletitle{Enhancing {Bitcoin} transactions with covenants}.
  In \bibinfo{booktitle}{\emph{Financial Cryptography Workshops}}
  \emph{(\bibinfo{series}{LNCS}, Vol.~\bibinfo{volume}{10323})}.
  \bibinfo{publisher}{Springer}.
\newblock
\urldef\tempurl%
\url{https://doi.org/10.1007/978-3-319-70278-0\_12}
\showDOI{\tempurl}


\bibitem[Pieter~Wuille(2020)]%
        {BIP341}
\bibfield{author}{\bibinfo{person}{Anthony~Towns Pieter~Wuille, Jonas~Nick}.}
  \bibinfo{year}{2020}\natexlab{}.
\newblock \bibinfo{title}{{Taproot: SegWit version 1 spending rules}}.
\newblock
\newblock
\newblock
\shownote{{BIP} 341,
  \url{https://github.com/bitcoin/bips/blob/master/bip-0341.mediawiki}}.


\bibitem[P{\^{\i}}rlea et~al\mbox{.}(2021)]%
        {Pirlea21pldi}
\bibfield{author}{\bibinfo{person}{George P{\^{\i}}rlea},
  \bibinfo{person}{Amrit Kumar}, {and} \bibinfo{person}{Ilya Sergey}.}
  \bibinfo{year}{2021}\natexlab{}.
\newblock \showarticletitle{Practical smart contract sharding with ownership
  and commutativity analysis}. In \bibinfo{booktitle}{\emph{{ACM} {SIGPLAN}
  International Conference on Programming Language Design and Implementation
  ({PLDI})}}. \bibinfo{publisher}{{ACM}}, \bibinfo{pages}{1327--1341}.
\newblock
\urldef\tempurl%
\url{https://doi.org/10.1145/3453483.3454112}
\showDOI{\tempurl}


\bibitem[Ramseyer and Mazi{\`{e}}res(2024)]%
        {Ramseyer24groundhog}
\bibfield{author}{\bibinfo{person}{Geoffrey Ramseyer} {and}
  \bibinfo{person}{David Mazi{\`{e}}res}.} \bibinfo{year}{2024}\natexlab{}.
\newblock \showarticletitle{Groundhog: Linearly-Scalable Smart Contracting via
  Commutative Transaction Semantics}.
\newblock \bibinfo{journal}{\emph{CoRR}}  \bibinfo{volume}{abs/2404.03201}
  (\bibinfo{year}{2024}).
\newblock
\urldef\tempurl%
\url{https://doi.org/10.48550/ARXIV.2404.03201}
\showDOI{\tempurl}
\showeprint[arXiv]{2404.03201}


\bibitem[Sergey et~al\mbox{.}(2019)]%
        {Sergey19pacmpl}
\bibfield{author}{\bibinfo{person}{Ilya Sergey}, \bibinfo{person}{Vaivaswatha
  Nagaraj}, \bibinfo{person}{Jacob Johannsen}, \bibinfo{person}{Amrit Kumar},
  \bibinfo{person}{Anton Trunov}, {and} \bibinfo{person}{Ken Chan~Guan Hao}.}
  \bibinfo{year}{2019}\natexlab{}.
\newblock \showarticletitle{Safer smart contract programming with {Scilla}}.
\newblock \bibinfo{journal}{\emph{Proc. {ACM} Program. Lang.}}
  \bibinfo{volume}{3}, \bibinfo{number}{{OOPSLA}} (\bibinfo{year}{2019}),
  \bibinfo{pages}{185:1--185:30}.
\newblock
\urldef\tempurl%
\url{https://doi.org/10.1145/3360611}
\showDOI{\tempurl}


\bibitem[{Solana Foundation}(2024)]%
        {solana-transactions}
\bibfield{author}{\bibinfo{person}{{Solana Foundation}}.}
  \bibinfo{year}{2024}\natexlab{}.
\newblock \bibinfo{title}{{Solana} documentation: Transactions and
  Instructions}.
\newblock
\newblock
\newblock
\shownote{\url{https://solana.com/docs/core/transactions\#array-of-account-addresses}}.


\bibitem[{SUI Foundation}(2024)]%
        {sui-parallelization}
\bibfield{author}{\bibinfo{person}{{SUI Foundation}}.}
  \bibinfo{year}{2024}\natexlab{}.
\newblock \bibinfo{title}{All About Parallelization}.
\newblock
  \bibinfo{howpublished}{\url{https://blog.sui.io/parallelization-explained}}.
\newblock


\bibitem[Torres et~al\mbox{.}(2021)]%
        {Torres21frontrunner}
\bibfield{author}{\bibinfo{person}{Christof~Ferreira Torres},
  \bibinfo{person}{Ramiro Camino}, {and} \bibinfo{person}{Radu State}.}
  \bibinfo{year}{2021}\natexlab{}.
\newblock \showarticletitle{Frontrunner {Jones} and the {Raiders} of the {Dark}
  {Forest}: An Empirical Study of Frontrunning on the {Ethereum} Blockchain}.
  In \bibinfo{booktitle}{\emph{{USENIX} Security Symposium}}.
  \bibinfo{pages}{1343--1359}.
\newblock


\bibitem[Vinogradova and Melkonian(2024)]%
        {Vinogradova24fc}
\bibfield{author}{\bibinfo{person}{Polina Vinogradova} {and}
  \bibinfo{person}{Orestis Melkonian}.} \bibinfo{year}{2024}\natexlab{}.
\newblock \showarticletitle{Message-Passing in the {Extended} {UTxO} Ledger}.
  In \bibinfo{booktitle}{\emph{Financial Cryptography Workshops}}
  \emph{(\bibinfo{series}{LNCS}, Vol.~\bibinfo{volume}{14746})}.
  \bibinfo{publisher}{Springer}, \bibinfo{pages}{150--169}.
\newblock
\urldef\tempurl%
\url{https://doi.org/10.1007/978-3-031-69231-4\_11}
\showDOI{\tempurl}


\bibitem[Xu et~al\mbox{.}(2021)]%
        {Cheng21pvldb}
\bibfield{author}{\bibinfo{person}{Cheng Xu}, \bibinfo{person}{Ce Zhang},
  \bibinfo{person}{Jianliang Xu}, {and} \bibinfo{person}{Jian Pei}.}
  \bibinfo{year}{2021}\natexlab{}.
\newblock \showarticletitle{{SlimChain}: Scaling Blockchain Transactions
  through Off-Chain Storage and Parallel Processing}.
\newblock \bibinfo{journal}{\emph{Proc. {VLDB} Endow.}} \bibinfo{volume}{14},
  \bibinfo{number}{11} (\bibinfo{year}{2021}), \bibinfo{pages}{2314--2326}.
\newblock
\urldef\tempurl%
\url{https://doi.org/10.14778/3476249.3476283}
\showDOI{\tempurl}


\bibitem[Yakovenko(2017)]%
        {solana-white-paper}
\bibfield{author}{\bibinfo{person}{Anatoly Yakovenko}.}
  \bibinfo{year}{2017}\natexlab{}.
\newblock \showarticletitle{Solana: A new architecture for a high performance
  blockchain v0.8.13}.
\newblock  (\bibinfo{year}{2017}).
\newblock
\urldef\tempurl%
\url{https://github.com/solana-labs/whitepaper/blob/master/solana-whitepaper-en.pdf}
\showURL{%
\tempurl}


\bibitem[Yakovenko(2019)]%
        {solana-parallelization}
\bibfield{author}{\bibinfo{person}{Anatoly Yakovenko}.}
  \bibinfo{year}{2019}\natexlab{}.
\newblock \bibinfo{title}{Sealevel - Parallel Processing Thousands of Smart
  Contracts}.
\newblock
  \bibinfo{howpublished}{\url{https://medium.com/solana-labs/sealevel-parallel-processing-thousands-of-smart-contracts-d814b378192}}.
\newblock


\bibitem[Zhang et~al\mbox{.}(2023)]%
        {Zhang23icpp}
\bibfield{author}{\bibinfo{person}{Haowen Zhang}, \bibinfo{person}{Jing Li},
  \bibinfo{person}{He Zhao}, \bibinfo{person}{Tong Zhou},
  \bibinfo{person}{Nianzu Sheng}, {and} \bibinfo{person}{Hengyu Pan}.}
  \bibinfo{year}{2023}\natexlab{}.
\newblock \showarticletitle{{BlockPilot}: {A} Proposer-Validator Parallel
  Execution Framework for Blockchain}. In
  \bibinfo{booktitle}{\emph{International Conference on Parallel Processing
  ({ICPP})}}. \bibinfo{publisher}{{ACM}}, \bibinfo{pages}{193--202}.
\newblock
\urldef\tempurl%
\url{https://doi.org/10.1145/3605573.3605621}
\showDOI{\tempurl}


\end{thebibliography}

\appendix

\section{Experimental evaluation: supplementary material}
\label{sec:app-experiments}

To make our experimental evaluation realistic, we use the same cryptographic primitives as Cardano, \ie 
BLAKE2b-512 for hashing, and Ed25519 for signatures. 
For simplicity, our simulator diverges from Cardano in some aspects, which however only affect the results of the experiments in a marginal way. 
First, \toolname only focusses on the execution of contracts, and does not include those functionalities of validators that are orthogonal to that aim (\eg, the consensus protocol).
Furthermore, we simplify the generation of transactions in our experiments:
\begin{itemize}
\item First, we use sequential numbers instead of hashes as transaction IDs, since it makes it easier to generate chains of transactions that refer to each other.
The actual cryptographic hash function is still used for all other purposes, including reading and updating the contract state as explained in~\Cref{sec:compilation}, and in particular in~\Cref{eq:sigma-var-definition,eq:sigma-map-definition} which define the expected hashes that occur in state outputs.
\item Second, we do not actually sign the transactions: instead, when a signature on a transaction would need to be checked, \toolname verifies a fixed signature on a fixed message. We ensure that this step is not optimized away when compiling our tool, so not to invalidate the performance measurements.
\end{itemize}


\paragraph{Crowdfund}

The first use case is the crowdfund contract of~\Cref{sec:overview}.
The experiment compares the centralized implementation of this contract (\ie, the one that keeps the whole donors map in the $\txf{datum}$ field, copying it each time a new donor is added) \wrt a distributed implementation that uses the technique of~\Cref{sec:compilation} to distribute the state across multiple outputs.


From~\Cref{tab:cf} we observe that the centralized contract executed by the parallel validator gets slower as the number of threads increases. This trend was expected, since the transactions in the centralized contract form a long chain where each one redeems the output of the previous one, preventing parallelization because of such a large number of soft conflicts. 
Increasing the number of threads only increases the contention between them.

The distributed contract instead benefits from the parallel validator, which becomes faster with more threads. Indeed, the experiment features many non-conflicting transactions since the state is distributed, allowing the parallel validator to exploit its threads. 
The percentage of (soft) conflicts, which only depends on the number of users,
drops steeply from $18.23\%$ in the experiments with 250 users, to $1.05\%$ in those with 50K users.
%
These conflicts are mostly due to the splitting and merging of map intervals (as per~\Cref{subsec:state-updates}) that occur during the $\ruleinline{donate}$ and $\ruleinline{refund}$ operations.
Effectively, these soft conflicts are caused by the implicit data dependencies as discussed in~\Cref{sec:parallelization}.

%
\Cref{fig:cf} (right) shows, for the distributed contract, the time speedup of the parallel validator (with 1+7 threads) over the sequential one. The speedup grows with the number of users, being only bounded by the number of available threads.
For instance, we have a $6.13$ speedup for $20K$ users and a $7.01$ speedup for $50K$ users.
This increase in the speedup is coherent with the reduced percentage of conflicts.


\paragraph{Map}

This use case involves a basic contract that stores a map (see~\Cref{sec:experiments}).
The contract has a single rule, which allows anyone to update the value of the map at any given key.
This allows us experiment with different rates of soft conflicts: we can perform a bunch of updates on completely distinct keys ($0\%$ soft conflicts), we can update the map always on the same key ($100\%$ soft conflicts), or fine-tune the conflict probability to any intermediate level.
In this way, we aim to measure how performance is affected by the presence of soft conflicts.

The goal here is to study the performance of the distributed Map contract under a varying conflicts probability.
Given a wanted probability $p$ of soft conflicts,
we construct a sequence of transactions~$\vec{T}$ with $50K$ \emph{random} calls of the rule $\ruleinline{inc}$, each time determining the actual parameter based on the toss of a biased coin. 
With probability $p$, we update $\ruleinline{m[0]}$;
with probability $1-p$, we update $\ruleinline{m[i]}$ where the index $i$ is incremented each time.
In this way, we repeatedly update the same map location $\ruleinline{m[0]}$ with probability $p$, ensuring a soft conflict each time.
By contrast, the other map updates access distinct locations $\ruleinline{m[i]}$, which is very unlikely to cause conflicts.
Indeed, in such cases a conflict happens only because of the implicit data dependencies discussed in~\Cref{sec:parallelization}, when 
two accesses $\ruleinline{m[i]}$ and $\ruleinline{m[j]}$ need to split the same interval.
Since the number of intervals increases after each $\ruleinline{m[i]}$ update, the probability of such conflicts quickly becomes very small, making the effective conflict probability be close to the wanted parameter $p$.

\Cref{fig:map} displays the results of our experiment with varying values of the conflict probability $p$ and with different validators.
We observe that the parallel validator is generally faster than the sequential one when using at least $2$ worker threads.
The sequential validator becomes faster when $p > 54\%$ (dotted vertical line in~\Cref{fig:map}), \ie when we have a very high conflict probability.
This corresponds to the highly unrealistic scenario where more than half of the transactions in the blockchain update the same map location. 
%
%
When $p$ assumes more realistic values such as $p < 15\%$, we observe a $1.71$ speedup comparing the sequential validator to the parallel one ($1+3$ threads).


\paragraph{Multisig wallet}

This use case implements a multisig wallet involving $N$ privileged users approved by the owner, where token transfers must be authorized by $K = N/2$ of them.
The contract (see~\Cref{fig:multisig-code}) has three rules.
The first rule allows the contract owner to add authorized users to the wallet.
The second rule allows anyone to deposit tokens within the wallet, without requiring any authorization.
The third rule allows to withdraw tokens from the contract: this is possible only when $K$ authorized users approve the withdrawal by signing the corresponding transaction.
The goal of this experiment is to measure the the impact of signatures, one of the most computationally expensive primitives of the blockchain.

We craft an experiment for the distributed version of the contract, varying the number $N$ of privileged users ($N = 2,4,6,10$) and the number of threads available to the validator.
In each experiment, the sequence of transactions $\vec{T}$ starts with the owner authorizing $N$ users, and then proceeds with $100K$ random operations. 
When the wallet holds no tokens, we perform a $\ruleinline{deposit}$ operation adding one token.
Otherwise, we flip an unbiased coin, and perform either 
a $\ruleinline{deposit}$ or a $\ruleinline{withdraw}$, both involving one token.
Each $\ruleinline{withdraw}$ operation is signed by $K=N/2$ users, so that it always succeeds. Consequently, to be executed it requires verifying $N/2$ signatures, plus an extra one to pay the transaction fees.
By contrast, $\ruleinline{authorize}$ operations only require verifying $2$ signatures (the owner's and the one for the fees), while $\ruleinline{deposit}$ operations require $2$ signatures (one for the deposited token and the one for the fees).
Overall, we measured \mbox{$\sim{1.5}\%$} of soft conflicts. This is not affected in any significant way by the number of users $N$, nor by the number of threads.

The results are displayed in~\Cref{fig:multisig}.
As expected, the average running time roughly scales with the average number of verified signatures. Assuming that the number of deposits is close to the number of withdraws, the number of verified signatures for $N$ participants can be estimated as $S(N) = 100K \cdot (0.5 \cdot 2 + 0.5 \cdot (N/2+1)) = 25K \cdot N + 150K$.
Observing the timings in~\Cref{fig:multisig} (left) we see that the time for each $N$ is roughly proportional to $S(N)$, confirming that the cost of signature verification is the main factor affecting performance.
\Cref{fig:multisig} (right) shows the speedups of the parallel validator over the sequential one for a varying number of threads. We observe that the speedup does not depend on the value of $N$ but only on the number of threads, confirming that contracts that make heavy use of signature verification can fully exploit parallel validation.


\paragraph{Registry}

This use case implements a registry contract, which allows users to register the hash of a document on the blockchain, and later to claim its ownership (\Cref{fig:registry-code}).

In each experiment, the input transaction sequence $\vec{T}$ starts with
$N$ distinct users invoking $\ruleinline{register}$ on a document of their own.
The users then perform a $\ruleinline{claim}$ on their documents, wait for the deadlines to expire, and finally 
invoke $\ruleinline{own}$ to assert their ownership on the document they registered.
\Cref{fig:registry} displays the timings and speedup for the distributed version of the contract.
We observe that the parallel validator is significantly faster than the sequential one. 
Using $1+2$ threads, the speedup is always greater than $1$ even for a small number of users (\eg, it $1.47$ is for $N = 250$). 
Using $1+7$ threads, the speedup is always greater than $1$, and it tends asymptotically to the bound $7$ as the number of users grows.
This is due to a low number of soft conflicts (less than $2\%$).

}

\end{document}